\newif\ifarxiv
\arxivtrue   

\documentclass[]{jfm}

\usepackage{graphicx}
\usepackage{epstopdf,epsfig}
\usepackage{floatrow}
\usepackage{subfig}
\usepackage{comment}
\usepackage{newtxtext}
\usepackage{newtxmath}
\usepackage{natbib}
\usepackage{siunitx}
\usepackage{upgreek}
\usepackage{hyperref}
\usepackage{gensymb}
\usepackage{multirow}
\hypersetup{
    colorlinks = true,
    urlcolor   = blue,
    citecolor  = black,
}

\usepackage{xcolor}
\usepackage{amsmath}
\usepackage{cancel}

\newcommand{\RomanNumeralCaps}[1]

\ifarxiv
\makeatletter 

\let\ps@titlepage\ps@plain

\def\@maketitle#1{%
  \newpage
  \thispagestyle{plain}%
  \vspace*{2em}
  {\centering
    {\LARGE\bfseries \@title \par}
    \vskip 1.5em
    {\large \@author \par}
    \vskip 1em
    {\small \@affiliation \par}
  }
  \vskip 2em
}

\makeatother
\fi

\ifarxiv
\makeatletter
\renewcommand{\email}[1]{}
\makeatother
\fi

\ifarxiv
\makeatletter
\renewcommand{\pagelimitfooter}{}
\makeatother
\fi

\title{Distributed Roughness-Induced Transition on a Blunt Body at Mach 6: a Numerical Investigation}

\author{Sean Dungan\aff{1},
  Mateus Braga\aff{2},
  Robyn Macdonald\aff{2}
 \and Christoph Brehm\aff{1}}

\affiliation{\aff{1}Department of Aerospace Engineering, University of Maryland, College Park, MD 20742, USA
\aff{2}Department of Aerospace Engineering Sciences, University of Colorado Boulder, Boulder, CO 80303, USA}

\begin{document}

\maketitle

\begin{abstract}
 Surface roughness significantly impacts transition to turbulence, especially over high-speed, blunt geometries where surface ablation is necessary to mitigate heat loads during atmospheric entry. Inspired by sand-grain roughness experiments performed by \citet{hollis17_NASATM}, we perform the first direct numerical simulation (DNS) of a blunt cylinder in Mach 6 cross-flow with roughness elements distributed along the entire surface. Such simulations aimed to uncover the precise means by which laminar-turbulent transition occurs given the limited measurements attainable from experiments and non-existent high-fidelity simulations. Element heights were held fixed at approximately 35\% boundary layer thickness, while the relative phasing between streamwise rows was varied. All configurations exhibited convective instabilities driving the transition process, with the mode type being set by the roughness configuration. A fundamental sinuous streak mode dominated the aligned roughness element case, whereas both the staggered and randomly phased cases saw 2D T-S waves dominating. These instability waves, when grown to sufficient amplitude, triggered the steady streaks seeded by the underlying roughness pattern to begin forming hairpin vortices and breakdown occurred soon thereafter. The roughness arrangement was found to dramatically influence the degree to which the waves were destabilised, as well as the strength of the underlying steady streaks, thereby combining to dictate the position along the surface where LTT occurred. Finally, exogeneous forcing was not required for the T-S dominated cases as the acoustics generated by the turbulence in the subsonic flow excited T-S waves on the other side -- a feedback mechanism hitherto unknown. 
\end{abstract}

\section{Introduction}
Prediction of laminar-turbulent transition (LTT) of flow over high-speed vehicles persists as an essential research topic due largely to  its implications for engineering design. In addition to the several factors increase of heat transfer and skin-friction drag experienced by the vehicle when subjected to a turbulent boundary layer versus a laminar one, the transition process between the two states often results in large overshoots in both quantities \citep{Hader_flare_2019}. Accurate prediction of heat transfer rates is particularly important for blunt atmospheric (re-)entry vehicles with thermal protection systems (TPS) on their surfaces designed to mitigate heating loads to the payload via surface ablation. Heat transfer rates at the surface dictate the rate of material loss due to ablation and, therefore, determine the TPS size and weight needed for vehicle survivability. The prediction of LTT in this case is challenging due in part to the small-scale (relative to the overall vehicle) surface roughness that can arise during surface ablation, which may be orderly or randomly distributed depending on the TPS material (see figure 2 in \citep{hollis25}). Consequently, direct numerical simulation (DNS) of the problem has remained prohibitively expensive, and the majority of research on roughness-induced LTT on blunt bodies has been experimental in nature.

Several roughness-induced transition correlations were proposed based on ballistic-range and wind-tunnel data for hypersonic blunt geometries, respective examples being \citep{Abbett75,Wool75} and \citep{reda1981}. These and others were reviewed by \citet{Schneider2008}. While representing the trends well within their respective data sets, the ballistic-range based correlation broke down when applied to the wind-tunnel data, and vice versa \citep{hollis19}. This incompatibility was later remedied \citep{hollis19}. Hollis also performed extensive wind-tunnel testing on hemisphere geometries with distributed sand-grain roughness \citep{hollis17_NASATM,hollis19} and spherical and sphere-cone geometries with hexcomb patterned roughness \citep{hollis21}. He formulated improved transition correlations \citep{hollis25} that can handle the resulting wide variety of data on roughness-induced transition on high-speed blunt bodies. These correlations, unfortunately, are purely empirical and lack a clear physical foundation.

Since the smooth wall laminar boundary layers are generally eigenmode stable, \citet{Reshotko2004} considered transient growth as the responsible flow instability for the case of distributed roughness over blunt bodies. The algebraic growth experienced by the coupling of highly oblique slightly damped boundary layer modes was computed therein for a range of self-similar, axisymmetric stagnation point flows and shown to reproduce the trends of \citet{Abbett75, Wool75} and \citet{reda1981}. The relations derived used assumptions of the velocity perturbations introduced by roughness scaling with roughness element height $k$, and transition occurring when that input disturbance energy amplified via transient growth exceeds some constant. It appears that constant was determined by fitting the derived relation to the PANT wind-tunnel database. Velocity perturbations scaling with roughness element height is typically deemed valid in the limit of small roughness height \citep{Choudhari1993}. This does not seem to be the case for many of the correlated data where $k/\theta>1$, $\theta$ being the boundary layer momentum thickness. In addition, transient growth alone as a linear mechanism cannot explain the constant that is needed to be exceeded for transition to occur. Despite this, the Reshotko \& Tumin (RT) correlation is the only mechanism-based correlation for blunt body roughness-induced transition prediction. In fact, several experimental campaigns since the inception of the RT correlation have confirmed its predictability. For example, \citet{Leidy2018} used uniformly distributed sand grit of variable height applied to the surface of an Orion Crew Exploration Vehicle (CEV) model and found that the RT correlation predicted transition locations within $25\%$ of the experimentally determined values. Similar experiments performed by \citet{Radespiel2019} in the Hypersonic Ludweig Tube Braunschweig (HLB) found agreement within $20\%$ of the values predicted with the RT correlation as well.

\citet{Paredes2018} sought to re-examine some of the RT correlation's underlying assumptions for the CEV geometry to see if it was somehow fortuitous in its success. Their improvements included allowing for non-parallel effects on disturbance growth, varying inflow/outflow optimisation locations, spanwise wavelength effects, surface curvature effects, and full Navier-Stokes basic states. These inclusions improved the theoretical foundations of the correlations, however, they were seen to make only a small difference in the final correlation functional form. In a similar previous work, \citet{Paredes2017} computed the optimal disturbances for transient growth on a hypersonic hemisphere. Here they highlight optimising for mean kinetic energy gain rather than total disturbance energy (i.e., including thermodynamic fluctuations) resulted in a shift of optimal energy gain factors from occurring near the stagnation point to closer to the sonic line where the basic state reaches its peak in wall-shear stress. This finding is thought to help explain some of the behaviour observed in experiments \citep{Kaattari1978} of transition occurring close to the sonic point. There, as in \citep{Paredes2018}, they note the low optimal amplification factors of roughly 10. Combining this knowledge with their previous work on the secondary instability of saturated optimal disturbances \citep{Paredes2017b}, they determined the initial streamwise perturbation introduced by the roughness would need to be larger than 1\% boundary layer edge velocity, $U_{e}$, in order for transient growth to amplify the disturbances past the $16\% U_{e}$ amplitude needed for secondary instabilities to become unstable and transition the flow to turbulence. Such high initial disturbance amplitudes make a linear analysis questionable and the authors point to non-linear effects as likely playing an important role.

Other numerical investigations of roughness-induced transition over high-speed blunt bodies simulated roughness patches. This drastically reduces the computational cost by focusing grid cells only around a limited streamwise extent of the overall domain. Such efforts by \citet{Hein2019} wherein DNS was performed over a hemisphere with a patch of roughness did not reveal evidence of modal or non-modal (i.e., transient growth) disturbances being amplified, therefore transition to turbulence was not observed. A series of investigations \citep{DiGiovanni2018,digiovanni2019,Ulrich2022,Ulrich2023}, analysed roughness patches in the supersonic region of flow over capsule-like geometries. Many interesting findings were discovered therein. However, since transition occurred in the wake downstream over the smooth wall, the implications of these studies to the case of fully distributed roughness over the entire surface where transition usually occurs in the subsonic region is not immediately clear. Another work \citep{Schilden2020} simulated deterministic roughness patches on the same Apollo-like geometry examined by \citet{Radespiel2019}. While transient growth was shown in the patch wake, the gain, $G$, achieved at $G=1.553$ versus $G=66.3$ promised by optimal transient growth theory at the fundamental roughness wavenumber was very suboptimal.

Given the inability of previous work to find a conclusive link between transient growth and roughness-induced transition over blunt bodies with fully distributed surface roughness, other instability mechanisms should be considered. Many previous works have shown mean flow three-dimensionality (e.g., `streaks') to have a significant impact on boundary instabilities \citep{Paredes2019,Caillaud2025,Cossu_2004}. Modal primary instabilities like the Tollmien-Schlichting (T-S) waves of low-speed and Mack modes (MM) of high-speed boundary layers have not been considered for the hypersonic blunt body for several reasons, however. First, the smooth wall laminar flow field over these blunt geometries is, generally, modally stable due to the favourable pressure gradient. Second, strong statements like, `The attempts to find a T-S explanation for three-dimensional roughness, both discrete and distributed, have failed,' made by \citet{Reshotko2004} rule out T-S waves based on previous work in low-speed boundary layers \citep{Reshotko1984,Morkovin1989}. It was shown by \citet{Corke1986}, however, that distributed sand-grain roughness on a flat plate can destabilise T-S waves. Since roughness heights were less than the boundary layer thickness (i.e., $k/\delta<1$) in that work, it may be presumed that below some threshold roughness height the transition is indeed dominated by roughness-destabilised modal instabilities. Such an idea is corroborated by work done more recently (albeit with 2D roughness) that showed significant destabilisation of modal instabilities by particular roughness configurations of height lower than the unperturbed boundary layer thickness in both low-speed \citep{Brehm2013JCP} and high-speed \citep{Saikia_Brehm_2025} boundary layers. Since the data correlated in \citep{hollis19} were for hypersonic hemispheres with distributed sand-grain roughness of size $k/\delta<1$, the present authors believe destabilisation of modal instabilities may be a critical, overlooked instability responsible for roughness-induced transition in this type of flow regime.

The present work seeks to provide new physical understanding of roughness-induced transition on hypersonic blunt bodies by performing the first DNS of a cylinder in Mach 6 free stream with sinusoidal-shaped roughness distributed over the entire surface. This builds on previous work done by the present authors \citep{braga_2024}. We first present the governing equations, numerical approach used to solve them, and simulation set-up in section \ref{sec:comp_approach}. Next, the mean flow field obtained over the roughness is analysed in section \ref{sec:mean_flow}, with careful attention being paid to the generation of streamwise streaks and their downstream development. Unsteadiness in the flow fields obtained from DNS is treated in section \ref{sec:unsteady_flow} and explained with linear stability calculations. Finally, concluding remarks are made and future work is proposed in section \ref{sec:conclusion}.

\section{Computational Approach} \label{sec:comp_approach}

\subsection{Governing Equations}
We consider a single component, ideal gas whose dynamics are governed by the compressible Navier-Stokes equations (CNSE). The CNSE written in conservative form for the Cartesian coordinate system are 

\begin{equation} \label{eq:cnse}
    \frac{\partial \mathbf{W}}{\partial t}+\frac{\partial \mathbf{F}_{i}}{\partial x_{i}} + \frac{\mathbf{F}_{v_{i}}}{\partial x_{i}}=0,
\end{equation}
where $x_i = (x,y,z)$ and a sum over repeated indices is understood. The definition of the conservative state vector $\mathbf{W}$, inviscid flux vector $\mathbf{F}_{i}$, and viscous flux vector $\mathbf{F}_{v_{i}}$ are
\begin{equation} \label{eq:flux_vectors_cnse}
\mathbf{W} =
\renewcommand{\arraystretch}{1.2} 
\begin{bmatrix}
\rho \\
\rho u_{j} \\
\rho E
\end{bmatrix}
;\quad
\mathbf{F}_{i} =
\renewcommand{\arraystretch}{1.2} 
\begin{bmatrix}
\rho u_i \\
\rho u_i u_j + P \delta_{ij} \\
(\rho E+P) u_i
\end{bmatrix}
;\quad
\mathbf{F}_{v_{i}} =
\renewcommand{\arraystretch}{1.2} 
\begin{bmatrix}
0 \\
-\tau_{ij} \\
q_i - \tau_{ij} u_j
\end{bmatrix}.
\end{equation}
In equation \ref{eq:flux_vectors_cnse} $\rho$ is the fluid density, $u_i=(u,v,w)$ are the velocity components in the Cartesian frame, $P$ is the thermodynamic pressure, $E=u_i u_i/2 + R_{gas}T/(\gamma-1)$ is the specific total energy (kinetic plus internal) in the fluid, $q_i$ is the heat flux vector, $\tau_{ij}$ is the stress tensor, $\gamma$ is the ratio of specific heats, $R_{gas}$ is the specific gas constant, and $\delta_{ij}$ is the Kronecker delta. Fourier's law is used to relate the heat flux vector to the temperature, $T$, gradient through the conductivity coefficient, $\kappa$, as: $q_i = -\kappa \partial T / \partial x_i$. Additionally, the shear-stress tensor, $\tau_{ij}$, is related to the velocity gradients as
\begin{equation}
    \tau_{ij} = \mu \left(\frac{\partial u_i}{\partial x_j} + \frac{\partial u_j}{\partial x_i}\right) + \lambda \frac{\partial u_k}{\partial x_k} \delta_{ij}. 
\end{equation}
Stoke's hypothesis is used to relate the second coefficient of viscosity, $\lambda$, to the first, $\mu$, as: $\lambda = -2/3\mu$. Sutherland's law is used to compute the first coefficient of viscosity as function of temperature, then the conductivity coefficient is computed from the definition of the Prandtl number, $\Pran = c_p \mu / \kappa$, assuming a constant $\Pran=0.71$. Since air is the working fluid under consideration, we take the specific heat capacity at constant pressure as $c_p=\qty{1005}{J.kg^{-1}.K^{-1}}$ and $R_{gas}=\qty{287.15}{J.kg^{-1}.K^{-1}}$.
The system of \ref{eq:flux_vectors_cnse} is closed through the ideal gas equation of state
\begin{equation} \label{eq:eos}
      P = \rho R_{gas} T.
\end{equation}
Simulations performed by integrating the discretisation of equations \ref{eq:cnse}-\ref{eq:eos} will be referred to as direct numerical simulation (DNS).

To understand how instabilities arise in the DNS, we also consider infinitesimally small disturbances evolving through the time-averaged flow field. Such dynamics of small disturbances in a compressible, single component, ideal gas are governed by the linear disturbance equations (LDE) written below
\begin{equation} \label{eq:lde}
    \frac{\partial \tilde{\mathbf{W}}}{\partial t}+\frac{\partial \tilde{\mathbf{F}}_{i}}{\partial x_{i}} + \frac{\tilde{\mathbf{F}}_{v_{i}}}{\partial x_{i}}=0,
\end{equation}
\begin{equation} \label{eq:flux_vectors_lde}
\tilde{\mathbf{W}} =
\renewcommand{\arraystretch}{1.2} 
\begin{bmatrix}
\tilde{\rho} \\
\bar{\rho} \tilde{u}_{j} +  \tilde{\rho} \bar{u}_{j}\\
\bar{\rho} \tilde{E} + \tilde{\rho} \bar{E}
\end{bmatrix}
,\quad
\tilde{\mathbf{F}}_{i} =
\renewcommand{\arraystretch}{1.2} 
\begin{bmatrix}
\bar{\rho} \tilde{u}_i + \tilde{\rho} \bar{u}_i \\
\bar{\rho} \bar{u}_i \tilde{u}_j + \bar{\rho} \tilde{u}_i \bar{u}_j + \tilde{\rho} \bar{u}_i \bar{u}_j + \tilde{P} \delta_{ij} \\
(\bar{\rho} \bar{E}+\bar{P}) \tilde{u}_i + (\bar{\rho} \tilde{E}+\tilde{P}) \bar{u}_i
\end{bmatrix}
,\quad
\text{and}
\end{equation}

\begin{equation} \label{eq:visc_flux_vector_lde}
\tilde{\mathbf{F}}_{v_{i}} =
\renewcommand{\arraystretch}{1.2} 
\begin{bmatrix}
0 \\
-\bar{\mu} \tilde{\mathcal{T}}_{ij} + -\tilde{\mu} \bar{\mathcal{T}}_{ij} \\
-\bar{\mu}\bar{\mathcal{T}}_{ij}\tilde{u}_j - \bar{\mu}\tilde{\mathcal{T}}_{ij}\bar{u}_j - \tilde{\mu}\bar{\mathcal{T}}_{ij}\bar{u}_j - \bar{\kappa}\frac{\partial \tilde{T}}{\partial x_i} - \tilde{\kappa}\frac{\partial \bar{T}}{\partial x_i}
\end{bmatrix}.
\end{equation}
Equation \ref{eq:lde}-\ref{eq:visc_flux_vector_lde} are the linearised version of equation \ref{eq:cnse}. They are obtained the typical way via decomposing variables into a mean, $\bar{(.)}$, and fluctuating, $\tilde{(.)}$, component, which is  assumed to be much smaller, i.e., $\phi = \bar{\phi} + \epsilon \tilde{\phi}$, $\epsilon<<1$. Then, the decomposition is substituted into \ref{eq:cnse}, terms purely involving mean variables are subtracted out (i.e., they satisfy \ref{eq:cnse}), and terms of $O(\epsilon)$ are retained. Readers are directed to the work of \citet{Browne2022} for more details. Simulations integrating equations \ref{eq:lde}-\ref{eq:visc_flux_vector_lde} will be referred to as `LDE'.

\subsection{Numerical Approach} \label{subsec:num_approach}
The DNS and LDE simulations presented herein were performed with the CHAMPS solver from the University of Maryland. While the development of relevant aspects of the solver for this work have been detailed elsewhere \citep{mcquaid2021_scitech,Browne2022,McQuaid2024}, the numerical schemes used here will be briefly enumerated for completeness. For the DNS (i.e., solution of \ref{eq:cnse}), the inviscid flux vector was computed using a modified Steger-Warming formulation \citep{McQuaid2024}. A fifth-order accurate, weighted, essentially non-oscillatory (WENO) flux reconstruction \citep{Brehm2015CAF} was used to obtain the inviscid flux on cell faces before differentiation. This high-order flux was blended with a first-order flux near the shock. This was needed to prevent the large aspect ratio cells near the shock from introducing spurious noise. The viscous fluxes were computed to second-order accuracy at cell faces before differentiation. The LDE were solved in a similar fashion in that fluxes were computed at cell faces before differentiation, but this time a linearised Rusanov inviscid flux scheme was deployed (see \citet{Browne2022}). Given the absence of strong shocks within the subdomain considered for the LDE, the optimal fifth-order WENO weights were used to reconstruct the inviscid fluxes at the faces before differentiation. Finally, linearised viscous fluxes were computed to second-order accuracy at the faces before differentiation.

While written in the Cartesian frame for brevity, the DNS and LDE equations were solved on body-conformal grids to facilitate clustering grid cells around the shock and roughness elements. After a careful grid convergence study (see appendix \ref{appB}), the grids used here had $(n_{\xi},n_{\eta},n_z) = (\num{10400},\num{300},\num{120})$ cells in the streamwise, wall-normal, and spanwise directions when both halves of the cylinder were simulated. Naturally, only half the streamwise points were used when only half the cylinder was simulated. This resulted in roughly 21 and 40 cells per roughness element in the streamwise and spanwise directions, respectively.

\subsection{Disturbance Generation} \label{subsec:dist_gen}
Numerical disturbances were introduced in several simulations to better explore the transition scenarios possible for each roughness configuration. A source term was added on the right-hand side of eq. \ref{eq:cnse} and eq. \ref{eq:lde} for DNS and LDE simulations, respectively, and took the form
\begin{equation}
    S(\xi,\eta,z,t) = A S_{\xi} S_{\eta} S_{z} S_{t}.
\end{equation}
Since the purpose of the disturbances differed between the two simulation approaches, so too did their defining shape functions. Applied to the wall-normal momentum equation in eq. \ref{eq:lde}, the LDE utilised the functions:
\begin{align} \label{eq:dist_eq}
    S_{\xi} &= \sin\left(2\pi\frac{\xi-\xi_{0}}{\xi_{1}-\xi_{0}}\right), & S_{\eta} &= \exp\left({-\left(\frac{\eta-\eta_{1/2}}{\sigma_{\eta}}\right)^{2}}\right), \\ S_{z} &= \sin\left(m\beta_{k}\frac{z-z_0}{z_1-z_0} + \phi_z\right),
    & S_{t} &= \exp\left({-\left(\frac{t-t_{1/2}}{\sqrt{2}\sigma_{t}}\right)^{2}}\right).
\end{align}
The `0' and `1' subscripts refer to the beginning and end of the forcing interval in that cylindrical coordinate direction, respectively. The `$1/2$' subscript used in the Gaussian functions is the mid-point of the interval. Everywhere outside the interval we set $S(\xi,\eta,z,t)=0$. Note, the $\phi_z=\pi$ phase shift was needed in the case of spanwise uniform forcing with $m=0$ but was otherwise null. More details on the set of parameters chosen for the LDE simulations are provided in section \ref{sec:lde}. As will be shown later, the aligned roughness case resulted in a more stable mean flow than the other two roughness patterns. Therefore, forcing was added to this case to enable LTT and provide another case for comparison. To make the comparison as fair as possible, the disturbances introduced attempted to mimic the background noise (characterised in section \ref{sec:background_noise}) present in the other two DNS. We took an approach similar to \citet{Hader_Fasel_2018} where grid cells are randomly perturbed to introduce a broad spectrum of disturbances. In this case, perturbations were added to the continuity equation in eq. \ref{eq:cnse} over 20 cells in the streamwise direction starting from $\Theta\approx6\degree$ and 60 cells in the wall-normal direction located near the upper part of the boundary layer. A single random number was drawn from a uniform distribution each simulation timestep and rescaled to be between -1 and 1. The random number was then rescaled by a factor $10^{6}$ and applied uniformly in the spanwise direction. The amplitude chosen aimed to place the LTT location close to those of the staggered and random case.

\subsection{Flow Conditions and Geometry} \label{subsec:conditions_geometry}
The present work models a $\qty{0.1524}{m}$ diameter cylinder in Mach 6 flow. The geometry and flow conditions are selected based on the experimental test campaign from \citet{hollis17_NASATM}, which investigated the effects of distributed surface roughness on boundary layer transition and turbulent heating for a $\qty{0.1524}{m}$ diameter hemisphere with a range of roughness heights and free-stream Reynolds numbers. The present work only considers Run 38 from Test 6975, corresponding to the 80-mesh (nominal sand grain surface roughness element diameter of $\qty{0.18}{mm}$), at high free-stream unit Reynolds number ($\Rey_\infty = \SI{2.74e7}{\per\meter}$). This roughness condition is selected as the primary focus because it corresponds to a moderately dense roughness distribution and the roughness-Reynolds number combination led to transitional flow. A cylindrical geometry was selected to ensure periodicity in the spanwise direction while exploring the impact of phasing of fixed-dimension roughness elements on LTT.

\begin{figure}[h]
    \centering
    \captionsetup{justification=centering}
    \includegraphics[width=0.95\linewidth]{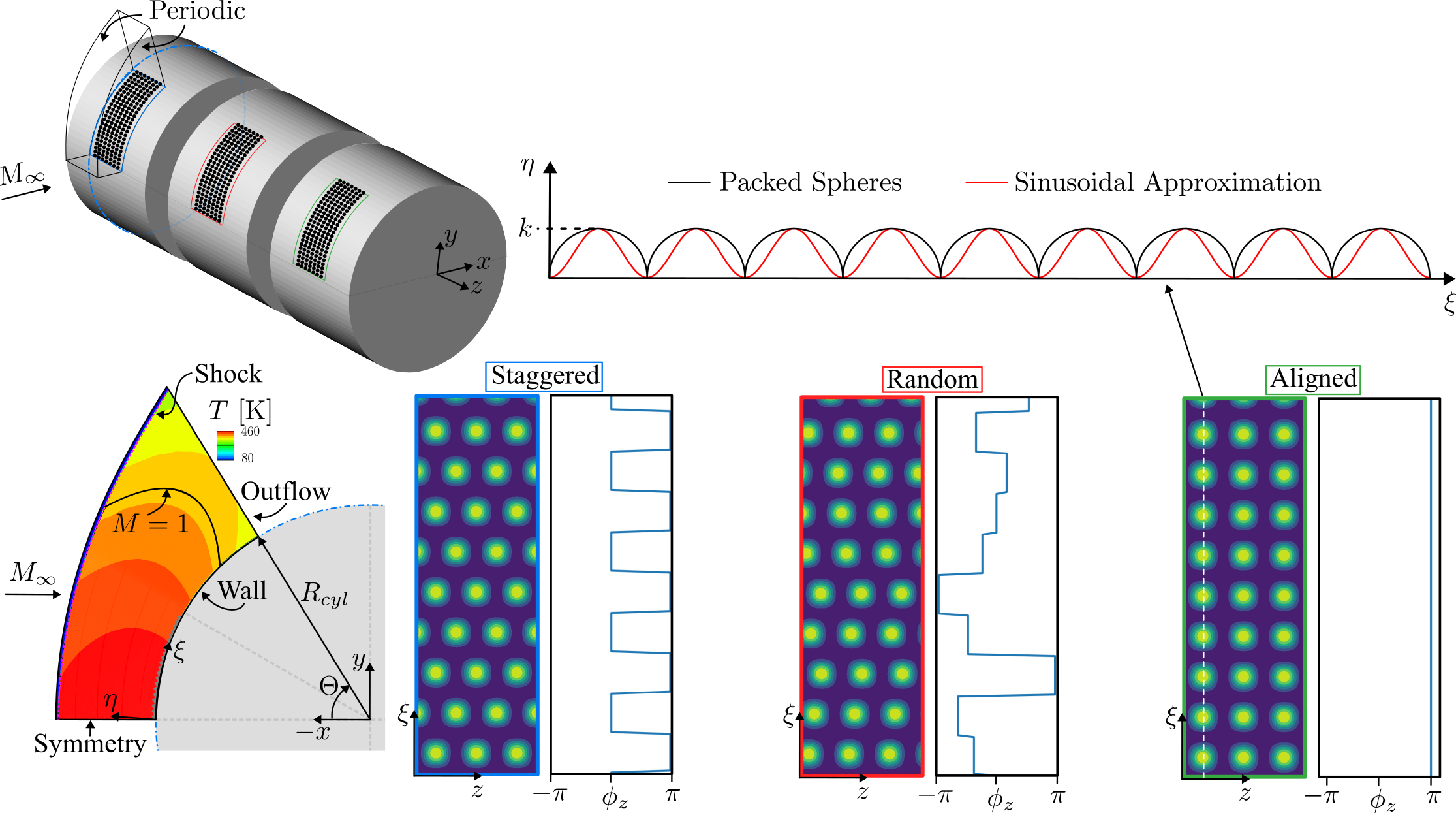}
    \caption{Schematic of cylinder geometry and roughness configurations. Only 9 out of 220 roughness element rows in the streamwise direction are shown for clarity.}
    \label{fig:schematic}
\end{figure}

The free-stream conditions are as follows: Mach number $M_\infty=5.98$, temperature $T_\infty=\SI{58.6}{\kelvin}$, and density $\rho_\infty=\SI{1.243e-1}{\kilogram\per\meter\cubed}$. At the surface of the cylinder the boundary is isothermal with $T_w=\SI{300}{\kelvin}$, in addition to no-slip and no-penetration at the wall. The boundary conditions are consistent with Test 6975 from \citet{hollis17_NASATM}. The free-stream conditions are prescribed via a supersonic inflow upstream from the bow shock. The cylinder grid is modelled using a symmetry plane at the centreline (if only half of the cylinder is modelled) and periodic boundary conditions at the spanwise bounds of the domain. The spanwise width of the domain is three nominal roughness element diameters. The domain is truncated with a supersonic outflow 60\textdegree{} from the stagnation line, after both the flow in the shock layer and at boundary layer edge become supersonic. A schematic of the cylinder geometry is provided in Figure \ref{fig:schematic}. The origin of the cylinder is centred at the Cartesian coordinates $x=\SI{0}{\meter}$, $y=\SI{0}{\meter}$ with the angle from the stagnation line (i.e., from the symmetry plane) defined as $\Theta=\tan^{-1}(y/(-x))$. The radius of the smooth cylinder is $R_{cyl}=\SI{0.0762}{\meter}$, where $R=\sqrt{x^2+y^2}$. In body-fitted coordinates, the streamwise, wall-parallel direction is $\xi=R_{cyl}\Theta$, and the wall-normal direction is $\eta=R-R_{cyl}$.

From the \cite{hollis17_NASATM} experimental test campaign, the distributed roughness is introduced by coating the hemispherical models in precision-manufactured spherical glass particles. Due to imperfections in the manufacturing process, the surface is not perfectly covered with uniform hemispheres and the cusps between adjacent roughness elements are smoothed. Therefore, in the present work roughness is introduced to the computational mesh by displacing the volume mesh cell centroids with a sinusoidal distribution. The grid displacement at the cylinder's surface is prescribed to achieve a maximum amplitude of $A=\SI{0.09}{\milli\meter}$ in the wall-normal direction (half the nominal spherical glass particle diameter). Likewise, the wavelength of the sinusoidal roughness distribution is $\lambda_k=\SI{0.18}{\milli\meter}$, again corresponding to the nominal diameter of the spherical glass particles. Strips of roughness elements are placed in streamwise rows from $\Theta=0\degree$ until $\Theta=30\degree$ (i.e., from stagnation to well past the expected location of LTT), spanning the width of the cylinder. Three different phasing approaches for the spanwise organisation of the roughness elements are considered. (1) Staggered phasing: the first distribution staggers the roughness elements to be 180\textdegree{} out of phase, leading to a peak-trough-peak pattern in the streamwise direction. (2) Random phasing: the second orientation randomly distributes the roughness elements across the span. (3) Aligned phasing: the final spanwise organisation aligns the roughness elements for each streamwise row such that all the peaks and troughs align (0\textdegree{} spanwise row-to-row phase shift). The idealised sinusoidal surface roughness profile is provided by Equation \ref{eq:idealized_sin_rough}. In the idealised sinusoidal configuration, $\Delta R$ is the distance in the wall-normal ($\eta$) direction to perturb the computational surface grid, $A_{k} = \lambda_k/2$ is the amplitude, $\kappa_k = 2\pi/\lambda_k$ is the roughness wavenumber, $\lambda_k$ is the roughness wavelength, $z$ is the spanwise coordinate, $\xi$ is the arc length in the streamwise direction, and $\phi$ is the spanwise phasing. Depending on the configuration, the phase is specified according to equations \ref{eq:span-phase-staggered}, \ref{eq:span-phase-random}, or \ref{eq:span-phase-aligned}, for the staggered, random, and aligned respectively. For the random phasing, a random variable is drawn from the continuous uniform distribution on the interval $[-\pi,\pi]$. The roughness parameters, following \citet{KADIVAR2021}, are tabulated in Table \ref{tab:roughness-parameters} and Figure \ref{fig:schematic} also illustrates the three phasing approaches. Additional computational grid details are provided in Appendix \ref{appB}.
\begin{equation}
	\Delta R(\xi,z) = \frac{A_{k}}{4}\left(1-\cos\left(\kappa_kz+\phi\right)\right)\left(1-\cos\left(\kappa_k\xi\right)\right)
	\label{eq:idealized_sin_rough}
\end{equation}

\begin{align}
    \phi_{\text{staggered}}(\xi) &=
    \begin{cases}
        \pi, & \text{if } \text{\textbf{mod}}\left(\left\lfloor \frac{\xi}{\lambda_k}\right\rfloor,2\right)=0\\
        0, & \text{otherwise}
    \end{cases} \label{eq:span-phase-staggered}\\
    \phi_{\text{random}}(\xi) &= \pi\epsilon_j,
    \; \epsilon_j \sim U(-1,1),\;j={\left\lfloor \frac{\xi}{\lambda_k} \right\rfloor},\; j\in\mathbb{Z} \label{eq:span-phase-random}\\
    \phi_{\text{aligned}}(\xi) &= \pi \label{eq:span-phase-aligned}
\end{align}

\begin{table}[h]
    \centering
    \begin{tabular}{llc}
        Parameter & Mathematical Description & Value \\ \hline
        Surface mean height & $R_m=\frac{1}{LzL_\xi}\iint\Delta R\;dzd\xi$ & $A_{k}/4.0$\\
        Peak-to-valley height & $R_{pv}=\max(\Delta R-R_m)-\min(\Delta R-R_m)$ & $A_{k}$   \\
        Arithmetic mean deviation & $R_a = \frac{1}{LzL_\xi}\iint|\Delta R-R_m|\;dzd\xi$ &  $A_{k}/4.3$  \\
        Root-mean-square & $R_{RMS}=\sqrt{\frac{1}{LzL_\xi}\iint|\Delta R-R_m|^2\;dzd\xi}$ &  $A_{k}/3.6$ \\
        Skewness & $s_k=\frac{1}{(R_{RMS})^3}\frac{1}{LzL_\xi}\iint|\Delta R-R_m|^3\;dzd\xi$ & 1.6\\
        Kurtosis & $k_u=\frac{1}{(R_{RMS})^4}\frac{1}{LzL_\xi}\iint|\Delta R-R_m|^4\;dzd\xi$ & 3.0  \\
        Effective slope & $ES=\frac{1}{LzL_\xi}\iint\left|\frac{\partial \Delta R}{\partial\xi}\right|\;dzd\xi$ & 0.5  \\
    \end{tabular}
    \caption{Roughness parameters. Physical heights are shown in normalised form with respect to the amplitude $A_{k}=\lambda_k/2$ and measured from the originally smooth surface $R_{cyl}$. The roughness parameters are identical for all phasing configurations.}
    \label{tab:roughness-parameters}
\end{table}

\section{Mean Flow Analysis} \label{sec:mean_flow}

\begin{figure}[h!]
    \centering
    \captionsetup{justification=centering}
    \includegraphics[width=0.9\linewidth]{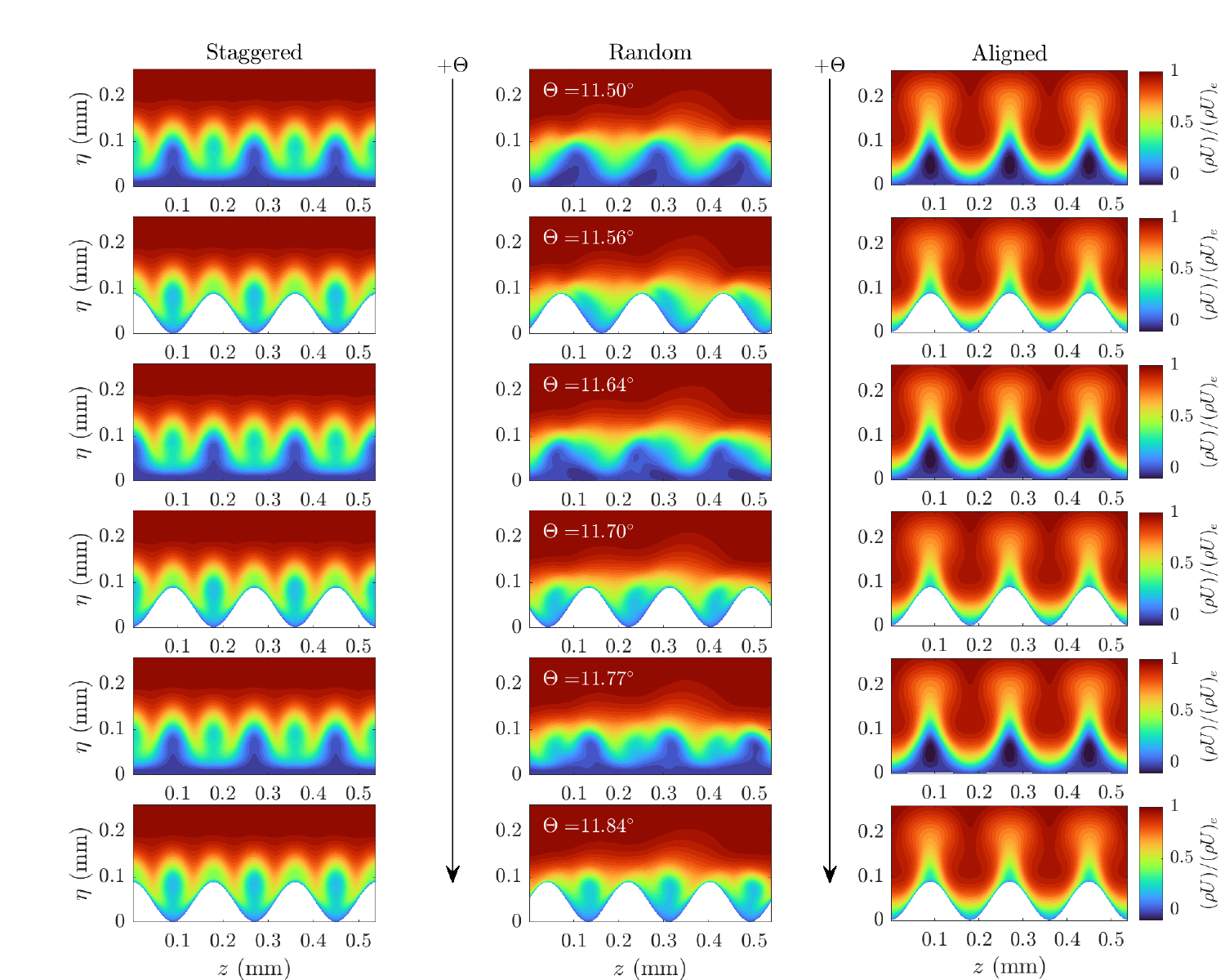}
    \caption{Contour plots of time-averaged, streamwise mass flux, $\rho U$, obtained from the DNS each roughness configuration. Angle from stagnation, $\Theta$, increases from top to bottom, and is constant for each row. Contour levels are the same for all plots.}
    \label{fig:dns_mass_flux_contours}
\end{figure}

An analysis of the mean flow field is performed first as it serves as the foundation for the unsteady dynamics dictating LTT, at least when perturbations remain `small'. Among other aspects, three-dimensional roughness elements introduce large variations in mean flow quantities in the spanwise direction. These gradients are known to play an important role in the dynamics of secondary instabilities \citep{Andersson2001,Paredes2019,Caillaud2025}. With this in mind, we visualise the spanwise variations with streamwise mass flux, $\rho U$, cross-planes in figure \ref{fig:dns_mass_flux_contours}. Even a brief glance at the contour planes reveals striking differences between the three roughness configurations considered. Beginning with the staggered case in the left column, 6 distinct peaks across the span are present and arise due to the consistent alternation of the roughness element rows by $180\degree$. Following the planes downstream (from top of left column down in figure \ref{fig:dns_mass_flux_contours}), one can see how the local minima in mass flux (hereafter referred to as `streaks') assume the spanwise position of the element peaks in the previous plane. These local minima pass unobstructed through the next row as the underlying roughness element phase shifts by $180\degree$ and generates three streaks following that spanwise location. The process repeats row-to-row downstream, hence, the 6 streak count. Alternatively, the aligned case possesses null phase shift from row to row (right column in figure \ref{fig:dns_mass_flux_contours}). Thus, the local mass flux minima reoccur in the same relative spanwise location (again, following element peaks) and only three streaks are generated. In addition, the consistent phasing appears to elevate low mass flux fluid high into the boundary layer (relative to the staggered case), perhaps by flow deflection by the roughness elements or lift-up due to streamwise vorticity.

The underlying symmetry of the staggered and aligned configurations is broken in the random case (middle column in figure \ref{fig:dns_mass_flux_contours}). Local minima do follow element peaks like the other two cases (e.g., $\Theta=11.56\degree\rightarrow11.64\degree$), but they occur irregularly and interact with the streaks produced upstream. Interestingly, the first plane ($\Theta=11.50\degree$) appears to have only three mass flux minima. After passing over the roughness element peak at $z=\SI{0.25}{\milli\meter}$ in the $\Theta=11.56\degree$ plane, the middle minimum is shifted slightly to the left. It is then able to pass largely unobstructed past the element peak at $z=\SI{0.3}{\milli\meter}$ in the $\Theta=11.70\degree$ plane. This phase shift, while not $180\degree$, is enough to produce 6 discernible peaks, much like the staggered case. In this way, the random case at times behaves much like the aligned (as the $\Theta=11.50\degree$ plane illustrates), but then at others behaves more like the staggered case. The decision is made by the underlying roughness pattern and how closely the elements from row to row are aligned, staggered, or de-tuned from either. The result is a more complicated mean flow field not described by simple integer streak counts like the other two roughness configurations.

\begin{figure}[h]
    \centering
    \captionsetup{justification=centering}
    \includegraphics[width=0.9\linewidth]{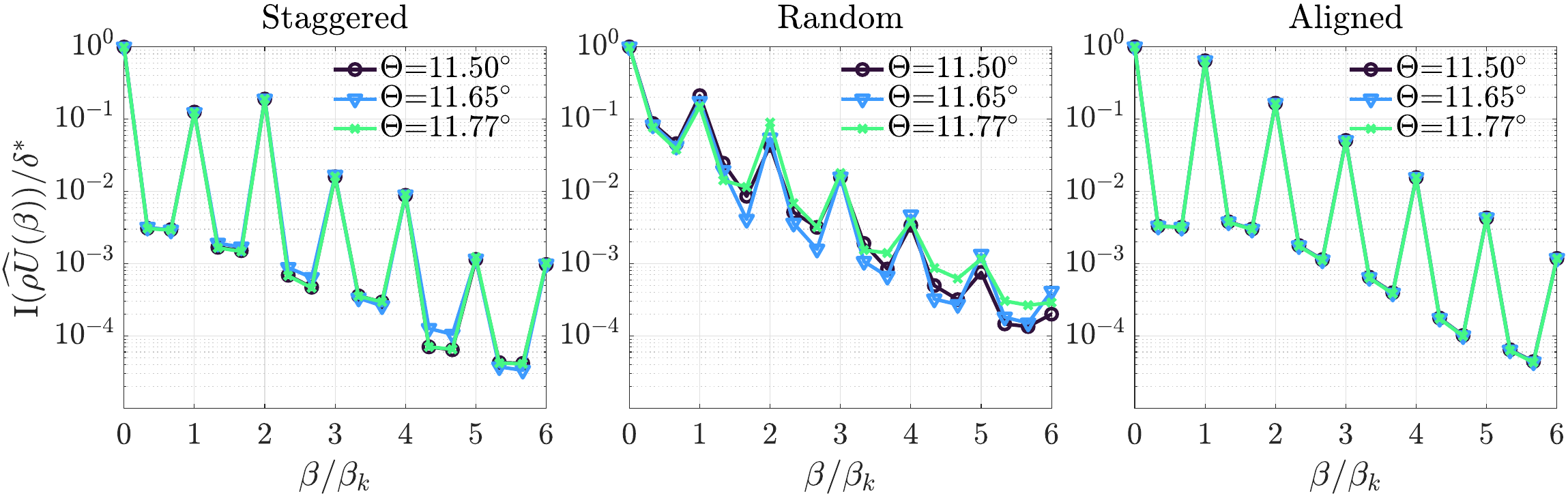}
    \caption{Wall-normal integrated spanwise harmonics of time-averaged streamwise mass-flux (see eq. \ref{eq:integral_rhoU_fourier}) at select planes for staggered (left), random (middle), and aligned (right) roughness configurations.}
    \label{fig:dns_mass_flux_fft}
\end{figure}

We confirm the above trends mathematically by computing the integral of the individual Fourier coefficients describing the streamwise mass flux planes located in the troughs in figure \ref{fig:dns_mass_flux_contours}, namely at $\Theta=\{11.50\degree,11.64\degree,11.77\degree\}$. These integrals, defined as 
\begin{equation} \label{eq:integral_rhoU_fourier} 
    \mathrm{I}(\widehat{\rho U}(\beta))=\int_{0}^{\infty} |\widehat{\rho U}(\beta)|/(\rho U)_e \,d\eta,
\end{equation}
are made dimensionless with the displacement thickness, $\delta^{*}$, to appreciate the relative contribution of each Fourier component. The trends described above are corroborated in figure \ref{fig:dns_mass_flux_fft} where the left panel for the staggered case shows more contribution from the $2\beta_{k}$ component than the $1\beta_{k}$ due to the constant $180\degree$ row-to-row phase shifting. Also, the right panel for the aligned case shows most amplitude in the $1\beta_{k}$, the roughness fundamental wavenumber. The shifting from $1\beta_{k}$ dominance at $\Theta=11.50\degree$ towards $2\beta_{k}$ as $\Theta=11.77\degree$ is approached is also seen for the random case in the middle panel. Also noteworthy for this case is the elevated contributions from off-fundamental roughness wavenumber harmonics. This showcases the more `complex' nature of the mean flow since more Fourier modes are needed to describe the distribution. It must also be pointed out the relative magnitude of the integrated components compared to the displacement thickness. With most components over $10\%$ the total displacement thickness, we demonstrate the non-linearity of the streak development.

\begin{figure}[h]
  \centering
  \subfloat[]{%
    \includegraphics[width=0.32\textwidth]{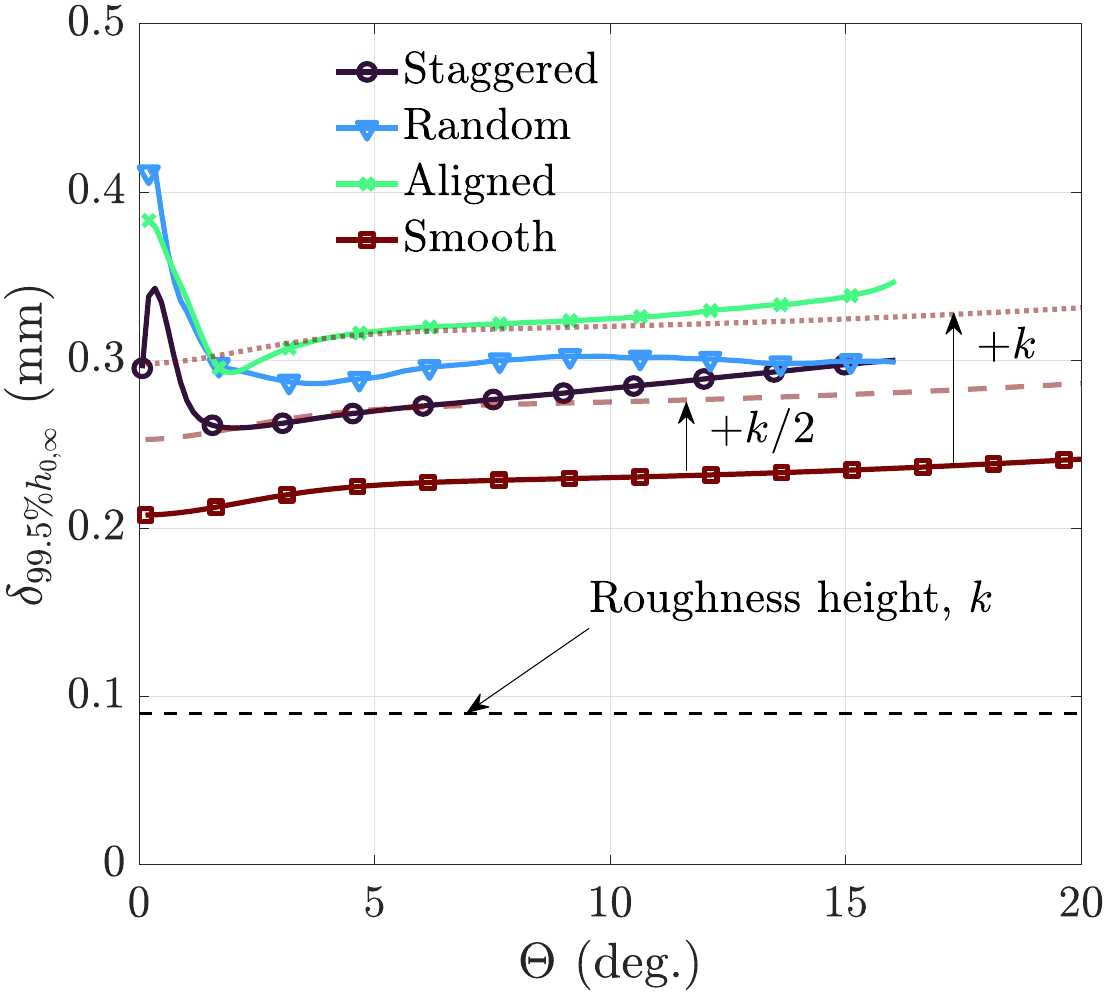}}
  \hfill
  \subfloat[]{%
    \includegraphics[width=0.32\textwidth]{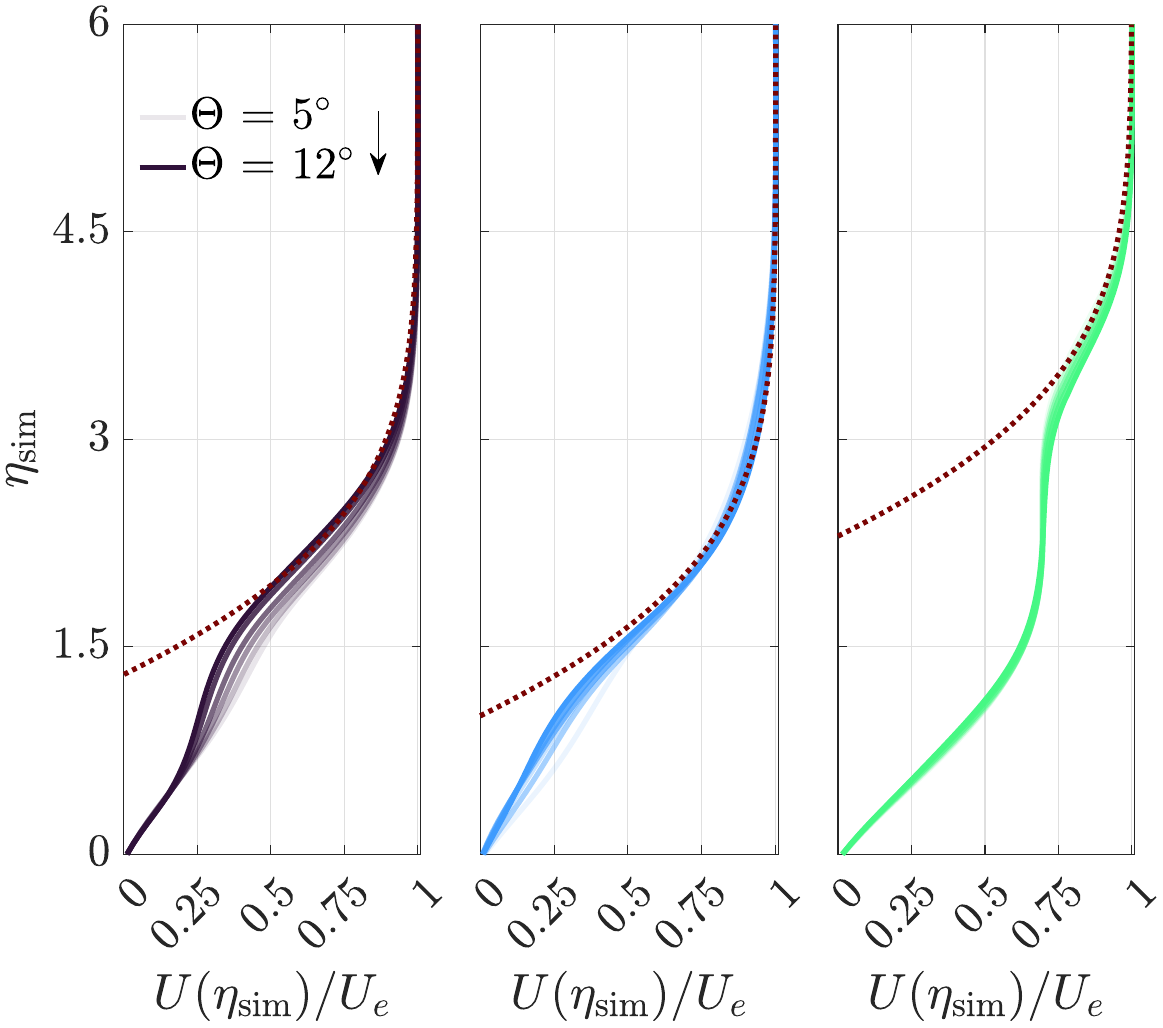}}
  \hfill
  \subfloat[]{%
    \includegraphics[width=0.32\textwidth]{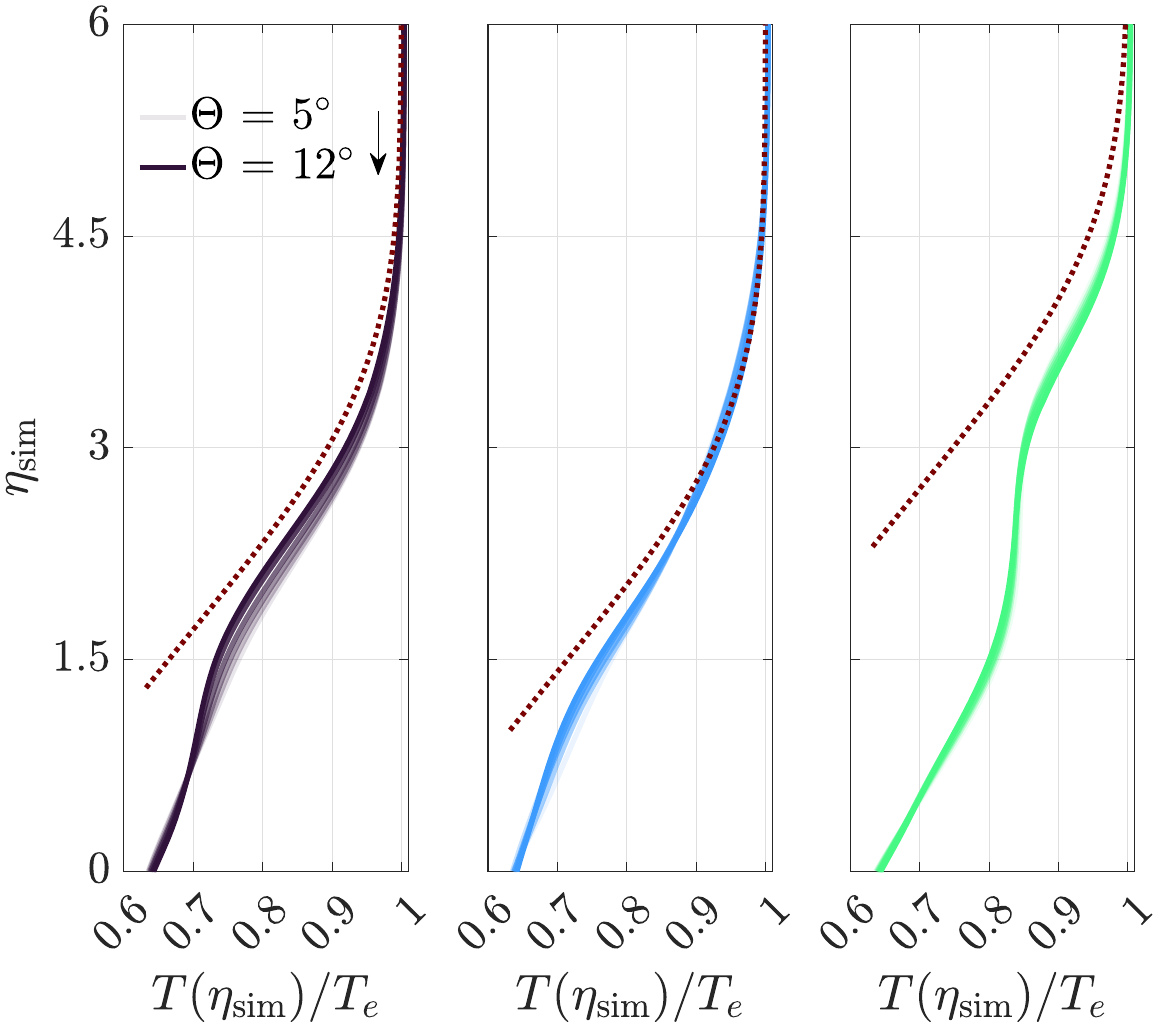}}
  \caption{Development of DNS time-averaged flow field in terms of (a) $99.5\%h_{0,\infty}$ boundary layer thickness, similarity variable transformed profiles of normalised (b) streamwise velocity and (c) temperature. Line colours denoting roughness pattern in (a) are maintained in (b) and (c).}
  \label{fig:dns_BL_development}
\end{figure}

To get a better sense of the development of the boundary layer in those regions not shown in figure \ref{fig:dns_mass_flux_contours}, profiles taken normal to the cylinder's surface were used to obtain a boundary layer thickness using the $99.5\%$ free-stream stagnation enthalpy, $h_{0,\infty}$, criterion. The extracted profiles (as in \citep{Saikia_Brehm_2025}) were then spanwise and streamwise averaged (over a roughness element wavelength) to provide smooth distributions and are shown in figure \ref{fig:dns_BL_development}(a). First, all roughness configurations display a thicker boundary layer than the smooth wall solution. The offset is very nearly two and four times the mean surface height, $R_{m}=k/4$ (see table \ref{tab:roughness-parameters}), for the staggered and aligned cases, respectively. The boundary layer thickness in the random case is bounded by the other two cases over the streamwise extent shown and tends more towards the staggered farther downstream. These outward shifts of the mean flow profiles are expected \citep{Gaster2016}. Knowing the local boundary layer thickness is useful in estimating what frequencies could be expected should there be any unstable boundary modes. 
\begin{table}[h]
    \begin{center}
    \def~{\hphantom{0}}
    \begin{tabular}{lcc}
        Roughness Type & $\lambda=2\delta_{99.5\%h_{0,\infty}}$ & $\lambda=10\delta_{99.5\%h_{0,\infty}}$ \\[3pt]
        Stg. & $f\in[42, 110]$ kHz & $f\in[8, 23]$ kHz \\
        Rnd. & $f\in[39, 110]$ kHz & $f\in[8, 23]$ kHz \\
        Alig. & $f\in[36, 100]$ kHz & $f\in[7, 20]$ kHz
    \end{tabular}
    \caption{Frequency estimates between $\Theta=[5\degree,15\degree]$ assuming $c_{r}\approx 0.5U_{e}$.}
    \label{tab:freq_est}
    \end{center}
\end{table}
Using the dispersion relation $f=c/\lambda$, a common estimation \citep{lastrac_manual} of the local disturbance phase speed of $0.5 U_{e}$, and assuming that the wavelength is between $2-10\delta_{99.5\%h_{0,\infty}}$, the range of frequencies for each case has been estimated and is tabulated in table \ref{tab:freq_est}. Even though the boundary layer is relatively thin (fractions of a millimetre), the slow edge velocity mandates a frequency range generally between $f\approx[10,100]$ kHz. It is worth noting the nearly constant boundary layer height that is present for the smooth wall and maintained in the rough wall solutions. Substantial growth for particular frequencies tuned to the boundary layer height could be expected as a result. 

Those mean flow velocity and temperature profiles used to search for $99.5\%h_{0,\infty}$ are plotted in figures \ref{fig:dns_BL_development}(b) and \ref{fig:dns_BL_development}(c) under the stagnation flow similarity variable transformation (see appendix \ref{appC}). Each case exhibits a remarkably good collapse under the similarity variables, indicating the flow in the presence of roughness still develops in a self-similar manner. Added to each plot is the smooth wall similarity solution shifted by a $\Delta \eta_{\mathrm{sim}}$ for each roughness case to match the outer velocity boundary layer shape. Interestingly, a non-uniform $\Delta \eta_{\mathrm{sim}}$ was required for each case, namely $\Delta \eta_{\mathrm{sim}}=\{1.3,1.0,2.3\}$ for Stg., Rnd., and Alig., respectively. Also, the thermal boundary layer, $T(\eta)/T_{e}$, would require a slightly larger shift for outer profile matching. The random roughness possess a thinner profile than staggered (when plotted under similarity variables) lower in the boundary layer, but a shallower approach to unity towards the upper boundary layer edge. This helps explain why a smaller $\Delta \eta_{\mathrm{sim}}$ is required, yet the $\delta_{99.5\%h_{0,\infty}}$ is larger. Despite these idiosyncrasies, the fact that all configurations simulated still possess self-similarity gives hope of finding approximate methods for incorporating distributed roughness effects into self-similar solutions for use with lower-order stability tools like linear stability theory (LST) to rapidly predict LTT on blunt bodies in a physics-based manner. This idea of course relies on knowledge that LST performed on spatially averaged profiles provides satisfactory agreement with results of DNS or LDE simulations, something to be explored later in section \ref{sec:lde}. 

\begin{figure}[h]
  \centering
    \subfloat[]{%
    \includegraphics[width=0.32\textwidth]{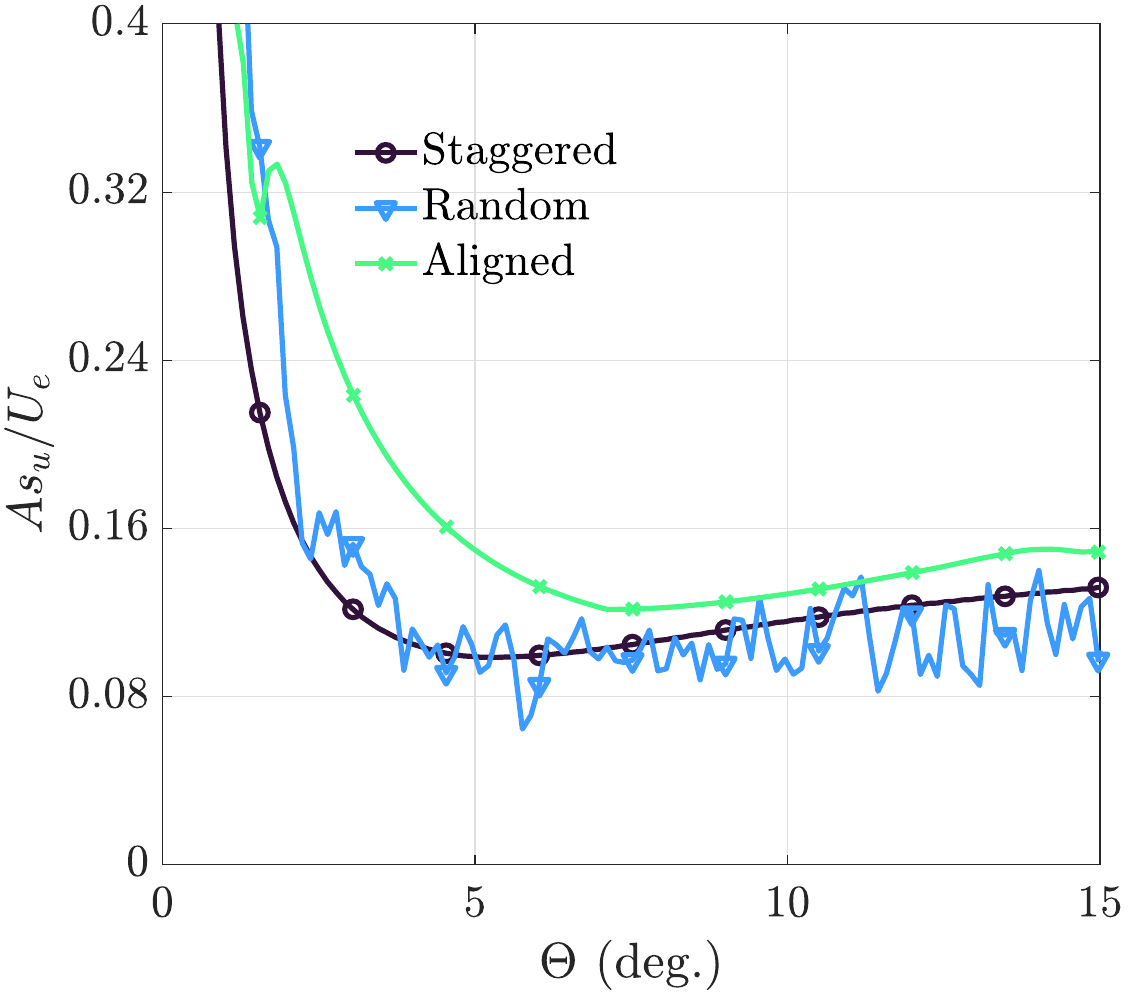}}
  \hfill
  \subfloat[]{%
    \includegraphics[width=0.32\textwidth]{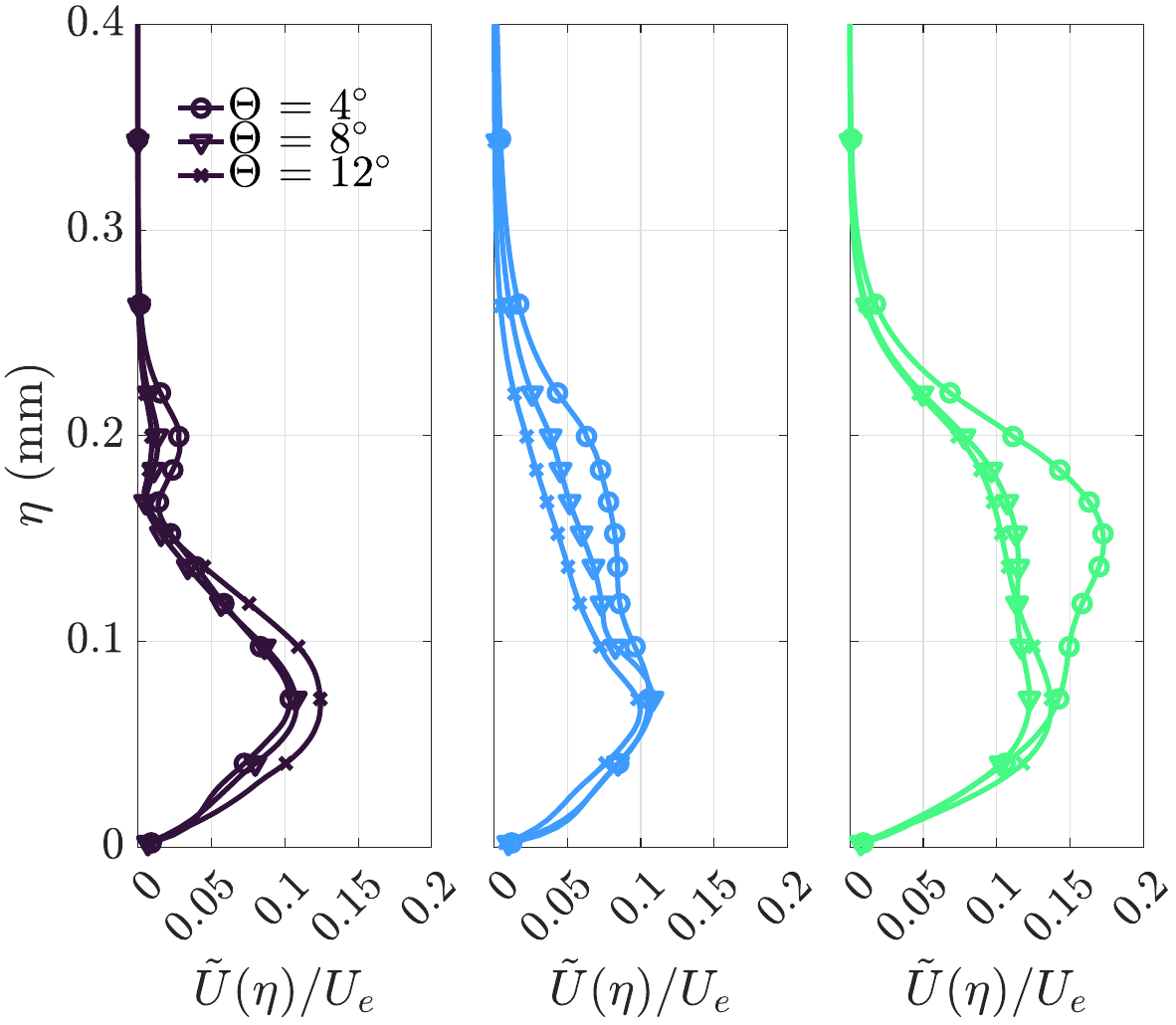}}
  \hfill
  \subfloat[]{%
    \includegraphics[width=0.32\textwidth]{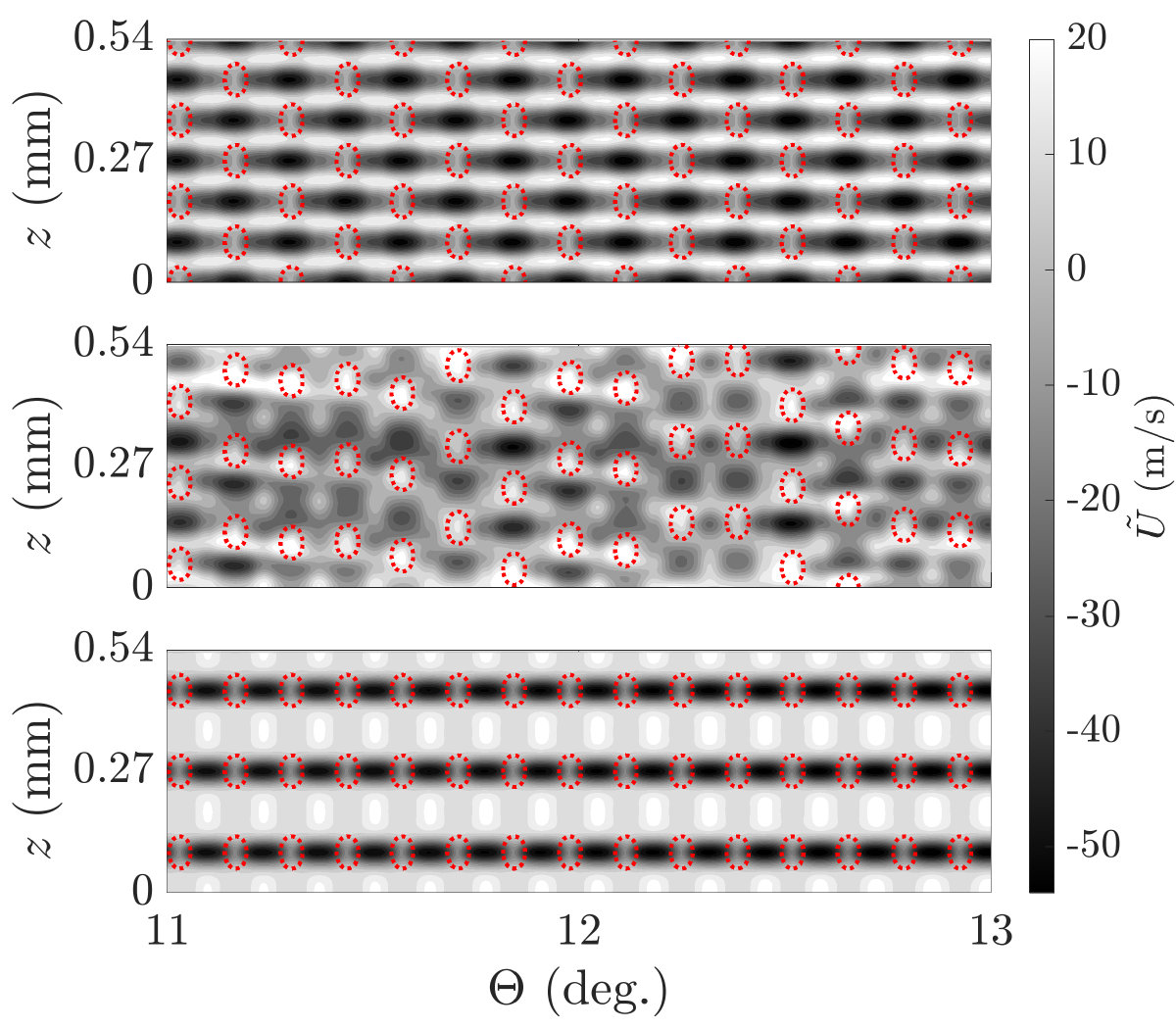}}
  \caption{Development of normalised DNS streamwise velocity streak amplitude in terms of (a) angle from stagnation point, (b) wall-normal profiles and (c) $\eta=0.1$ mm plane just above roughness for staggered (top), random (middle), and aligned (bottom) configurations. Line colours denoting roughness configuration in (a) are maintained in (b).}
  \label{fig:dns_streak_development}
\end{figure}

The spanwise averaging used to obtain the profiles in figure \ref{fig:dns_BL_development} was then used to define a spanwise perturbation in the mean flow quantities, i.e., a `streak,' denoted with a $\tilde{(.)}$. These are computed with the formula for any mean quantity, $\bar{Q}(x,y,z)$ as 
\begin{equation} \label{eq:streak_calc_eq}
    \tilde{Q}(\xi,\eta,z)=\bar{Q}(\xi,\eta,z)- \bar{Q}_{z}(\xi,\eta),
\end{equation}
where $\bar{Q}_{z}(\xi,\eta)$ is the spanwise averaged mean flow quantity. Computing this for the streamwise velocity, $U$, one can then compute the streak amplitude in common use in the literature \citep{Andersson2001,Cossu_2004,Paredes2017b,Caillaud2025} as
\begin{equation}
    As_{u}(\xi) = \frac{1}{2}\left(\max_{\eta,z} \tilde{U}(\xi,\eta,z)-\min_{\eta,z} \tilde{U}(\xi,\eta,z)\right).
\end{equation}
To mitigate increases in streak amplitude due purely to the accelerating external flow, the streak amplitude is presented as normalised by the local edge velocity, $U_{e}$. The result, after roughness wavelength averaging in the streamwise direction, is presented in figure \ref{fig:dns_streak_development}(a). It should first be noted that, with the above definition, the streak amplitude decays more slowly than the edge velocity as the stagnation line is approached and an increase in the ratio is observed. A breakdown criteria based purely on streak amplitude relative to local edge velocity may therefore not be helpful since it would predict two locations where, for example, $As_{u}/U_{e}=10\%$ is reached. Beyond this initial region of decay near the stagnation line, the streak amplitude begins to increase, and at nearly linear rates for the staggered and aligned cases. Generally, the aligned case has higher streak amplitudes, followed by staggered, and finally by random, although this case surpasses the staggered sporadically. Quantitatively, all cases exhibit less than a factor 2 relative growth of streaks over the regions of increase ($\Theta\approx[5\degree,15\degree]$). This is much less than the factor 10 found by \citet{Paredes2017} for a smooth wall hypersonic hemisphere using optimal transient growth theory. 

Figure \ref{fig:dns_streak_development}(b) shows the distribution of the streak amplitude before the wall-normal maximum was sought and provides a sense of the streak shape in that coordinate direction. Beginning with the aligned case in the rightmost panel, one can see how the wall-normal maximum switches from being located at approximately $\eta=0.15$ mm at $\Theta=4\degree$, to around $\eta=0.075$ mm at $\Theta=8\degree$ and $\Theta=12\degree$. Recall, the roughness height $k$ is 0.09 mm. It is near this height that the profiles examined for staggered and random exhibit their wall-normal maximum. A second, smaller peak higher in the boundary layer around $\eta=0.2$ mm is visible for the staggered case ($\Theta=4\degree$) and seen to decay quickly in the further downstream profiles. The random case also has this secondary peak, but it decays more slowly and is spread more evenly through the outer boundary layer. This broader shape bears a stronger resemblance to the aligned case. It is thus most likely that this outer peak is being tracked in the streak amplitudes reported in figure \ref{fig:dns_streak_development}(a) for those locations close to the stagnation point before switching to the peak that occurs close to the roughness element height, which exhibits relative growth downstream.

What is lost in the roughness wavelength streamwise averaging of the profiles is regained in figure \ref{fig:dns_streak_development}(c) where a constant $\eta=0.1$ mm plane is taken through the volume near the streak maximum (see figure \ref{fig:dns_streak_development}(b)). In this way, the streamwise development of the streaks can be appreciated. Beginning with the staggered configuration in the top panel, one can clearly see streak amplitudes peaking in the mid-wake following element peaks (shown with red dotted circles in figure \ref{fig:dns_streak_development}). These peaks decay slightly before impinging the next in-phase row of elements where the process repeats. The aligned case (bottom panel) has even less streamwise extent for the steaks to develop, yet their peak extends more fully from element peak to peak. This explains why the streak amplitude came out larger, on average, for the aligned case than the staggered. The occurrence of sporadic peaks seen in figure \ref{fig:dns_streak_development}(a) for the random case becomes clear after referencing the middle panel of figure \ref{fig:dns_streak_development}(c). Interestingly, one can see the strongest streaks occur whenever the phase shift between element rows is most nearly $180\degree$, i.e., staggered. See, for example, $\Theta\approx11.85\degree$. In fact, the strongest streaks shown in the figure occur when a null phase shift from row to row is followed immediately by a $180\degree$ phase shift (see $\Theta\approx12.3\rightarrow12.5\degree$). In this scenario, the two extremes of row phase shifting work together to produce strong streaks. This happens rarely, though, and the result is generally weaker, more sporadic streaks than either staggered or aligned roughness configurations.

\begin{figure}[h]
    \centering
    \captionsetup{justification=centering}
    \includegraphics[width=0.95\linewidth]{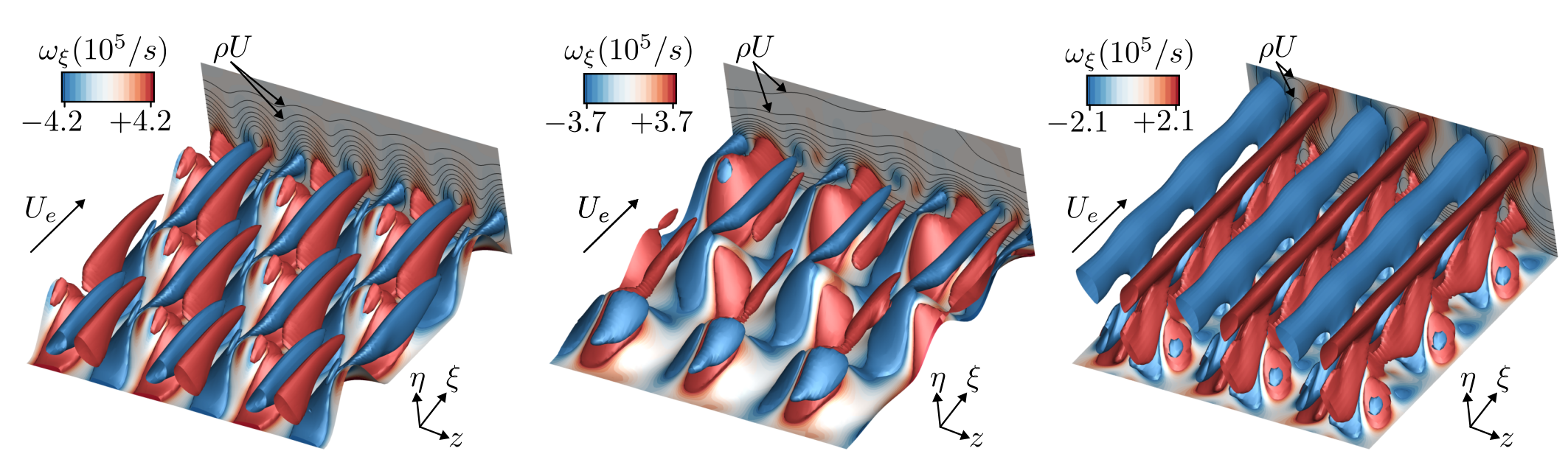}
    \caption{Iso-surfaces of streamwise vorticity of the time-averaged flow field for the staggered (left), random (middle), and aligned (right) roughness configurations between $\Theta=[11.5\degree,11.9\degree]$. Vorticity contour planes are provided as a back-plane where lines of constant streamwise mass flux $\rho U$ are overlaid.}
    \label{fig:mean_vorticity}
\end{figure}

To help understand how the streaks are being produced, we next show perspective views of iso-surfaces of streamwise vorticity $\omega_{\xi}$ for all three roughness configurations between the locations $\Theta=[11.5\degree,11.9\degree]$ in figure \ref{fig:mean_vorticity}. Included in the back plane is a $\eta-z$ plane coloured by streamwise vorticity and lined by fractions of streamwise mass flux (analogous to figure \ref{fig:dns_mass_flux_contours}). As shown in the left part of figure \ref{fig:mean_vorticity}, staggering the elements results in clean counter-rotating vortex pairs (CVP) wrapping around the sides of each roughness element. Their positive (clockwise looking downstream) and negative (counter-clockwise looking downstream) rotation coincides with the right and left flanks of low mass flux streaks shown in the back plane, respectively. This illustrates the lift-up mechanism working to help generate the streak profile by pulling low speed fluid upward and pushing high speed fluid downward. The CVPs decay slightly as they pass through the alley of the next row but then are re-energised by the following in-phase elements. The aligned roughness configuration in the right part of figure \ref{fig:mean_vorticity} shows similar behaviour in that CVP are produced, however, the proximity of another in-phase row of elements close behind seems to deflect the CVPs upward. This consistent dynamic, row after row, results in the accumulation of vorticity above the elements and helps explain the relatively tall streak profiles seen previously. Again, the back plane visualises the lift-up mechanism aiding in generation of the mass-flux streaks. The randomised roughness arrangement in the middle of figure \ref{fig:mean_vorticity} exhibits mixed behaviour. Referring briefly back to the middle panel of figure \ref{fig:dns_streak_development}(c), $\Theta=11.5\degree$ is located in between element rows that are very nearly aligned. This is why on the up-slope on the first row visualised there is a vertical stacking of $\pm \omega_{\xi}$ as in the aligned case. Moving past this first row, one can see there is a slight phase shift to the next row and it is enough to show (at this iso-surface level) positive $\omega_{\xi}$ convecting off the tip of the elements. The final row shown in the figure has an even larger relative phase shift, and CVPs more similar to the staggered case can be seen. Note the non-uniformity in the span at this location. The right most CVP is strongest, indicating some bias in the accumulation of vorticity despite the row phases being selected randomly. It should be pointed out that different levels of vorticity were needed in each case to most clearly illustrate the dynamics just described, as indicated in the $\omega_{\xi}$ legend in each subplot in figure \ref{fig:mean_vorticity}.

Given the use of transient growth theory as the only physics-based model for LTT on hypersonic blunt bodies, the present authors feel obligated to attempt a reconciliation of the preceding results with that theory. While figures \ref{fig:dns_streak_development}(c) and \ref{fig:mean_vorticity} show characteristics of transient growth in the wake of the roughness elements, namely decaying vorticity giving rise to velocity streaks, the reality is more complicated than can be predicted by linear transient growth calculations. The first issue is the non-linearity of the streak developments. Our results show streak amplitudes in all roughness arrangements at all times remaining above $8\%U_{e}$ (see figure \ref{fig:dns_streak_development}(a)), which most would agree is outside the amplitude regime for linear description. In addition, the close proximity of subsequent roughness elements results in the relative phasing of roughness elements rows having a significant influence on the streak's development. Recall the work of \citet{Denissen2009}, which demonstrated the ability of linear stability analysis to \textit{reconstruct} the development of transient growth downstream of a single spanwise array of roughness elements in an incompressible flat plate boundary layer. Good agreement between the linear calculation and DNS for the evolution of individual spanwise Fourier modes could only be realised 35 boundary layer thicknesses downstream of the roughness elements. The present simulations have roughness elements spaced approximately unit boundary thickness downstream. Also, their analysis is a reconstruction, meaning the DNS data is needed at some initial streamwise location in order for the analysis to reproduce the perturbation evolution. This is helpful as an \textit{a posteriori} analysis, however, it is not clear how this could be used as a predictive theory provided no DNS data is available for the disturbance field behind the roughness elements. 

Furthermore, real roughness-induced transient growth is, as a rule, suboptimal -- and usually very suboptimal at that. For example, \citet{Choudhari2005} through DNS of a roughness array in a flat plate boundary layer found growth ratios of $\mathcal{O}(300)$ versus the optimal value of $\mathcal{O}(2255)$ (inferred from \citep{Levin2003}). The results were similar from \citet{Schilden2020} where transient growth behind roughness patches of aligned and staggered configurations on an Apollo-type capsule geometry resulted in small amplification ratios of 1.74 and 1.553 versus the 66.3 found for optimal disturbances. Finally, as a linear calculation, transient growth alone cannot predict the secondary instabilities required to transition the flow to turbulence. Indeed, it has been shown \citep{Denissen2013} that actual roughness mean flow profiles significantly alter the secondary instability characteristics when compared to using the profiles obtained from optimal streaks. Taking our current results and those just quoted from the literature together, it remains questionable how valid transient growth is as a physical mechanism for explaining transition on hypersonic blunt geometries like the one considered here.

\begin{figure}[h]
    \centering
    \captionsetup{justification=centering}
    \includegraphics[width=0.5\linewidth]{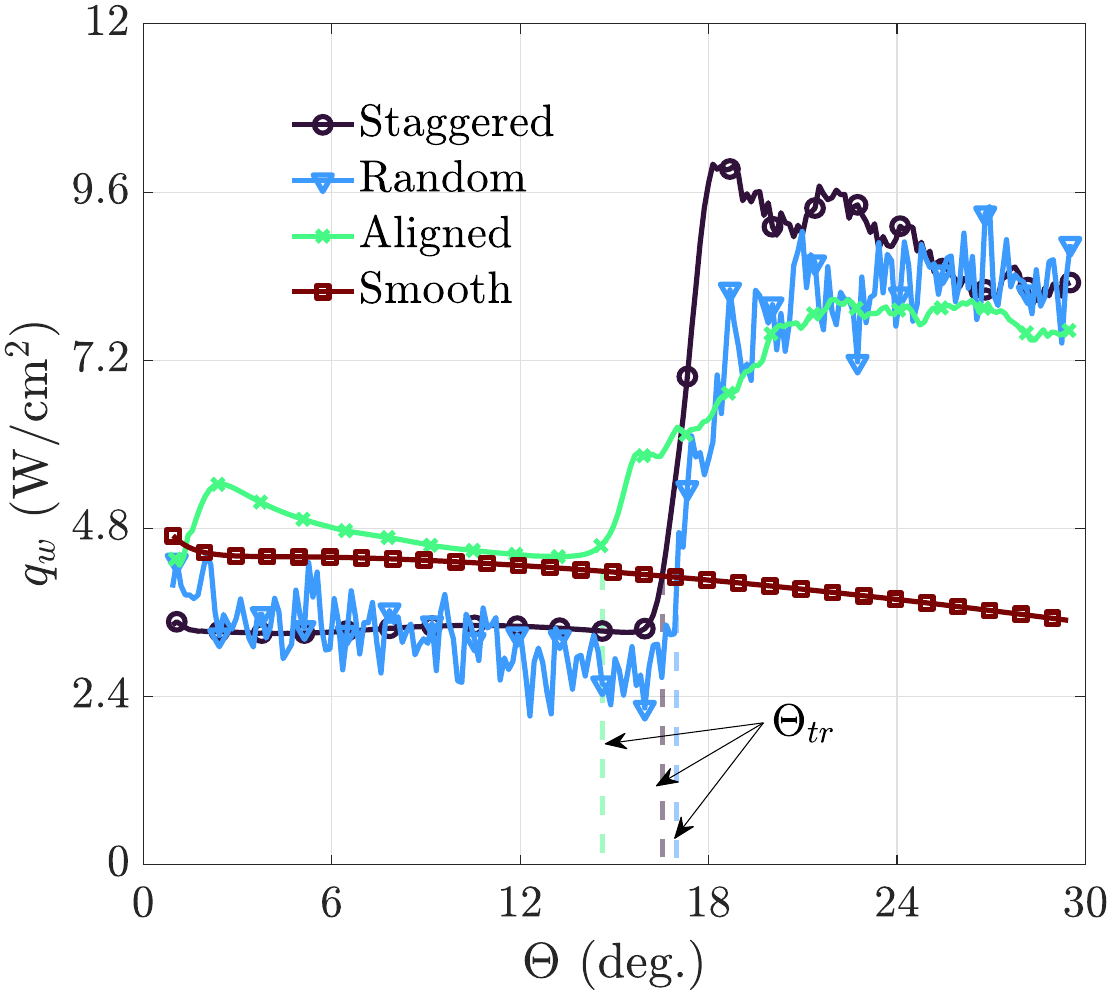}
    \caption{Time, span, and streamwise wavelength-averaged surface heat flux for all roughness configurations. The smooth wall solution is provided for reference.}
    \label{fig:dns_heat_flux}
\end{figure}

As a segue to the next sections, we now present the time-, span-, and streamwise roughness wavelength-averaged surface heat flux for all roughness configurations in figure \ref{fig:dns_heat_flux}. All cases exhibit the sudden rise in heat flux indicative of laminar-turbulent transition, marked by dashed lines and labelled $\Theta_{tr}$. Prior to transition at $\Theta_{tr}$, the aligned configuration has laminar heating values slightly elevated relative to the smooth wall near the stagnation point but approaches that solution later downstream. The staggered case, on the other hand, shows a laminar heating reduction relative to the smooth wall. Interestingly, the random case exhibits a similar reduction as the staggered case, but it also shows an overall steeper decrease in heat flux throughout the laminar region that looks similar to the aligned case. The tangent slope method of \citet{hollis17_NASATM} was used to consistently extract the streamwise locations of LTT for each case and are tabulated in table \ref{tab:corr-trans-locs}. Also included are the predictions from the state-of-the-art correlations for predicting LTT on hypersonic blunt bodies. These correlations predict transition in the range $\Theta\approx[7\degree,13\degree]$ for the hemisphere, and $\Theta\approx[6\degree,11\degree]$ for the cylinder geometry. In addition to the large variability, the correlations predict transition significantly further upstream than the present simulations.

\begin{table}[h] 
\centering
\resizebox{\textwidth}{!}{%
\begin{tabular}{lcccccccccc}
\hline
 & \multicolumn{7}{c}{Correlation} & \multicolumn{3}{c}{DNS} \\
\cline{2-8} \cline{9-11}
Geometry
 & P-AW & P-R & P-RT
 & $\Rey_k$-CR & $\Rey_\theta$-LR & SAND & HEX-PAT
 & Stg. & Rnd. & Alig.$^*$ \\
\hline
Cylinder
 & 9.41$^\circ$ & 8.98$^\circ$ & 10.06$^\circ$
 & 7.79$^\circ$ & 11.03$^\circ$ & 10.06$^\circ$ & 6.27$^\circ$
 & 16.53$^\circ$ & 16.97$^\circ$ & 14.63$^\circ$ \\

Hemisphere
 & 9.41$^\circ$ & 7.57$^\circ$ & 9.30$^\circ$
 & 7.14$^\circ$ & 12.98$^\circ$ & 10.93$^\circ$ & 7.68$^\circ$
 &  &  &  \\
\hline
\end{tabular}}
\caption{Transition locations, $\Theta_{tr}$, predicted by correlations (see appendix \ref{appA}), and those found in the present DNS. PANT correlations were abbreviated further to `P'.}
\label{tab:corr-trans-locs}
\end{table}

\section{Unsteady Flow Analysis} \label{sec:unsteady_flow}

\subsection{Direct Numerical Simulation} \label{sec:DNS}
\begin{figure}[h]
    \centering
    \captionsetup{justification=centering}
    \includegraphics[width=0.95\linewidth]{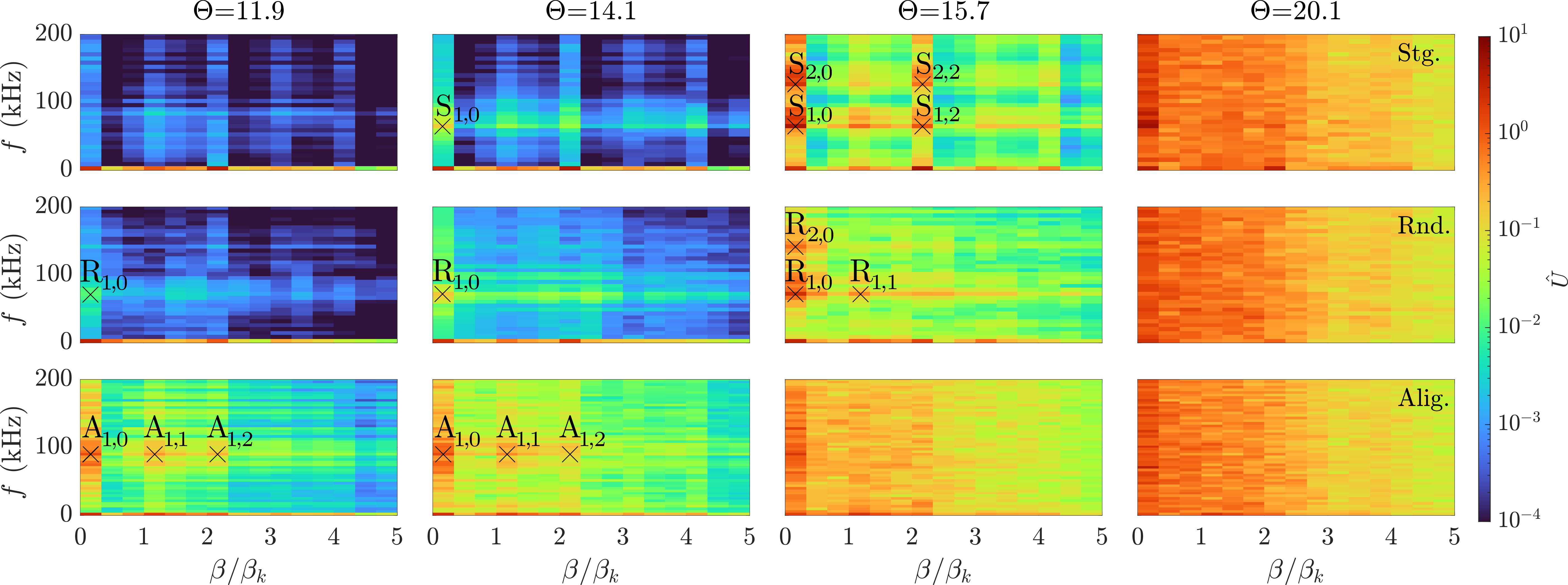}
    \caption{Fourier-transformed streamwise velocity frequency-spanwise wavenumber spectra for several streamwise locations for staggered (top), random (middle), and aligned (bottom) roughness distributions.}
    \label{fig:dns_spectra}
\end{figure}

We turn our attention now to the flow unsteadiness leading to LTT observed in the DNS of each roughness configuration. The streamwise velocity spectra shown in figure \ref{fig:dns_spectra} were obtained by first sampling the unsteady flow field with $\eta-z$ planes located in the roughness troughs. A fast Fourier transform (FFT) was then performed in time and span for each spanwise row of probes in the plane and the maximum amplitude for each $(f,\beta)$ pair in the wall-normal direction was reported for the given plane location. It is understood this FFT decomposition will not isolate instability waves evolving independently from one another given the mean flow's underlying spanwise periodicity \citep{Andersson2001}. A spectral proper orthogonal decomposition considering the spanwise periodic mean flow, as was done in \citet{Caillaud2025}, could have been performed, however, the FFT used here are primarily for determining dominant frequencies and spanwise wavenumbers present in the signals. 

The top row of figure \ref{fig:dns_spectra} corresponds to the staggered roughness case. The first streamwise location plotted ($\Theta=11.9\degree$) shows a fairly rich spectrum of relatively low level fluctuations with no discernible peaks. At $\Theta=14.1\degree$, though, a distinct frequency band centred about $f=62$ kHz begins to dominate the spectrum and is marked by $S_{1,0}$. The subscripts on the first letter of the roughness configuration (S,R, or A) denote the integer multiple, $m$, of the dominant frequency $f_{1}$ and integer multiple, $n$, of the roughness spanwise wavenumber, $\beta_{k}$. While most amplitude is in the spanwise uniform component ($\beta=0\beta_{k}$), there are also contributions from $2\beta_{k}$, $1\beta_{k}$, and $4\beta_{k}$ (in that order of importance). Referring back to figure \ref{fig:dns_mass_flux_fft}(a), these spanwise wavenumbers are likely the imprint of the mean flow. One might expect then to see more contribution from $\beta = 3\beta_{k}$, which is not the case and may indicate more complicated disturbance dynamics at play. The amplitude in that frequency band and those wavenumbers appears to grow downstream and generate strong harmonics like $S_{2,0}$ and $S_{2,2}$ seen at $\Theta=15.7\degree$. The last position shown in this row, $\Theta=20.1\degree$, exhibits a broad spectrum, indicating the flow is at least in the later non-linear stages of LTT if not fully turbulent. 

Spectra obtained from the random roughness configuration (middle row of figure \ref{fig:dns_spectra}) bears some similarity to the staggered case, except the frequency band becomes visible by $\Theta=11.9\degree$ and the frequency centre is slightly higher at $f_{1}=63$ kHz. While again the $0\beta_{k}$ wavenumber hold the most amplitude (marked as $R_{1,0}$), this case is different in that amplitude is more evenly distributed between wavenumbers $\beta\in[0,2]\beta_{k}$. This is expected given the more `complex' mean flow profiles shown previously in section \ref{sec:mean_flow} where wakes were seen to develop irregularly (see figure \ref{fig:dns_mass_flux_fft}). From $\Theta=14.1\degree$ to $15.7\degree$ we see that 63 kHz frequency band grows substantially and generates some temporal harmonics, for example the marked $R_{2,0}$. In addition, the $R_{1,1}$ that was not very dominant at $\Theta=14.1\degree$ increases substantially by $\Theta=15.7\degree$. The $\Theta=20.1\degree$ spectrum (rightmost middle row) is virtually identical to the staggered case and indicates full breakdown has occurred for this roughness pattern as well. 

Finally, the spectra from the aligned case are shown in the bottom row of figure \ref{fig:dns_spectra}. Appreciable amplitude is contained in spanwise wavenumbers $1\beta_{k}$ and $2\beta_{k}$ as marked by $A_{1,1}$ and $A_{1,2}$, respectively. This distribution in $\beta$ can be expected when recalling the right panel in figure \ref{fig:dns_mass_flux_fft} where the $\beta=1\beta_{k}$ and $\beta=2\beta_{k}$ are on the order of the spanwise uniform mass flux. The $\Theta=14.1\degree$ shows slight amplification of this frequency band, and the spectrum becomes relatively full by $\Theta=15.7\degree$, indicating an earlier onset of LTT for this case. Finally, by $\Theta=20.1\degree$ the spectrum is full and the flow is presumed turbulent. It is also worth noting how disturbance amplitude primarily builds up in spanwise wavenumber preceding breakdown for the aligned case. This contrasts the staggered and random cases where breakdown occurs first through excitation of temporal then spatial harmonics. Hence, differences in the responsible instability mechanisms may be at hand between the various cases.

\begin{figure}[h]
  \centering
  \subfloat[Staggered]{%
    \includegraphics[width=0.32\textwidth]{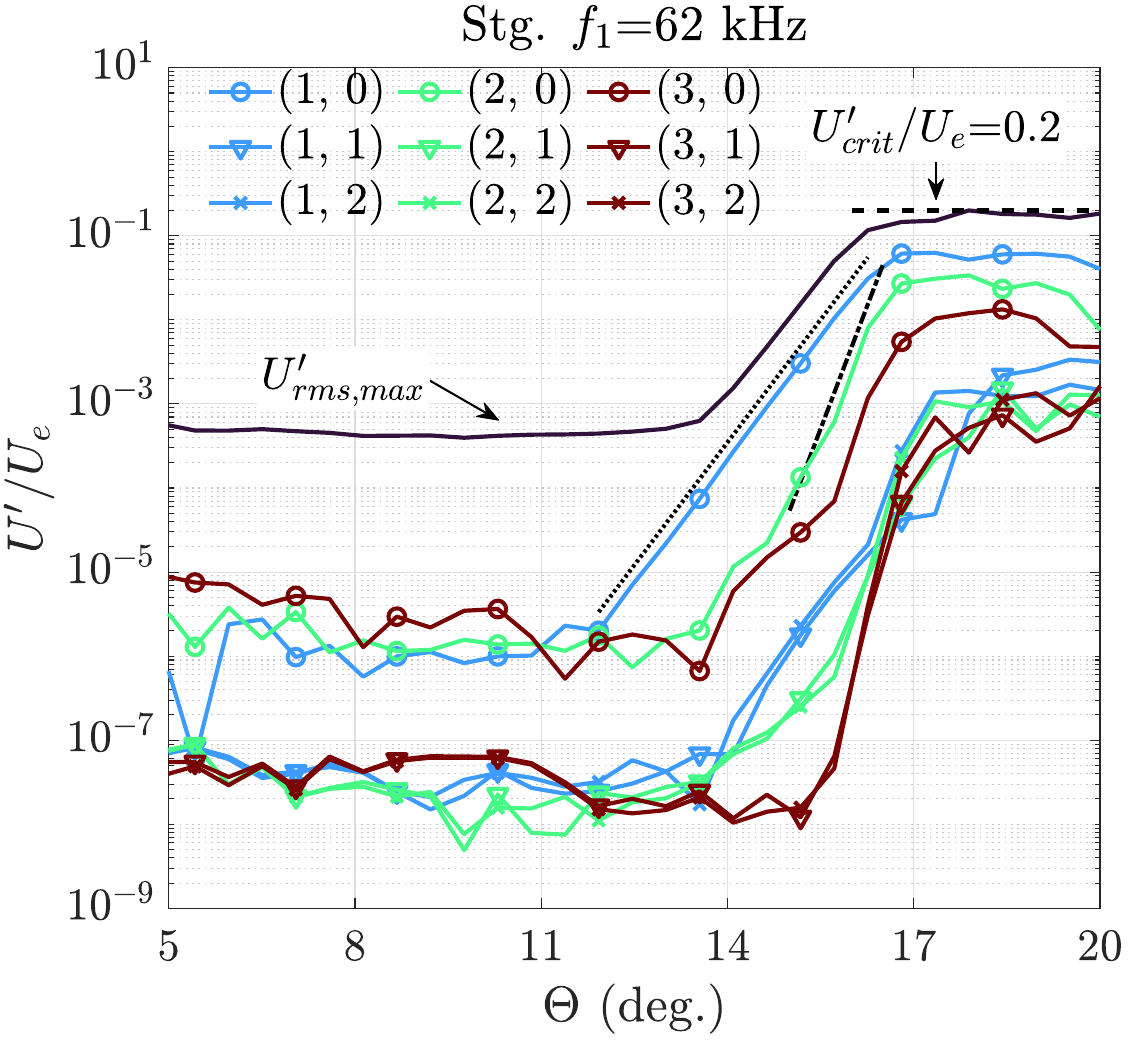}}
  \hfill
  \subfloat[Random]{%
    \includegraphics[width=0.32\textwidth]{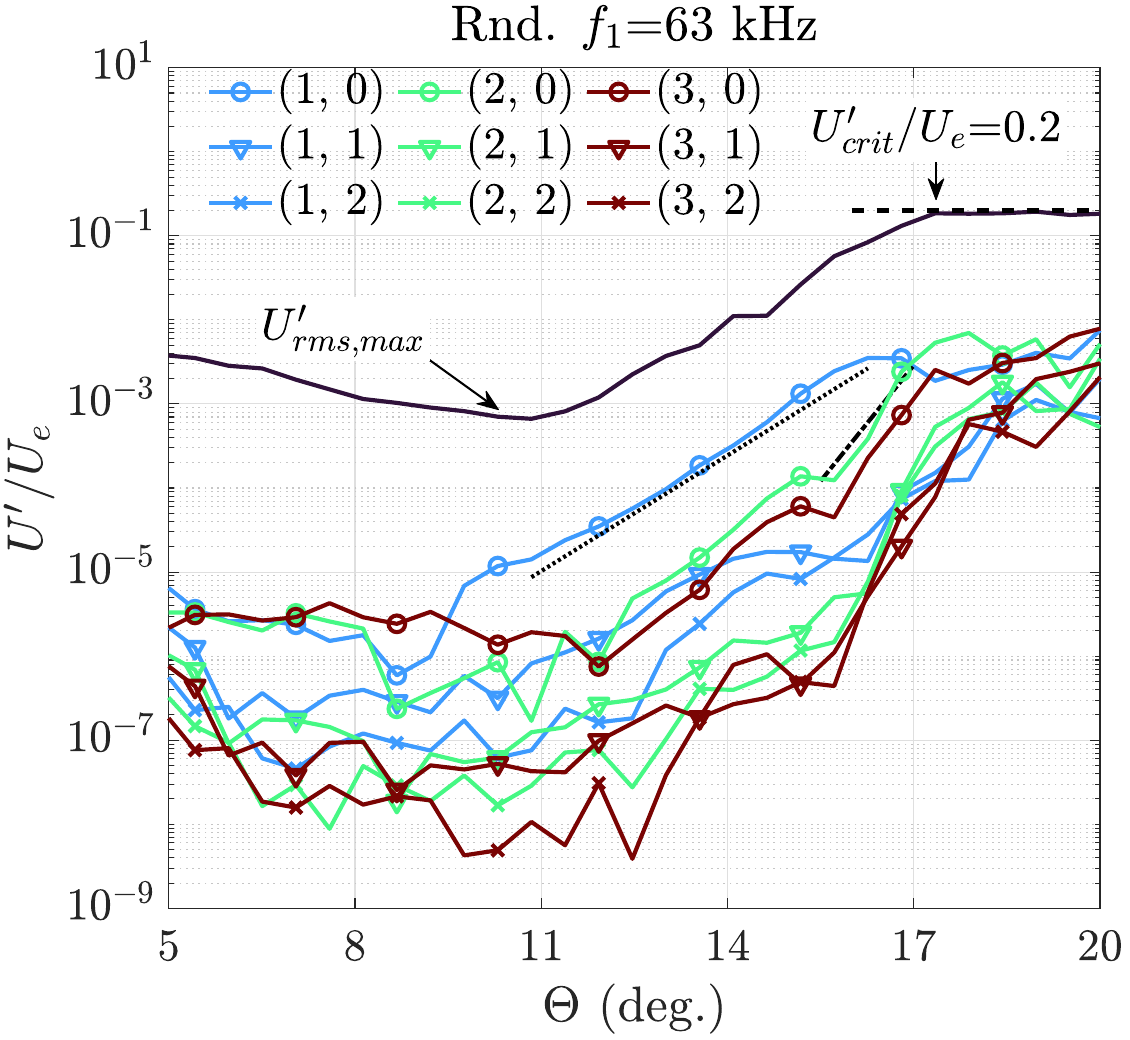}}
  \hfill
  \subfloat[Aligned]{%
    \includegraphics[width=0.32\textwidth]{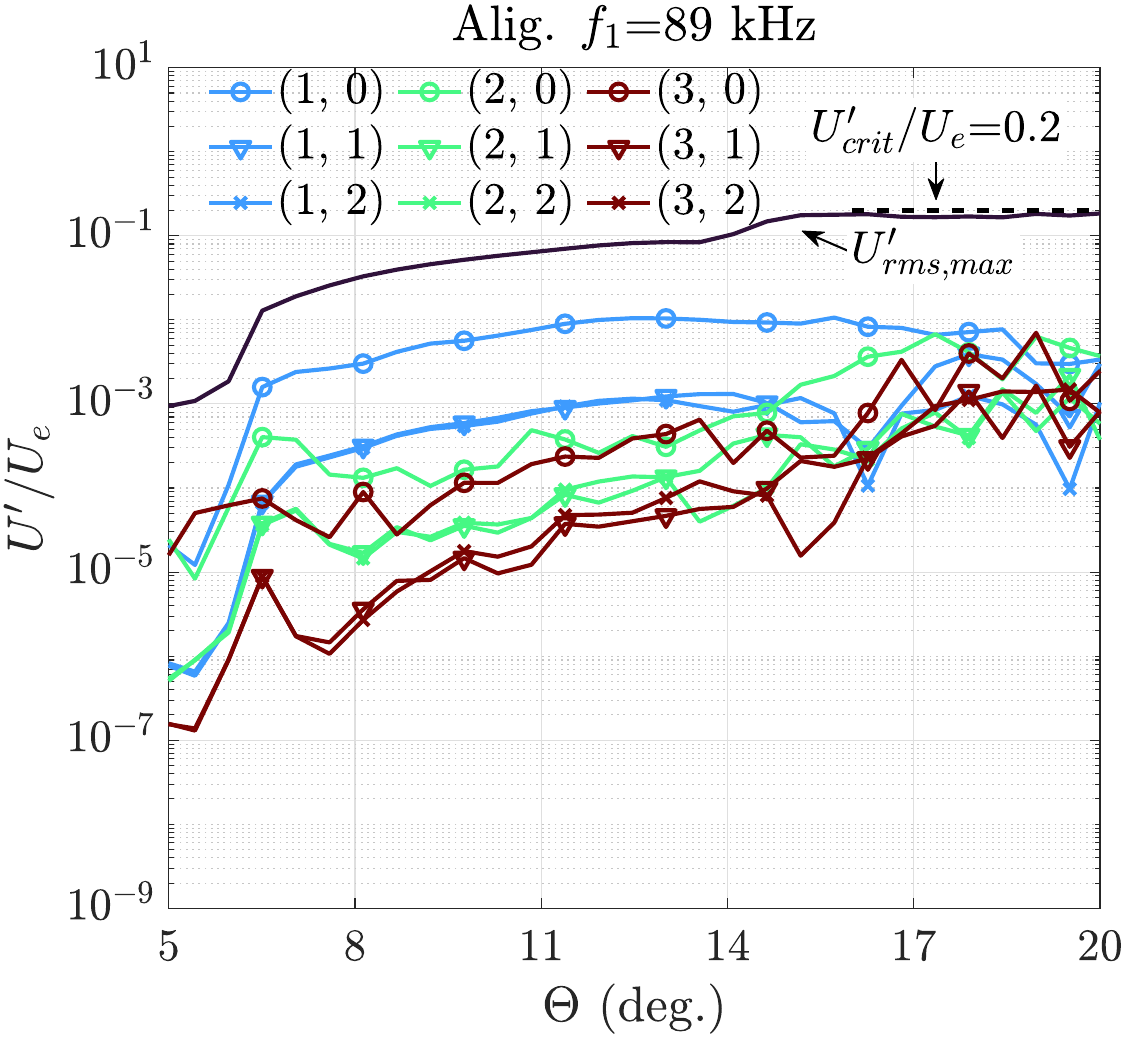}}
  \caption{Normalised amplitude development downstream of Fourier-transformed streamwise velocity for select frequency-spanwise wavenumber combinations for (a) staggered, (b) random, and (c) aligned, roughness elements.}
  \label{fig:dns_amplitude_curves}
\end{figure}

With the dominant frequencies and spanwise wavenumbers identified from the $f-\beta$ spectra, the amplitudes obtained from all $\eta-z$ cross-planes sampled between $\Theta=[5\degree,20\degree]$ by increments of $0.5\degree$ are shown in figure \ref{fig:dns_amplitude_curves}. The amplitudes reported are for integer multiples of the dominant frequency $f_1$ and the roughness wavenumber $\beta_{k}$. As was noted previously, the random case exhibited appreciable amplitude in non-integer multiples of $\beta_{k}$, however, were not plotted here for the sake of comparison. Beginning with the staggered configuration in figure \ref{fig:dns_amplitude_curves}(a), very large exponential growth of the $S_{1,0}$ pair can be seen beginning around $\Theta=11\degree$ until saturation at $\Theta=17\degree$. The relative growth between these two locations is roughly 4 orders of magnitude or, in terms of $N$-factor, $N=\ln(A/A_0)=9.2$. Spanwise harmonics at the primary frequency, i.e., $S_{1,1}$ and $S_{1,2}$, are about an order of magnitude lower, suggesting dominance of 2D flow structures. Their growth rates (i.e., slope) are very nearly the same as the $S_{1,0}$, which emphasises the incompleteness of the $f-\beta$ decomposition. The higher temporal harmonics like $S_{2,0}$ and $S_{3,0}$ grow at twice the rate (see black dash-dot line in \ref{fig:dns_amplitude_curves}(a)) beginning from $\Theta\approx13\degree$, confirming they are non-linearly generated by the dominant $S_{1,0}$. Also plotted is the streamwise velocity fluctuation root mean square (RMS) maximum in each plane (black solid line), which remains relatively flat at $0.1\%U_e$ until $\Theta=13.5\degree$ where the in-plane maximum begins tracking the $S_{1,0}$, evidenced by the similar slopes between those two curves.

The randomly arranged rows result in amplitude curves that behave somewhat similarly to the staggered, as shown in figure \ref{fig:dns_amplitude_curves}(b). Here, the $R_{1,0}$ for $f_{1}=63$ kHz begins its exponential growth earlier around $\Theta=9.0\degree$ and at a slower rate. The relative growth, while lower, is still large at roughly 3 orders of magnitude or $N=6.9$. Dominance of 2D structures is observable again by the fact that higher spanwise harmonics remain about an order of magnitude lower until near saturation around $\Theta=17\degree$ when the RMS reaches $20\%U_{e}$. The RMS curve also corroborates this with its slope matching very nearly the $R_{1,0}$ curve. 

Lastly, the amplitudes from the aligned case are shown in figure \ref{fig:dns_amplitude_curves}(c). Like the other two cases, there is dominance of the $A_{1,0}$ at $f_{1}=89$ kHz, however, the relative growth is significantly lower at roughly only one order magnitude or $N=2.3$. The presence of the forcing at $\Theta=6\degree$ for this case is evident in all curves by the sudden increase in amplitude near that location. The background noise before $\Theta=6\degree$ indicated by the RMS curve is roughly the same as the other cases at $0.1\%U_{e}$, however, given the low N factor, higher background fluctuations were needed for LTT to occur. Continuing to examine the RMS, there is a change in slope near $\Theta=13\degree$ that takes the amplitude from slightly below $10\%U_{e}$ up to $20\%U_{e}$. This change in slope could be a different instability mode in the flow taking over the RMS max search, suggesting there may be a more unstable disturbance than the 2D one biased by the introduced forcing. We re-emphasise the fact that all cases saturate at $U_{rms,max}=20\%U_{e}$, which is spatially very close to the sharp rise in heat flux (recall table \ref{tab:corr-trans-locs} and figure \ref{fig:dns_heat_flux}) typically taken as the location of LTT. This criterion combined with a model for disturbance growth could be used in an amplitude-based LTT prediction \citep{Mack1977,Fedorov2022}.

\begin{figure}[h]
    \centering
    \captionsetup{justification=centering}
    \includegraphics[width=0.9\linewidth]{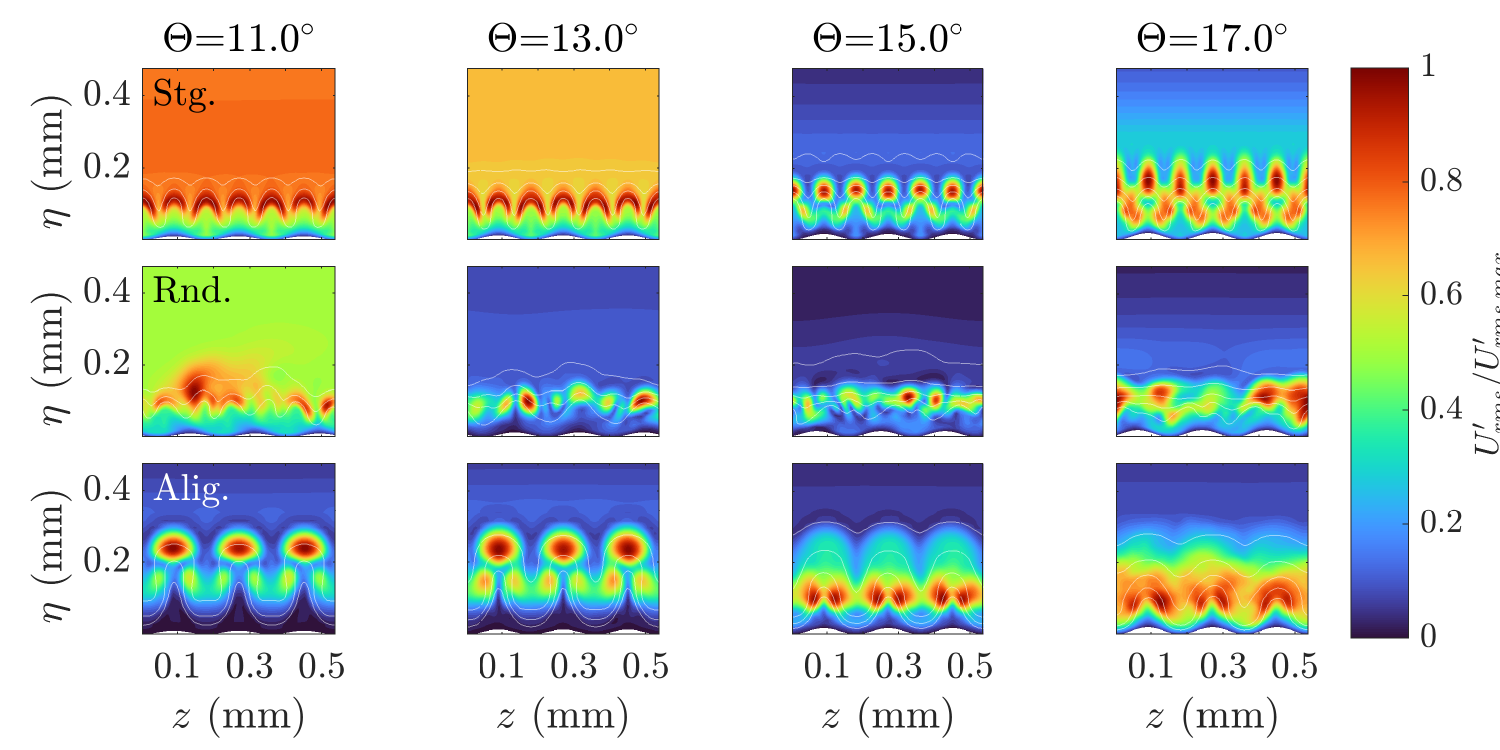}
    \caption{Streamwise velocity fluctuation root-mean-square (RMS) in the $\eta-z$ plane for several streamwise locations ($\Theta$) for staggered (top), random (middle), and aligned (bottom), roughness elements. White contour lines are fractions of time-averaged streamwise velocity in increments of 10\%.}
    \label{fig:dns_rms_planes}
\end{figure}

The character of the disturbances being tracked by the previously examined amplitude curves is better understood by considering the distribution of streamwise velocity RMS in several $\eta-z$ planes, as is shown in figure \ref{fig:dns_rms_planes}. Recalling figure \ref{fig:dns_amplitude_curves}(a), the first two streamwise locations of $\Theta=11\degree$ and $13\degree$ for the staggered case show maxima in the RMS distribution that do not grow downstream and are not of interest to the present discussion. The next position, $\Theta=15\degree$, is where growth of the dominant 62 kHz wave was observed in the RMS curve. The distribution shown at this location in figure \ref{fig:dns_rms_planes} (top row, third column from left) suggests a varicose streamwise velocity disturbance since the fluctuations are symmetric with respect to the centreline of the low speed streaks (recall mass flux planes in figure \ref{fig:dns_mass_flux_contours} with 6 minima across the span) located approximately at $\eta=0.15$ mm. The varicose distribution, while still present, becomes distorted with other peaks by $\Theta=17\degree$ as the flow begins to breakdown to turbulence. We also note the fluctuations that are spanwise uniform (i.e., 2D) in the upper boundary layer near $\eta=0.25$ mm visible in the $\Theta=15\degree$ plane. 

Similar to the staggered arrangement, the random case had decaying RMS maximum at $\Theta=11\degree$ (see figure \ref{fig:dns_amplitude_curves}(b)), and the distribution is not of interest to the present analysis. In the $\Theta=13\degree$ plane we see modes take hold not too much lower in the boundary layer than the staggered case at $\Theta=15\degree$. The peaks shift around moving to $15\degree$ as the underlying steady streaks shift according to the changing roughness element pattern. The fluctuations above the boundary layer are showing a wavelength of about the domain width. Moving down to $\Theta=17\degree$ saturation of these peaks is observed with peaks becoming more spread out across the span and activation lower in the boundary layer. The more complicated spatial structure of the peaks in planes $\Theta=13\degree$ and $15\degree$ are not easily categorised by the classic Floquet modes of varicose or sinuous nature and will be momentarily referred to as `mixed'.

The distributions of RMS for the aligned case shown in the bottom row of figure \ref{fig:dns_rms_planes} already exhibit varicose looking fluctuations at $\Theta=11\degree$. These fluctuations appear symmetric with respect the low speed streaks and, hence, have 3 peaks across the span. They are situated much higher at $\eta=0.25$ mm than the staggered case. This distribution is maintained in $\Theta=13\degree$ plane, however, a different distribution dominates the $\Theta=15\degree$ plane. The peak fluctuations are now straddling either side of the steady low speed streaks. Assuming there is a $180\degree$ phase shift between the peaks on either side, this mode dominating the signal would be considered a sinuous instability. It is this fluctuation shape that saturates in the $\Theta=17\degree$ plane and would appear to be taking over the breakdown to turbulence. This switch in distribution explains the kink previously highlighted for the RMS max amplitude curve in figure \ref{fig:dns_amplitude_curves}(c) and suggests a different mode other than the one primarily excited by the forcing is more important for the aligned case.

\subsection{Linear Stability Analysis} \label{sec:lde}

Time-averaged flow fields obtained from each DNS were then used as the mean flow for linear disturbance calculations (i.e., solving eq. \ref{eq:lde}) on subdomains. These domains were chosen to begin near the start of growth seen in the amplitude curves in figure \ref{fig:dns_amplitude_curves} and ended before saturation of the instabilities. Specifically, these domains extended $\Theta=[10.4\degree, 17\degree]$ for the staggered and random cases, and $\Theta=[7.9\degree,15\degree]$ for the aligned case. These were chosen judiciously, ensuring the residual from evaluating the governing CNSE (eq. \ref{eq:cnse}) using the time-averaged flow field remained below appropriate values. Specifically, the energy component of the system remained $L_{\infty,E}<4.0(10^{-5})$ for all roughness configurations. These calculations aimed to better understand the nature of the fluctuations found in the DNS by first introducing short duration `pulse' disturbances in the boundary layer, then tracking its evolution in time and space. The spatial parameters in eq. \ref{eq:dist_eq} were set to position the forcing interval approximately in the middle of the boundary layer $0.6\degree$ ($0.3\degree$) downstream of the start of the domain for the staggered and random (aligned) cases. The temporal parameters were chosen such that a broad range of frequencies that included the ones observed in DNS would be excited. Specifically, the pulse width was set to 1 $\mu s$. Given the linear nature of the analysis, the pulse amplitude, $A$, was arbitrarily set to $A=1$ for all cases. Three forcing wavenumbers in the spanwise direction were utilised, namely $m=\{0,1,2\}$, and will be referred to in the figures by their multiple of the roughness spanwise wavenumber. Recalling the clean streak counts in the mean flow fields (see figure \ref{fig:dns_mass_flux_contours}), pulses with a $1\beta_{k}$ spanwise shape would theoretically favour a subharmonic for the staggered and fundamental mode for the aligned cases. Likewise, forcing $2\beta_{k}$ would favour a fundamental mode for the staggered case. Given the 3 roughness element arrangement across the span and the periodic boundary condition, capturing a subharmonic for the aligned case was not possible. Finally, buffer zones were added at the inflow and outflow to ramp disturbances down to zero and mitigate wave reflections. 

\begin{figure}[h]
    \centering
    \captionsetup{justification=centering}
    \includegraphics[width=0.9\linewidth]{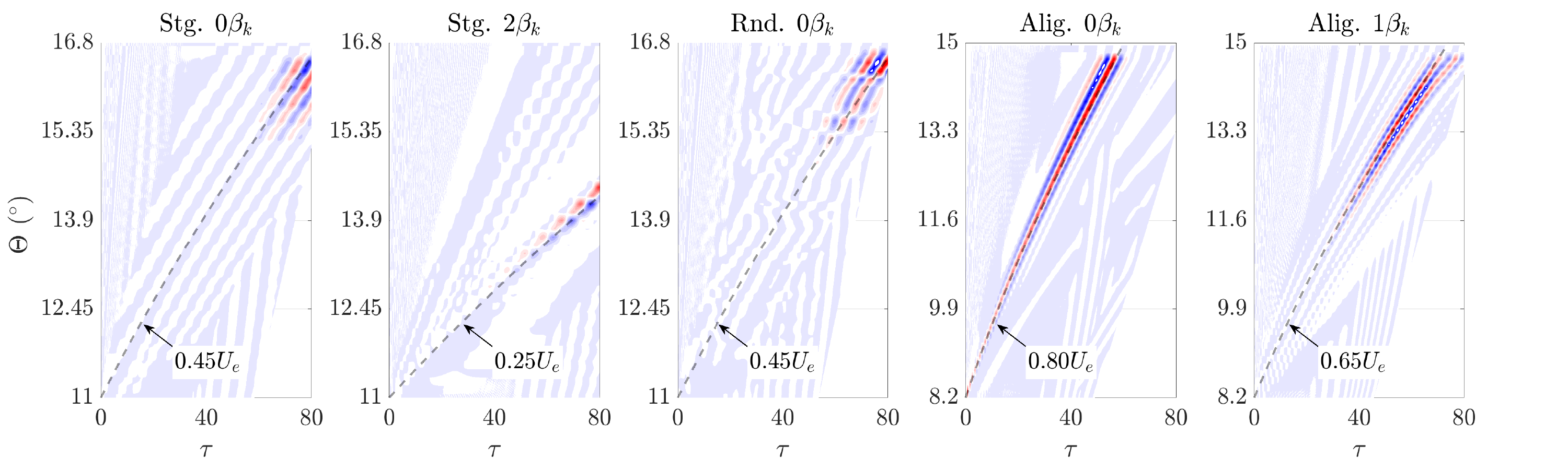}
    \caption{Contour plots of normalised (based on maximum) streamwise velocity disturbance probed from LDE pulse simulations at locations of maximum r.m.s. within each probe plane. The roughness configuration and forcing function initiating the pulse title each subplot.}
    \label{fig:lde_xt_diagrams}
\end{figure}

We first examine the space-time development of the streamwise velocity perturbation obtained from the LDE simulations in figure \ref{fig:lde_xt_diagrams}. These were obtained by examining the RMS distribution in each probe plane (similar to figure \ref{fig:dns_rms_planes}), then taking the space-time signal at the locations where the maximum consistently occurred. Time was made dimensionless using the mean boundary layer thickness from figure \ref{fig:dns_BL_development} of $\delta_{99.5\%h_{0,\infty}}\approx0.25$ mm and the local boundary layer edge velocity as $\tau = t (U_{e}/\delta_{99.5\%h_{0,\infty}})$. This was done to straighten lines of constant phase that were otherwise curved given the accelerating boundary layer. In all cases, the disturbances introduced by the pulse convect fully through the domain. Absolute instabilities were initially considered a possible mechanism given the increasing Reynolds number and fixed roughness height. These are clearly not at hand for the examined regions as the induced wave-packets propagate fully through the domain and no local absolute disturbance growth was observed to grow in time. Only 5 out of 9 simulations are shown since: Stg. $1\beta_{k}$ and Stg. $2\beta_{k}$ were identical; Rnd. $0\beta_{k}$, $1\beta_{k}$, and $2\beta_{k}$ were identical; and Alig. $1\beta_{k}$ and $2\beta_{k}$ were identical.

The $0\beta_{k}$ forcing in the staggered case (first panel in figure \ref{fig:lde_xt_diagrams}) excited a wave-packet that travelled very nearly $45\%U_{e}$ (marked with a black dashed line) consistently through the subdomain and amplified as it travelled downstream. Forcing with $2\beta_{k}$ (second panel in figure \ref{fig:lde_xt_diagrams}) (or $1\beta_{k}$) primarily excited an unstable wave-packet with a group velocity of about $25\%U_{e}$, which naturally took much longer to propagate fully through the subdomain. This longer time was allowed for in the simulation but truncated in the figure for the sake of comparison with other cases. Waves of similar speed as the $0\beta_{k}$ case are visible, but their amplitude are much lower. The random case with $0\beta_{k}$ (middle panel in figure \ref{fig:lde_xt_diagrams}) excited a wave-packet with near identical group velocity at $45\%U_{e}$ as Stg. $0\beta_{k}$, indicating a seemingly unlikely similarity in the dominant instability between the two roughness patterns. We re-emphasise the dominant wave-packet's group velocity for the random case is independent of the spanwise shape of the initiating pulse disturbance. The use of a single probe location in each $\eta-z$ plane explains the occasional wiggles in contours of constant phase as the mode peak shifts around with the underlying steady streaks, similar to what was seen in the RMS planes from DNS in figure \ref{fig:dns_rms_planes}. The aligned configuration, like the staggered, showed dependence on the initiating disturbance shape, which is illustrated in the fourth and fifth panels in figure \ref{fig:lde_xt_diagrams} for $0\beta_{k}$ and $1\beta_{k}$ forcing shapes, respectively. Specifically, $0\beta_{k}$ initiates a wave-packet that travels at $80\%U_{e}$, which is faster than the group velocity with $1\beta_{k}$ forcing of $65\%U_{e}$. Again, forcing $2\beta_{k}$ yielded the same behaviour as $1\beta_{k}$ and is omitted for brevity.

\begin{figure}[h]
    \centering
    \captionsetup{justification=centering}
    \includegraphics[width=0.95\linewidth]{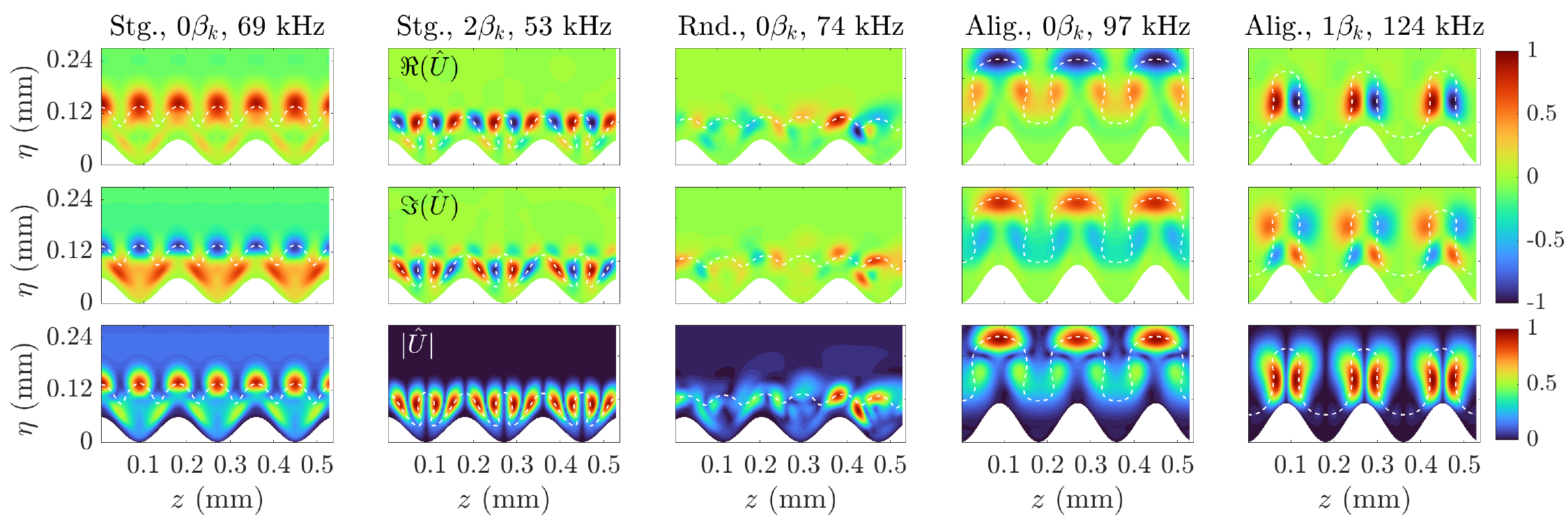}
    \caption{Contour plots of normalised (based on max./min.) Fourier-transformed streamwise velocity disturbance components across the $\eta-z$ plane at $\Theta=14\degree$ for Stg. and Rnd., and $\Theta=12\degree$ for Alig. White dashed lines represent contour lines of the closest fraction of boundary layer edge velocity that matches wave-packet group velocity. The roughness configuration, forcing function, and selected frequency title each column.}
    \label{fig:lde_fft_planes}
\end{figure}

A FFT (in time only) was then performed on each $\eta-z$ plane to examine the spatial structure of the instabilities dominating the wave-packets shown in the space-time diagrams. Each frequency's FFT distribution, $\hat{U}(\eta,z)$, was compared to the RMS distribution through a normalised inner product
\begin{equation}
    \psi = \frac{\int \hat{U}(\eta,z) U_{rms}(\eta,z) \,d\eta dz}{\int U_{rms}(\eta,z)^{2} \,d\eta dz}.
\end{equation}
The frequency with highest $\psi$ for each simulation was plotted in figure \ref{fig:lde_fft_planes}. The Stg. $0\beta_{k}$ at 69 kHz portrays a varicose mode shape peaking at the tops of the mean flow streamwise velocity contour line (chosen at the same level as that shown in the space-time diagrams). The real and imaginary parts illustrate that the peaks oscillate not only symmetrically about the centreline of the low speed steady streaks, but also in phase together across the span. This suggests the disturbance is predominantly 2D with modulation due to the streaks. With $2\beta_{k}$ at 53 kHz, the streamwise velocity fluctuations are concentrated at the sides of the $25\%U_{e}$ mean flow contour line. The real and imaginary parts confirm these fluctuations are anti-symmetric with respect to the low speed streaks, therefore deserving the classification `sinuous'. The resemblance of the varicose mode to the RMS distribution in the plane from DNS (figure \ref{fig:dns_rms_planes}) is noteworthy and suggests dominance of this mode in the DNS. The aligned cases exhibit similar distinction between a varicose mode at 97 kHz that rides the mean flow high above the roughness elements (hence the high group velocity of the wave-packet), and a sinuous mode at 124 kHz that peaks lower in the boundary layer and rides the sides of the low-speed streaks. Likewise, the real and imaginary parts are useful to distinguish the modal character (i.e., symmetry or anti-symmetry with respect to the low speed streak centre lines). The varicose mode has lower frequency than the sinuous in this roughness configuration, which is opposite in the staggered arrangement. The FFT distributions from the random case is not as easily categorised as varicose/sinuous modes. In fact, some regions show fluctuations that look anti-symmetric with respect to the underlying streak ($z=0.2$ mm), and some look symmetric ($z=0.3$ mm). The two largest peaks in magnitude are near $z=0.45$ mm and could be interpreted as sinuous, though that classification is likely to change from plane to plane. As previously seen in the DNS RMS planes (figure \ref{fig:dns_rms_planes}), we note the fluctuations are higher in the boundary layer for the $0\beta_{k}$ cases. The space-time diagrams from these fluctuations higher in the boundary layer match those taken from the lower, more dominant peak, showing they are of the same boundary layer mode.

\begin{figure}[h]
    \centering
    \captionsetup{justification=centering}
    \includegraphics[width=0.9\linewidth]{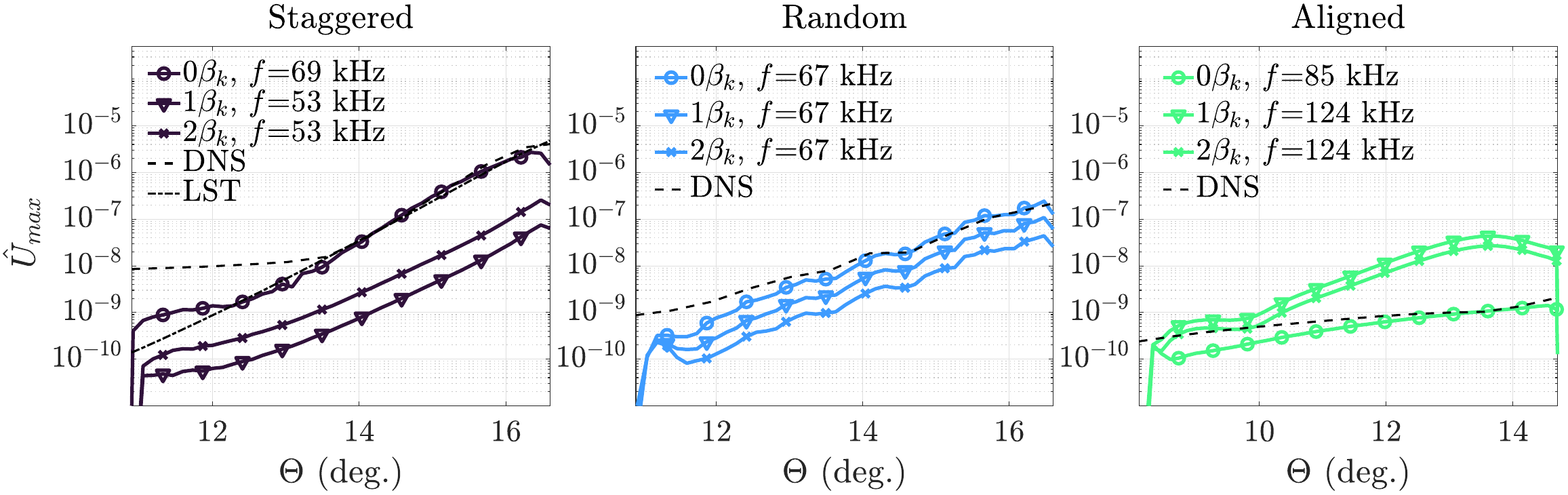}
    \caption{Downstream evolution of (temporal only) FFT streamwise velocity perturbation maximum from LDE pulse simulations. The roughness configuration titles each subplot, and the legend therein signifies the initiating pulse shape and the frequency being tracked. Rescaled DNS RMS and LST (Stg. at $(f,\beta)=(69 \text{kHz},0)$) are included for reference.}
    \label{fig:lde_fft_max_curves}
\end{figure}

To quantify the previously alluded to growth of the frequencies dominating the wave-packets, the maximum in (temporal only) FFT streamwise velocity perturbation in the $\eta-z$ plane was tracked downstream for each roughness case and forcing shape and is shown in figure \ref{fig:lde_fft_max_curves}. Subjected to the $0\beta_{k}$ forcing (i.e., 2D), the staggered case's varicose mode had a longer receptivity region but achieves a higher growth rate than the sinuous modes for the $1\beta_{k}$ and $2\beta_{k}$ sinuous modes. They all experience tremendous relative growth, though, of roughly 3 orders of magnitude or $N=6.9$ within the LDE subdomain. The agreement in growth rate of the varicose mode from the LDE simulation and the DNS RMS maximum curve rescaled from figure \ref{fig:dns_amplitude_curves}(a) confirms the dominance of the varicose mode in the DNS. In addition, it was found that LST \citep{lastrac_manual} performed on the spanwise and streamwise wavelength averaged profiles (see figure \ref{fig:dns_BL_development}) for this case yielded unstable modes that agree remarkably well with the LDE and DNS simulations in terms of amplitude (black dash-dot line in figure \ref{fig:lde_fft_max_curves}(left)) and phase speed. This was not the case for the random and aligned cases where the phase speeds did not match and the results were omitted.  

The wave-packets over the randomly phased roughness elements (middle panel of figure \ref{fig:lde_fft_max_curves}) all grow significantly, albeit less smoothly than the staggered cases, by about 3 order of magnitude or $N=6.9$. The `mixed mode' takes hold most quickly when initiated from a pulse with $0\beta_{k}$ shape, and less so as $m\beta_{k}$ increases. Besides this initial difference in receptivity, all forcing shapes result in wave-packets dominated by the same `mixed mode' as indicated by the identical growth rates. The agreement with the DNS RMS maximum is good starting downstream of $\Theta=14\degree$ and illustrates this mode dominating LTT in the DNS for that case as well. Note, a slightly different frequency was plotted here than figure \ref{fig:lde_fft_planes} to facilitate agreement with the DNS amplitude. The need of a slightly lower frequency simply highlights a bias towards higher frequencies in the wave-packet than the background noise in the DNS.

Finally, the varicose mode arising from $0\beta_{k}$ forcing undergoes very little growth at just one order of magnitude or $N=2.3$ (right panel in figure \ref{fig:lde_fft_max_curves}). On the other hand, there is much more growth for the $1\beta_{k}$ and $2\beta_{k}$ induced sinuous modes with about 2 orders of magnitude relative growth or $N=4.6$. Similar to the staggered arrangement, forcing a non-fundamental wavenumber of the steady streaks (i.e., $2\beta_{k}$ for the aligned case) only affects initial receptivity, but once the fundamental sinuous mode takes hold it grows at the same rate as the $1\beta_{k}$ case. The lack of agreement here with the DNS RMS maximum indicates the forcing introduced upstream at $\Theta=6\degree$ for the DNS favours slightly different frequencies than when initiating at $\Theta=8.2\degree$ as in the LDE simulations. That said, the slow growth observed in the DNS RMS maximum is more similar to the $0\beta_{k}$ varicose mode than the sinuous modes that grow more quickly.

\begin{figure}[h]
    \centering
    \captionsetup{justification=centering}
    \subfloat[]{%
    \includegraphics[width=0.43\textwidth]{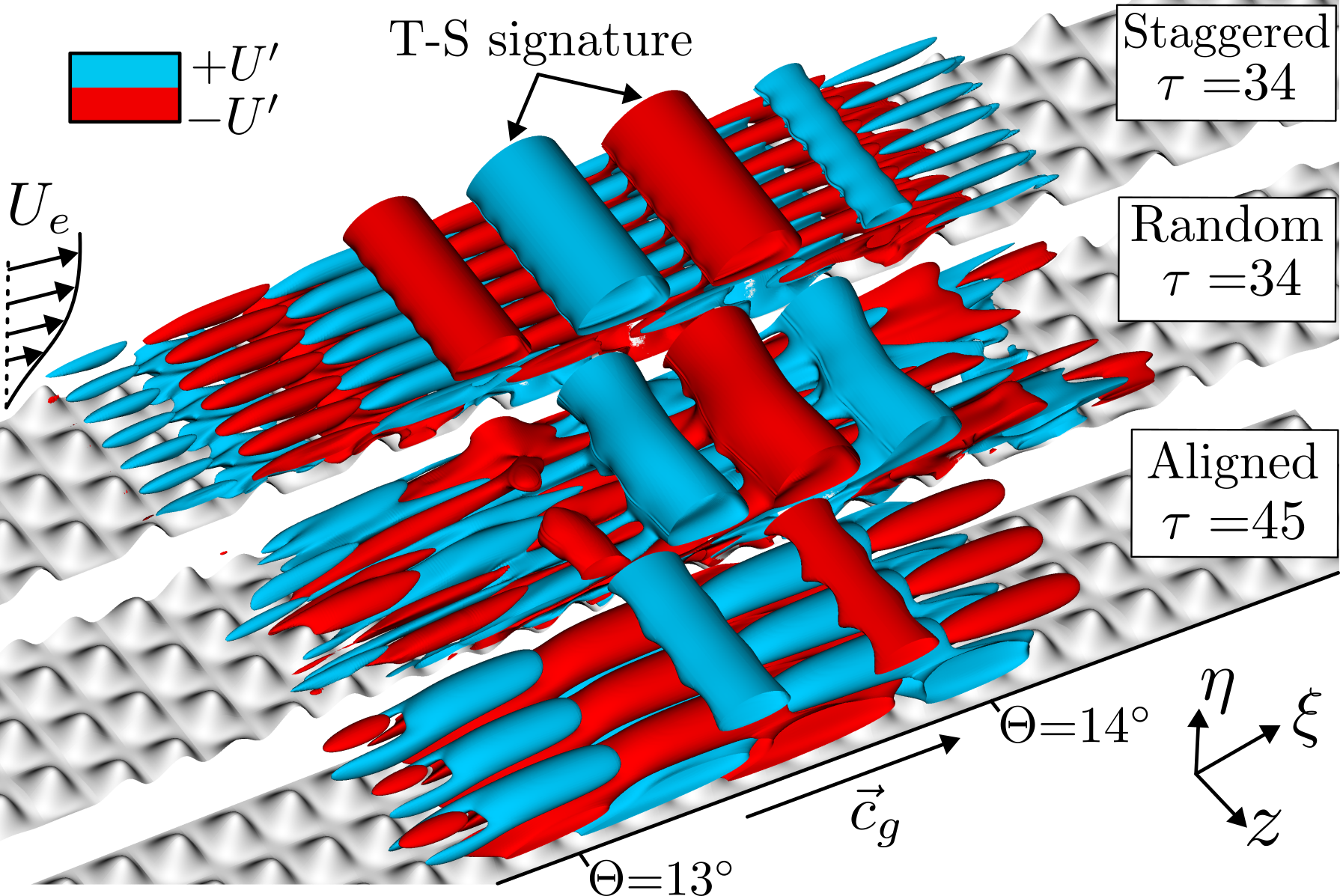}}
    \subfloat[]{%
    \includegraphics[width=0.21\textwidth]{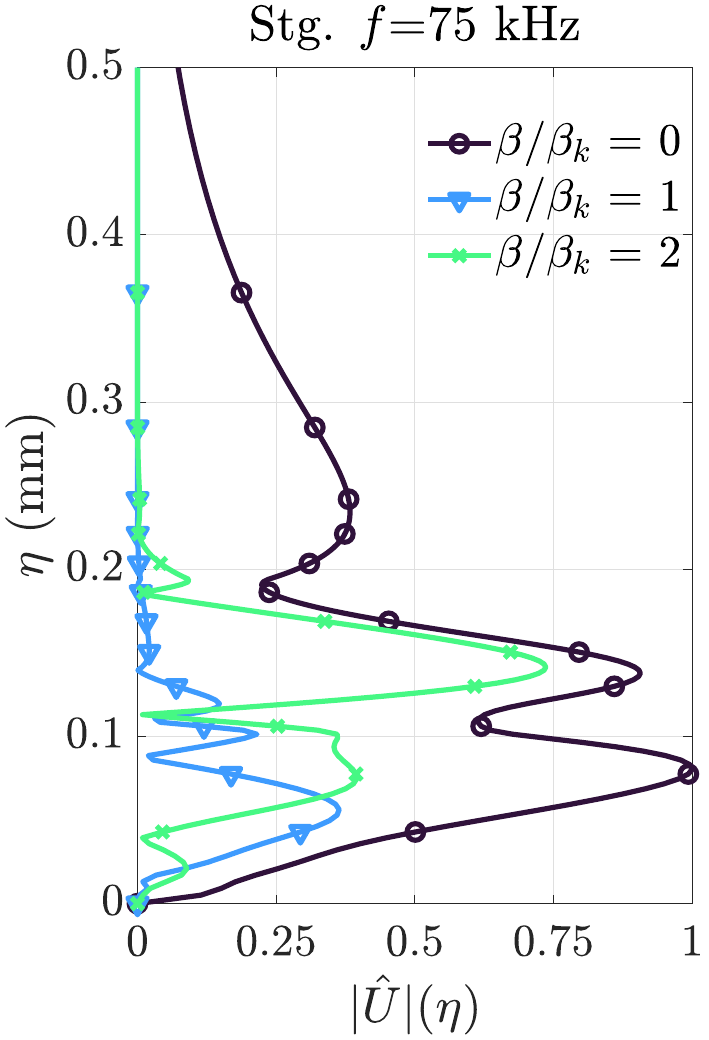}}
    \subfloat[]{
    \includegraphics[width=0.33\textwidth]{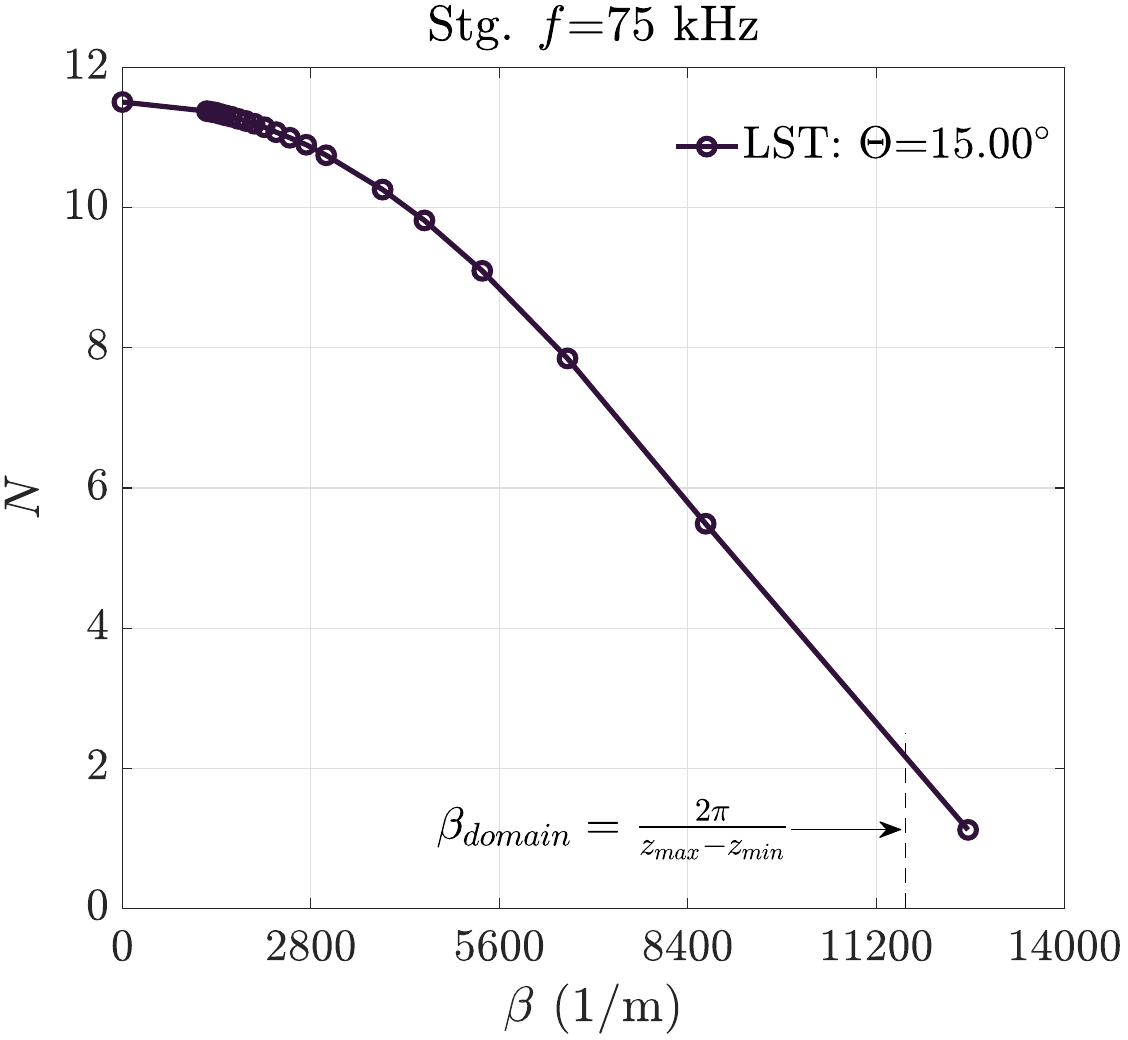}
    }
    \caption{(a) Iso-contours of streamwise velocity disturbance from LDE simulations using spanwise uniform initiating pulse disturbances. Each case is saturated to highlight the 2D T-S spatial structure. (b) Wall-normal distribution of normalised streamwise velocity perturbation Fourier transformed in time and spanwise direction. (c) T-S $N$-factor variation with respect to spanwise wavenumber from LST at $f=75$ kHz for the staggered case.}
    \label{fig:lde_wavepackets}
\end{figure}

Given the similarity between the modes obtained by $0\beta_{k}$ initiating pulse shapes in the LDE simulations and those found in the DNS, we take a closer look at the spatial structure of the wave-packet at locations where the varicose mode dominates in the LDE simulations. Since the pulse was initiated more upstream in the aligned case, time instant corresponding to $\tau=45$ was used to compare with the staggered and random cases. All three wave-packets are visualised in figure \ref{fig:lde_wavepackets}(a) by iso-surfaces of streamwise velocity perturbation. The levels were chosen such that the structure of the fluctuations above the boundary layer may be appreciated. We find 2D wave fronts above the peak fluctuations shown lower in the boundary layer in the FFT planes (see figure \ref{fig:lde_fft_planes}). Very clean shapes are observed for the staggered and aligned wave-packets, while the random one has some slight undulations in span of the otherwise largely 2D fronts. Indeed, these 2D structures are the signature of the Tollmien-Schlichting (T-S) instability typically found in incompressible boundary layers and here, for the first time, are exposed as a relevant instability for LTT over a hypersonic blunt body's rough surface. Keep in mind the smooth wall boundary layer is modally stable and no unstable T-S wave can be found. 

Finally, to confirm the T-S signature, a $0\beta_{k}$ monochromatic continuous (in time) volume forcing at $f=75$ kHz replaced the Gaussian pulse in the staggered case so the entire flow field could be Fourier-transformed at this frequency and the wall-normal disturbance distribution could be inspected further from the wall. Plotting the distributions after an FFT in $z$ (see figure \ref{fig:lde_wavepackets}(b)) showcases the classic T-S eigenfunction with the M-shape modification previously reported by \citet{Cossu_2004} for the streaky Blasius flat plate boundary layer. Since LST performed on the staggered mean flow profiles in figure \ref{fig:dns_BL_development}(b-c) provided such good agreement with the DNS and LDE (as was shown in figure \ref{fig:lde_fft_max_curves}), LST was used to perform a $\beta$ sweep for values much smaller than the spanwise domain width permits. As figure \ref{fig:lde_wavepackets}(c) shows, the $N$-factor decreases monotonically while $\beta$ increases from 0, thereby supporting the T-S classification rather than as first Mack modes where $N$ achieves its maximum at $\beta\ne0$.

\begin{figure}[h]
    \centering
    \captionsetup{justification=centering}
    \includegraphics[width=0.9\linewidth]{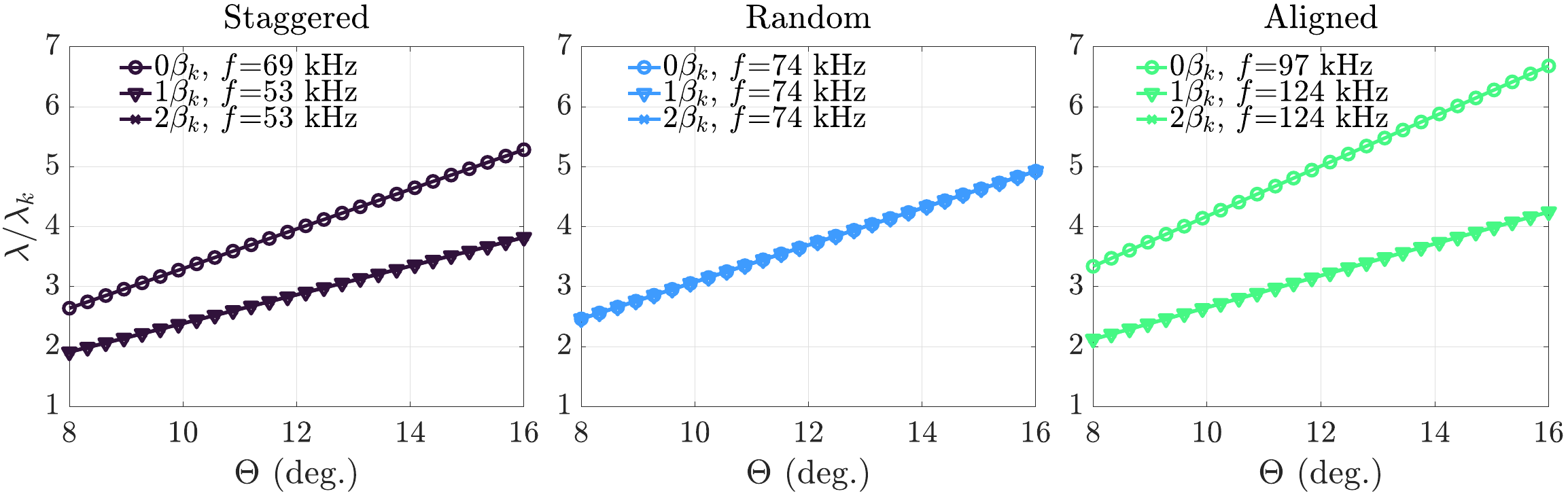}
    \caption{Disturbance streamwise wavelength evolution for most amplified modes identified in linear (i.e., LDE) pulse simulations for staggered (left), random (middle), and aligned (right), roughness configurations.}
    \label{fig:lde_wavelengths}
\end{figure}

We conclude the linear analyses with inspection of the streamwise wavelengths, $\lambda = c/f$, of the dominant varicose T-S waves and sinuous instabilities from each case, as shown in figure \ref{fig:lde_wavelengths}. Interestingly, the streamwise disturbance wavelengths for the varicose waves are very similar between the three roughness configurations in the way they start at roughly $\lambda/\lambda_{k}=[2.5,3.5]$ near $\Theta=8\degree$ and increase to $\lambda/\lambda_{k}=[5.0,6.5]$ by $16\degree$. The wavelengths of the sinuous instabilities are similar among the cases (where they were detectable), starting at roughly $\lambda/\lambda_{k}=2$ near $\Theta=8\degree$ and increased more slowly up to $\lambda/\lambda_{k}=4$ by $\Theta=16\degree$. These wavelengths were reported with division by the streamwise wavelength of a single roughness element row since it was suspected that the dominant waves may be phase-locked to this wavelength. This does not appear to be the case; however, it is noteworthy that the frequencies and phase speeds of the two types of instabilities are such that the streamwise wavelengths come out very similar between all roughness patterns. 

\subsection{Breakdown to Turbulence}

Returning back to the DNS results with newfound understanding of the dominant instabilities, the late stages of LTT are now examined. The following section has been included to demonstrate that latter stages of LLT are occurring, including breakdown and the beginnings of turbulent flow. Some flow visualisation and turbulence analysis have been included to a limited extent; however, turbulence analysis is not within the scope of this paper is left as future work. The final stages, also called `breakdown', is commonly visualised through the Q-criterion. This quantity is helpful in identifying regions of strong rotation with low shear, which are synonymous with vortices. A key vortex structure in the breakdown process is the formation of hairpin vortices. These are important as their heads localise strong shear, thereby enabling tertiary instabilities to grow and for the energy cascade of turbulence to commence.

\begin{figure}[h]
    \centering
    \captionsetup{justification=centering}
    \includegraphics[width=0.94\linewidth]{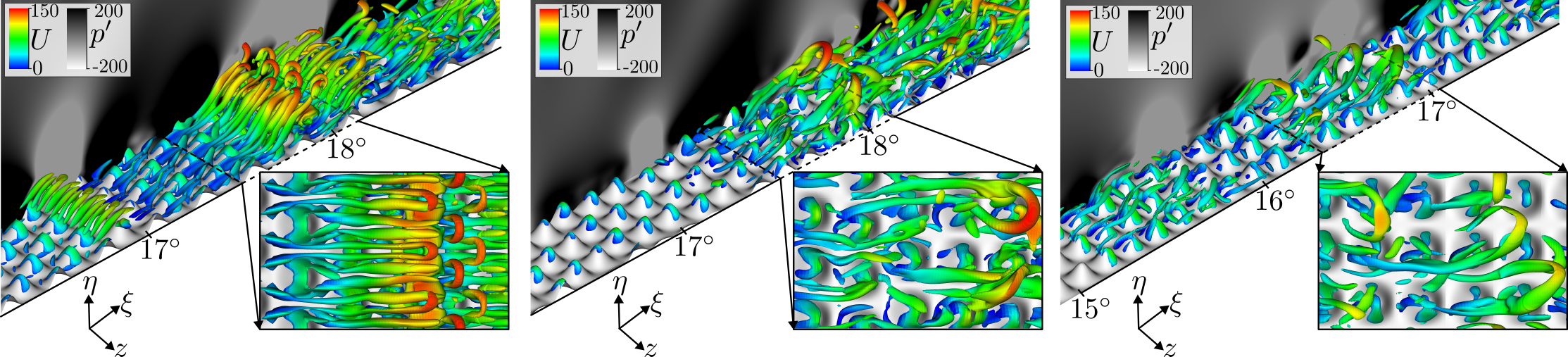}
    \caption{Iso-surfaces of $Q=5(10^{11})$ coloured by streamwise velocity for staggered (left), random (middle), and aligned (right), roughness configurations. Instantaneous pressure fluctuation contours are also shown in the $z=0$ plane.}
    \label{fig:q_criterion}
\end{figure}

Figure \ref{fig:q_criterion} shows iso-surfaces of $Q=5(10^{11})$ coloured by streamwise velocity with a slice of the instantaneous pressure disturbance field provided in the background at the $z=0$ plane. The first panel in figure \ref{fig:q_criterion} corresponding to the staggered case showcases a packet of hairpin vortices forming right in front of a high pressure fluctuation front. The packet, when viewed from the top, clearly has 3 rows of 3 hairpins across the span. Each row is staggered with respect to the previous one. The legs of the hairpins extend down to the sides of an element and show how they are being seeded by the steady vorticity field already examined in section \ref{sec:mean_flow} (see figure \ref{fig:mean_vorticity}). The aligned element arrangement shown in the right panel of figure \ref{fig:q_criterion} shows the influence of the more dominant sinuous instability in that case. Around $\Theta=15.5\degree$ the vortices originating off the fronts of the elements meander first right then left. Further downstream around $\Theta=18\degree$, we can see a vortex originating on the right side of the middle column of elements reach over to the vortex formed on the left side to form a hairpin vortex structure. This happens simultaneously on either side of the middle column, hence, 3 hairpins are visible across the span. The action of the sinuous instability to move the vortices right and left makes the formation process appear less efficient compared to the staggered case where the flow is accelerated over the element rows uniformly across the span. Finally, the random case (middle panel in figure \ref{fig:q_criterion}) is similar to the aligned in that vortices form less compactly, however, the most clearly identifiable hairpin has legs narrowly spaced and coming off a single element much like the staggered case. It is much wider towards the top and its development seems interrupted by the surrounding vortices forming. In all cases, one can see the hairpins forming behind a high pressure fluctuation instability wave. In this way, it is easy to see how the dominant instability wave, when it reaches critical amplitude, triggers the formation of hairpin vortices seeded by the underlying steady vorticity field.

\begin{figure}[h]
    \centering
    \captionsetup{justification=centering}
    \includegraphics[width=0.9\linewidth]{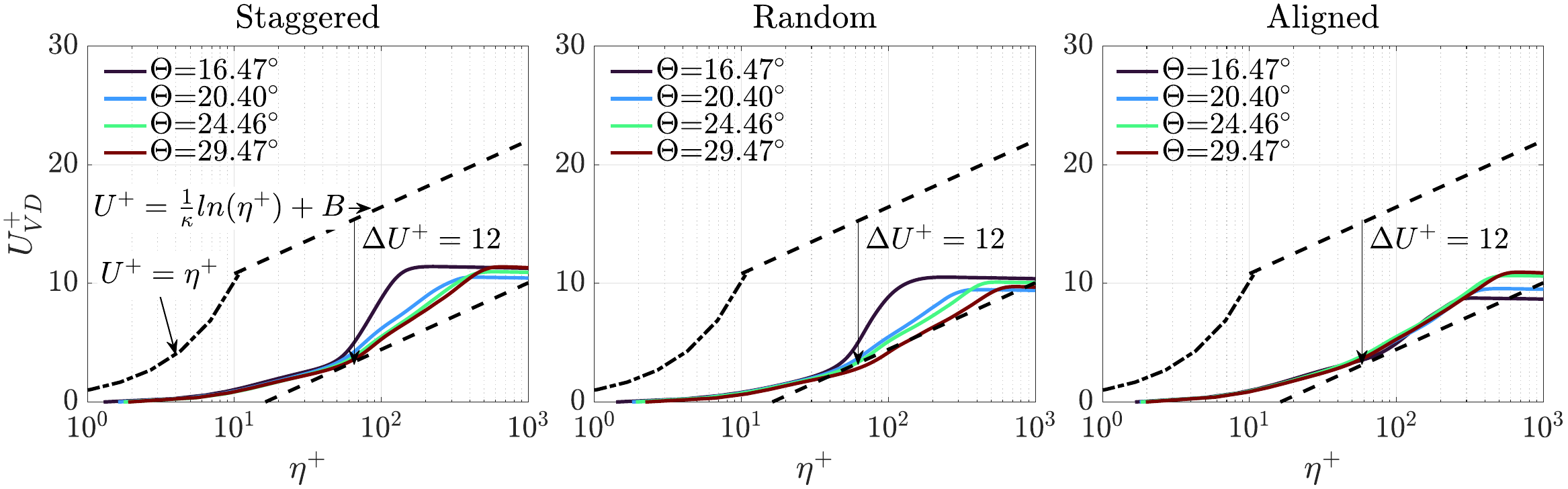}
    \caption{van Driest transformed time-span-streamwise roughness wavelength-averaged profiles for staggered (left), random (middle), and aligned (right), roughness configurations.}
    \label{fig:turb_profiles}
\end{figure}

To get an idea of how far the flow has advanced towards a turbulent state, the \cite{VanDriest1951} transformed velocity profiles are shown in figure \ref{fig:turb_profiles} for all three cases. Recall, the roughness ends at $\Theta=30\degree$ and the grid stretching commences thereafter. Thus, the last profile for which valid data were available is $\Theta=29.47\degree$. The wall shear stress, $\tau_w$, necessary for the velocity transformations in inner-scaled (plus) units was obtained by summing both the viscous shear and pressure drag effects, integrated across the entire span and over one roughness wavelength in the streamwise direction. The viscous component is given by $\frac{1}{A}\int_S\tau_vdS$, where S is the wetted surface area, A is the projected planform area, $\tau_v=\left.\mu\frac{\partial \vec{u}}{\partial n}\right|_w\cdot\hat{e}_\xi$ is the local tangential viscous wall shear stress (specifically the streamwise component), $\mu$ is the dynamic viscosity, $\vec{u}$ is the Cartesian velocity vector, $n$ is the wall-normal direction (distance), subscript $w$ indicates a `wall' quantity, and $\hat{e}_\xi$ is the streamwise unit vector. The pressure component is given by by $\frac{1}{A}\int_S-p_w(\hat{n}\cdot\hat{e}_\xi)dS$, where $p_w$ is the surface pressure, and $\hat{n}$ is the wall-normal unit vector. 

The profiles from the staggered case shown in the left panel of figure \ref{fig:turb_profiles} shows a progression towards the $\Delta U^{+}$ shifted log-law, however, the slope remains steeper than the classical curve given by $U^{+}=\frac{1}{\kappa}\ln(y^{+})+B$, where $\kappa=0.41$ and $B=5.0$. The random case behaves similarly but achieves a slope more closely following the log-law. Finally, the aligned case by $\Theta=17\degree$ already exhibits a log-layer forming, which persists in the downstream locations. Like the staggered case, the slope in the log-layer is steeper than the $\Delta U^{+}$ shifted log-law, a consequence, perhaps, of the symmetries imposed by the roughness pattern. In addition, larger Reynolds numbers downstream lead to the velocity profiles remaining linear (in semi-logarithmic axes) for larger $\eta^+$, i.e., the profile aligns with the log-law for longer. Based on the wall shear stress calculation $\tau_w$  and $u_\tau$ increase downstream, the vertical shift due to the roughness also becomes larger the further downstream, indicating as expected that the shift is proportional to the roughness Reynolds number $k^+=\frac{ku_\tau}{\nu_w}$, where $k$ is the roughness height, $u_\tau$ is the friction velocity, and $\nu_w$ is the kinematic viscosity at the wall.

\subsection{Upstream waves and Feedback Mechanism} \label{sec:background_noise}

\begin{figure}[h]
    \centering
    \captionsetup{justification=centering}
    \subfloat[]{
    \includegraphics[width=0.35\linewidth]{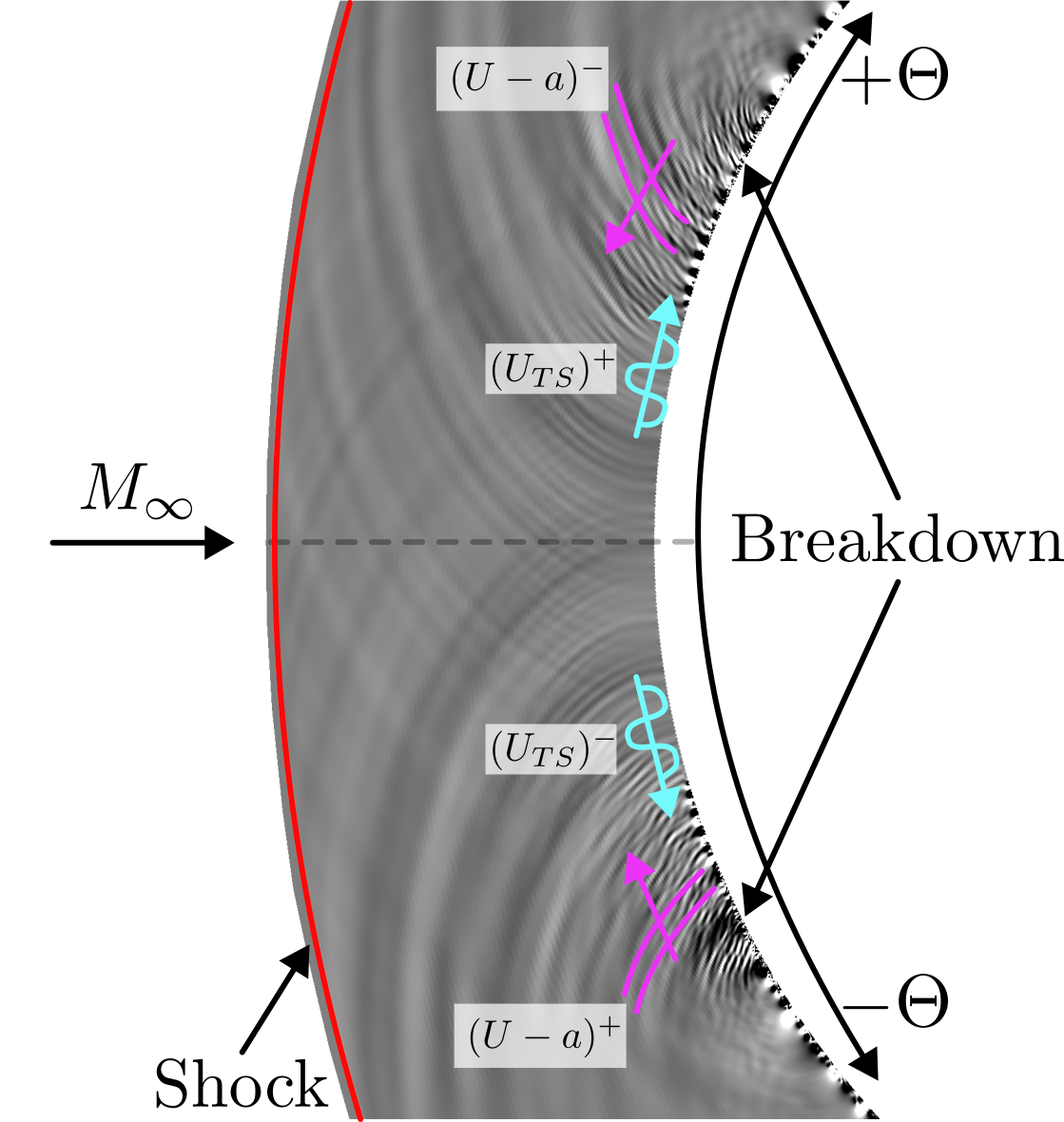}}
    \subfloat[] {
    \includegraphics[width=0.55\linewidth]{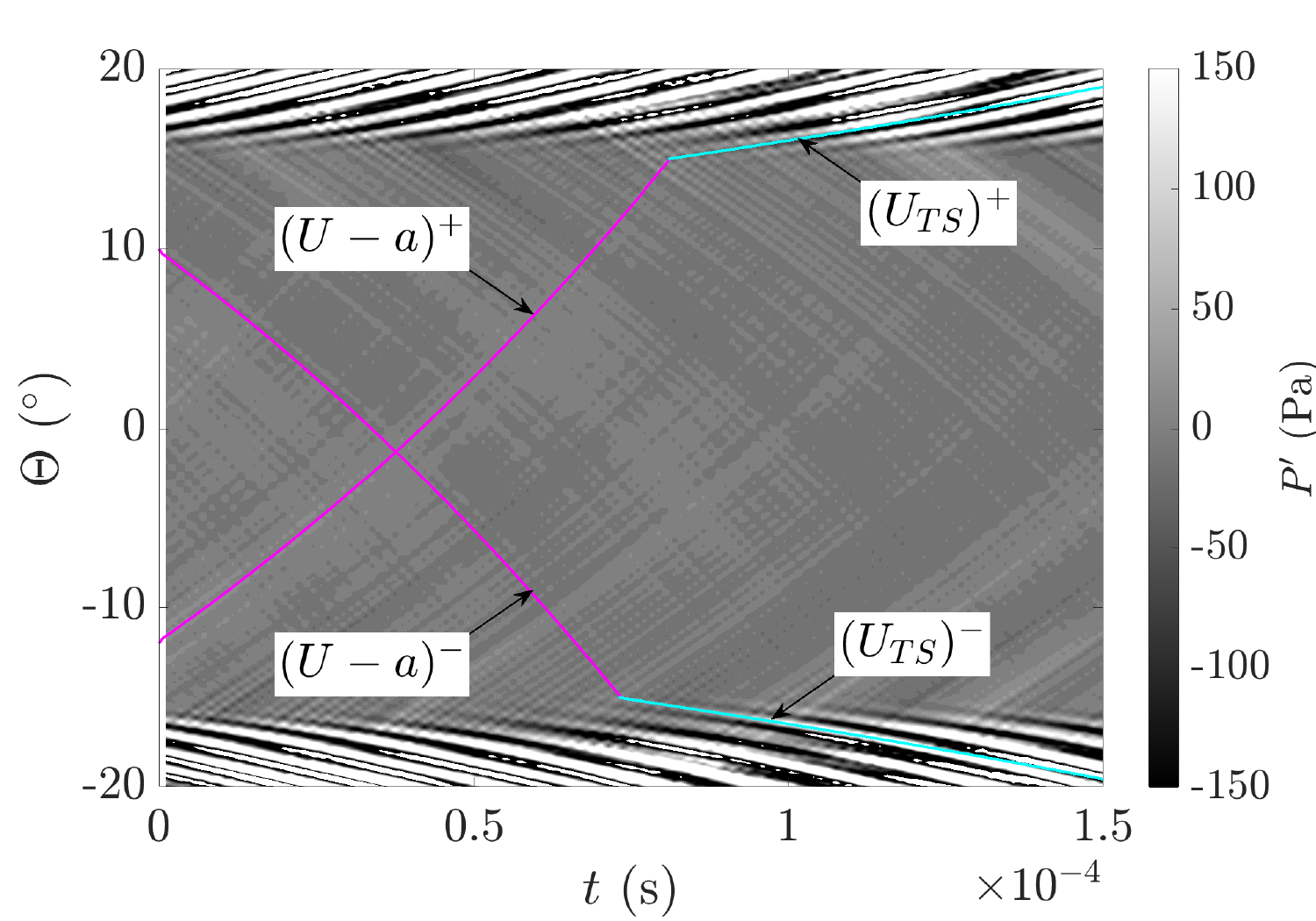}}
    \caption{Pressure disturbance field at $z=0$ plane (a) and temporal signal at boundary layer edge (b) supporting proposed feedback mechanism. Data shown from the staggered roughness case.}
    \label{fig:upstream_waves}
\end{figure}

We now formally address the nature of `background noise' present in the staggered and random roughness simulations. First, the disturbance pressure field at plane $z=0$ for the staggered case is shown in the left part of figure \ref{fig:upstream_waves}. One can see disturbances emanating from the portion of the cylinder's surface where the flow has broken down to turbulence. Since the flow is subsonic behind the shock and within the entire flow field shown, these disturbances form wave fronts (magenta lines) in the shock layer that propagate \textit{upstream}, extending from the near wall region up to the bow shock. Underneath these upstream waves, the roughness-destabilised T-S waves propagate downstream as the cyan arrow illustrates. 

Absent a movie, the right portion of figure \ref{fig:upstream_waves} shows the temporal disturbance pressure signals along the boundary layer edge as a $\Theta-t$ diagram (similar to figure \ref{fig:lde_xt_diagrams}). The contours between $\Theta = [-15\degree,15\degree]$ illustrate an interference pattern of low amplitude wave fronts propagating upstream from both ends of the cylinder. Trajectories of slow acoustic waves originating from $\Theta<0$ and propagating towards $\Theta>0$ and those originating from $\Theta>0$ and propagating towards $\Theta<0$ at speeds denoted with $(U-a)^{+}$ and $(U-a)^{-}$, respectively, are shown with magenta lines. The agreement between the wave fronts and the magenta lines suggest these are indeed slow acoustic waves originating from the turbulent flow over the cylinder's rough wall. Also shown are the trajectories of the T-S waves with their much slower speed of roughly $U_{TS}\approx0.5 U_{e}$ (recall figure \ref{fig:lde_xt_diagrams}). It is not until about $|\Theta|=15\degree$ that these waves grow to sufficient amplitude to dominate the pressure signal, after which the agreement between the cyan line and their wave fronts can be appreciated.

We now recall the great work performed previously on the generation of Tollmien-Schlichting waves via acoustic waves interacting with local \citep{Ruban1984,Goldstein1985} and distributed \citep{zavolskii1983,Choudhari1993} small-scale surface inhomogeneities (e.g., roughness). For example, \citet{Choudhari1993} showed how 2D roughness with a narrow band of Fourier components can lead to effective generation of an unstable eigenmode through acoustic input. This result, done under the assumption of roughness heights much smaller than the boundary layer thickness to permit asymptotic analysis, was demonstrated for the case of deterministic and randomly distributed roughness. Despite the present roughness heights being larger than permitted by such an analysis, and the fact that the mean flow absent roughness is eigenmode stable, the present authors propose the following feedback mechanism as it occurred in the staggered and random roughness simulations (sides 1 and 2 arbitrarily denote positions on the cylinder above and below the stagnation line, respectively): 
\begin{enumerate}
    \item breakdown to turbulence over the rough wall on side 1 in the subsonic zone scatters acoustic waves, \label{feedback_step1}
    \item slow acoustic waves travel upstream to side 2 of the cylinder and excite roughness destabilised T-S waves at their neutral point, \label{feedback_step2}
    \item excited T-S waves grow to critical amplitude and transition the flow to turbulence on side 2 (also within subsonic zone), \label{feedback_step3}
    \item acoustic waves generated on side 2 propagate back upstream to side 1, \label{feedback_step4} 
    \item and the feedback loop repeats from step (\ref{feedback_step2}) to sustain LTT.
\end{enumerate}
From this proposed feedback mechanism, we suggest the following requirements for the process to occur in a real flight scenario:
\begin{enumerate}
    \item critical roughness Reynolds number achieved within subsonic zone for absolute instability, \label{feedback_req1}
    \item sufficient integrated growth of roughness-destabilised boundary layer modes (e.g., T-S waves) within subsonic zone, \label{feedback_req2}
    \item and roughness pattern contains wavenumbers close enough \citep{Choudhari1993} to disturbance spectrum for effective receptivity. \label{feedback_req3}
\end{enumerate}

Requirement (\ref{feedback_req1}) represents the need for scattered upstream acoustic waves to be present for the feedback loop to begin. These could be provided by an event like particulate impingement \cite{al_hasnine_russo_tumin_brehm_2023}, however. In a wind tunnel it could be a transient event related to start-up. Requirement (\ref{feedback_req2}) is supported by the fact that the aligned roughness simulation, with its weakly destabilised boundary layer modes, remained laminar everywhere until the external forcing was introduced. Finally, requirement (\ref{feedback_req3}) is only assumed given the available literature \citep{zavolskii1983,Choudhari1993}. Our proposed mechanism could help explain the sonic flow requirement in the PANT correlation (see eq. \ref{eq:PANT-AW}) Also, the feedback loop could render the LTT process independent of other disturbance sources provided the initial amplitudes supplied by the turbulence-generated acoustic waves are largest. Rigorous proof of the proposed feedback mechanism was unfortunately out of reach given the computational resources available and is left for future work.

\section{Conclusion} \label{sec:conclusion}
Despite numerous previous works showing very suboptimal amplification ratios when fully resolving roughness elements through DNS \citep{Choudhari2005,Schilden2020}, the optimal transient growth theory remains the current explanation for the so-called `blunt body paradox'. The present work sought to provide new physical understanding to this problem of roughness-induced transition to turbulence over hypersonic blunt bodies in the presence of distributed roughness. To this end, the first direct numerical simulation of a cylinder in Mach 6 free stream with roughness elements distributed over the entire surface were performed. Fixing the roughness element height and varying the relative phasing between streamwise rows allowed for a systematic comparison of the distinct LTT scenarios possible within this narrow subspace of roughness parameters.

An analysis of the time-averaged, laminar flow preceding LTT in each case focused on, in addition to other things, the generation and development of streamwise velocity streaks. The simulations demonstrated the lift-up effect at work in generating the streaks, however, the close proximity of roughness elements resulted in non-linear development governed entirely by the roughness pattern. This combined with the low (less than factor two) relative gain in streak amplitude highlights the lack of utility of optimal transient growth theory for predicting the actual development of roughness-induced steady perturbations. The success of the theory in correlating the available data, in the authors' view, stems from the assumption of roughness-induced perturbation velocities being proportional to $k/\theta$, which lacks physical justification and is inconsistent with the optimal input to transient growth being in the form of streamwise aligned vortices with null streamwise velocity perturbation.

On the other hand, unsteady analysis of the DNS found disturbance frequency bands undergoing tremendous growth and saturating near the location of sharp rise in time-averaged heat flux at the surface. Regardless of the roughness configuration, saturation occurred where $U^{'}_{rms,max}$ reached $20\%U_{e}$, which may be of use as a breakdown criteria in an amplitude-based LTT prediction methodology \citep{Mack1977,Marineau2017,Fedorov2022}. The combination of the DNS and LDE pulse simulations on the time-averaged flow fields exposed the convective, modal nature of the disturbances causing LTT in each case. Streak modes and streaky T-S waves were found to dominate instability growth, the relative importance of which was governed by the roughness pattern. Consistent with previous work on streaky flat plate boundary layers \citep{Cossu_2004}, the aligned roughness elements with their strongest streaks saw weak destabilisation of T-S waves and dominance of the fundamental sinuous streak modes. The staggered and randomly phased roughness configurations were remarkably similar in their high degree of T-S destabilisation and growth. All cases saw the steady vorticity introduced by the roughness pattern directly seeding hairpin vortices once the most dominant instability reached critical amplitude. We thus demonstrated how distributed roughness elements play a complex role in LTT, namely serving to destabilise exponentially growing modal instabilities while also providing the initial vorticity needed in the late stages to fully transition the flow to turbulence.

Exposure of the rich set of roughness-induced boundary layer instabilities shown here, which required DNS of fully resolved roughness elements, may initiate a fundamental shift in the thinking surrounding the types of disturbances causing LTT on hypersonic blunt bodies. We also hope it motivates future efforts to study whether these instabilities persist on axisymmetric bodies and for more realistic roughness topologies through high-fidelity simulation. Again, we note how simply changing the relative phasing of roughness element rows resulted in dramatic changes to the stability characteristics of the laminar flow, which highlights the extreme sensitivity of the problem. Nonetheless, results from the randomly phased roughness arrangement fell between the aligned and staggered patterns for many quantities of interest. So, although sensitive, the behaviour of even more complicated roughness patterns may be able to be bounded and understood by analysing more simple topologies, as was demonstrated here.

\backsection[Acknowledgements]{Resources supporting this work were provided by the NASA High-End Computing (HEC) programme through the NASA Advanced Supercomputing (NAS) Division at Ames Research Center. In addition, the authors acknowledge the University of Maryland supercomputing resources (https://hpcc.umd.edu) made available for conducting the research reported in this paper. We also appreciate helpful discussions with Bijaylakshmi Saikia.}

\backsection[Funding]{This work has been supported under a NASA Space Technology Research Institute Award (ACCESS, grant number
80NSSC21K1117). The authors would also like to thank the National Science Foundation (NSF) for supporting part of this work
under CBET-2146100 with Dr. R. Joslin as program manager.}

\backsection[Declaration of interests]{The authors report no conflict of interest.}

\backsection[Author ORCIDs]{S. Dungan, https://orcid.org/0009-0005-6286-241X; M. Braga, https://orcid.org/0000-0002-5060-7295; R. Macdonald, https://orcid.org/0000-0001-5678-7736; C. Brehm, https://orcid.org/0000-0002-9006-3587;}

\begin{appendix}
\section{Transition Correlation Definitions}\label{appA}
The first set of correlations to consider are the various PAssive Nosetip Technology (PANT) rough-wall transition criteria, first outlined by \cite{Abbett75} and \cite{Wool75}. The PANT method was developed as a correlation of sand-grain rough-wall blunt body data taken in hypersonic wind tunnels at free-stream Mach number of $M_\infty=5$. The formulation of the correlation involves an amplification parameter ($\Rey_\theta$), as a function of a disturbance parameter $\left(\frac{k}{\psi\theta}\right.$, $\left.\psi = \frac{T_w}{T_e}\right)$, where $\Rey_\theta$ is the momentum thickness Reynolds number, $\theta$ is the momentum thickness, $M_e$ is boundary layer edge Mach number, $T_e$ is boundary layer edge temperature, $T_w$ is wall temperature, and $k$ is roughness peak-to-valley height. From \cite{Abbett75} and \cite{Wool75} the form of the correlation is as follows:

\begin{equation}
    \Rey_\theta\left(\frac{k}{\psi\theta}\right)^{0.7} = \Rey_\theta\left(\frac{T_ek}{T_w\theta}\right)^{0.7} = \begin{cases}
    255 \text{ by } M_e=1, \text{ onset} & \\
    215, \text{ location}
    \end{cases}
    \label{eq:PANT-AW}
\end{equation}

Equation \ref{eq:PANT-AW} indicates that the transition parameter must exceed 255 at the sonic point before the transition location is determined as the position where the parameter equals 215. If 255 is not reached by the time $M_e=1$, then there is no transition. The correlation in Eq. \ref{eq:PANT-AW} is denoted as \textit{PANT-AW}. \cite{reda1981} updated the PANT correlation based on ballistic range experiments, finding a different coefficient and exponent from the original data fit. The Reda updated PANT correlation from eq. \ref{eq:PANT-R} is denoted \textit{PANT-R}.

\begin{equation}
    \Rey_\theta\left(\frac{T_ek}{T_w\theta}\right)^{1.30} = 574
    \label{eq:PANT-R}
\end{equation}

For self-similar, axisymmetric, stagnation point flow, assuming parallel flow and roughness-induced disturbance velocities proportional to the roughness height, \cite{Reshotko2004} used transient growth calculations to get a similar PANT based correlation, \textit{PANT-RT}:
\begin{align}
    &\Rey_{\theta,tr} = 180\left(\frac{k}{\theta}\right)^{-1}\left(\frac{T_e}{2T_w}\right)^{-1.27}\\
    &\;\;\;\;\;\;\;\;\;\;\;\;\;\;\;\;\;\;\;\;\;\;\;\;\;\;\;\text{or}\nonumber\\
    &\Rey_{\theta}\left(\frac{k}{\theta}\right)\left(\frac{T_e}{T_w}\right)^{1.27} = \frac{180}{2^{-1.27}}\approx434
    \label{eq:PANT-RT}
\end{align}

Another class of correlation separate from the PANT variants are (roughness) Reynolds number correlations that do not explicitly depend on the edge-wall temperature ratio. Based on models of hemispherical and blunt large-angle conical geometries ﬂown in the NASA Ames hypersonic ballistic range, \cite{Reda2008} lists transition correlations for all three roughness regimes potentially encountered by an ablating blunt-body heat shield undergoing atmospheric entry. The first is the smooth-wall asymptote for the quiet-flow regime:

\begin{equation}
    \left[\frac{\rho_eu_e\theta}{\mu_w}\right]_{\text{tr}}=500 \pm20\%\;\;\;\text{ when}\; \left[\frac{\rho_ku_k\overline{k}}{\rho_eu_e\theta}\right]\leq0.5
\end{equation}

The next is the critical-roughness Reynolds number regime:
\begin{equation}
    \left[\frac{\rho_ku_k\overline{k}}{\mu_w}\right]_{\text{tr}}=250 \pm20\%\;\;\;\text{ when}\; 0.5\leq\left[\frac{\rho_ku_k\overline{k}}{\rho_eu_e\theta}\right]\leq2.5
    \label{eq:Rek-CR}
\end{equation}

Finally, the large-roughness/low-Reynolds-number asymptote regime:
\begin{equation}
    \left[\frac{\rho_eu_e\theta}{\mu_w}\right]_{\text{tr}}=100 \pm20\%\;\;\;\text{ when}\; \left[\frac{\rho_ku_k\overline{k}}{\rho_eu_e\theta}\right]\geq2.5
    \label{eq:Retheta-LR}
\end{equation}

$\overline{k}$ is average roughness height, $\theta$ is laminar boundary layer momentum thickness, $\rho_e$ is the boundary layer edge density, $u_e$ is the boundary layer edge velocity, $\rho_k$ is the density evaluated at the roughness height, $u_k$ is the streamwise velocity at the roughness height, and $\mu_w$ is the dynamic viscosity at the wall. The focus is for rough-wall conditions, so the two roughness conditions are considered and denoted as $\Rey_{k}$-CR and $\Rey_{\theta}$-LR, for the critical-roughness and large-roughness correlations respectively.

Hollis \citep{hollis12,hollis17_NASATM,hollis19,hollis21,hollis25} performed a number of experiments at the 20-Inch Mach 6 Air and 31-Inch Mach 10 Air Tunnels of the NASA Langley Aerothermodynamic Laboratories (LAL) for hemisphere and sphere-cone geometries with sand-grain and patterned distributed surface roughness. That body of work led to updated correlations (similar to the original PANT correlations) for both sand-grain roughness and hexcomb pattern roughness. The correlation originally for sand-grain roughness, denoted as \textit{SAND}, is as follows:
\begin{equation}
    \Rey_\theta\left[\left(\frac{T_ek}{T_w\theta}\right)\left(\frac{H_e}{H_k}\right)^{-1}M_e^{-0.5}\right]^{0.5} = 165
    \label{eq:SAND}
\end{equation}

The hexcomb pattern roughness correlation, \textit{HEX-PAT}, necessitated an extra pressure gradient component via the Pohlhausen parameter:
\begin{align}
\lambda_{\text{POHL}} &= \frac{dU_e}{ds}\frac{\delta^2}{\nu_e}\\
\beta_{\text{POHL}} &= \left\{10^{\left[\min\left(10,\max\left(-10,\lambda_{\text{POHL}}\right)\right)\right]/10}\right\}^{-1}
\end{align}
\begin{equation}
    \Rey_\theta\left[\left(\frac{T_ek}{T_w\theta}\right)^{0.45}\left(\frac{H_e}{H_k}\right)^{-1.8}M_e^{-0.6}\beta_{\text{POHL}}^{-0.5}\right]^{0.6299} = 171.4
    \label{eq:HEX-PAT}
\end{equation}

where
\begin{equation}
    k = \begin{cases}
        k_{\text{PV50}} &\text{sand-grain}\\
        h_{\text{hex}} &\text{hexcomb pattern }
    \end{cases}
\end{equation}

$U_e$ is the boundary layer edge velocity magnitude, $s$ is the streamline distance from stagnation region along the edge, $\delta$ is the boundary layer height, $\nu_e$ is the edge kinematic viscosity, $\lambda_{\text{POHL}}$ is the Pohlhausen parameter, $\beta_{\text{POHL}}$ is the modified Pohlhausen parameter, $\Rey_\theta$ is the momentum thickness Reynolds number, $\theta$ is the momentum thickness, $M_e$ is boundary layer edge Mach number, $T_e$ is boundary layer edge temperature, $T_w$ is wall temperature, $H_e$ is the boundary-layer edge total enthalpy, $H_k$ is the total enthalpy at the roughness height (for the present work $H_k$ is defined based on the wall temperature and velocity magnitude at the roughness height $H_k=c_pT_w+0.5U_k^2$), $k_{\text{PV50}}$ is the 50\% exceedance value of the actual peak-to-valley roughness heights, and $h_{\text{hex}}$ is the hexcomb cell height. For the present work, the sinusoidal roughness is based on the 80 mesh \citep{hollis17_NASATM}, according to \citep{hollis25} the corrected $k_{\text{PV50}}$ for the 80 mesh is $k_{\text{PV50}}=0.0668$ mm.

For consistency and comparison across all correlations, velocities are considered to be velocity magnitudes, and unless otherwise specified the roughness heights, $k$, are taken as the peak-to-valley height. In summary, the various correlations (with results tabulated in section \ref{sec:mean_flow}) are:
\begin{itemize}
    \item PANT-AW \citep{Abbett75,Wool75} 
    \item PANT-R \citep{reda1981}
    \item PANT-RT \citep{Reshotko2004}
    \item $\Rey_k$-CR \citep{Reda2008}
    \item $\Rey_\theta$-LR \citep{Reda2008}
    \item SAND \citep{hollis25}
    \item HEX-PAT \citep{hollis25}
\end{itemize}

\section{Grid Resolution Study}\label{appB}

\begin{figure}
  \centering
  \subfloat[]{%
    \includegraphics[width=0.32\textwidth]{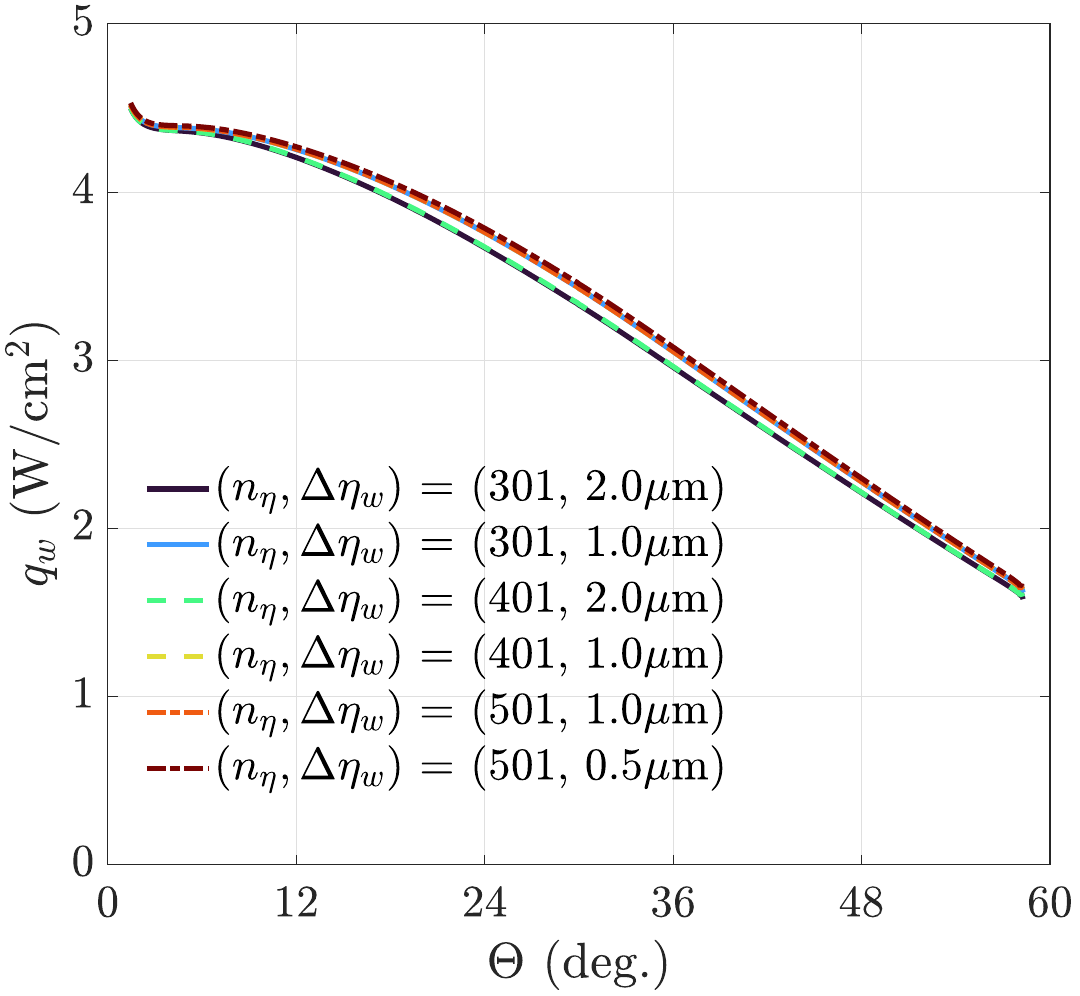}}
  \hfill
  \subfloat[]{%
    \includegraphics[width=0.32\textwidth]{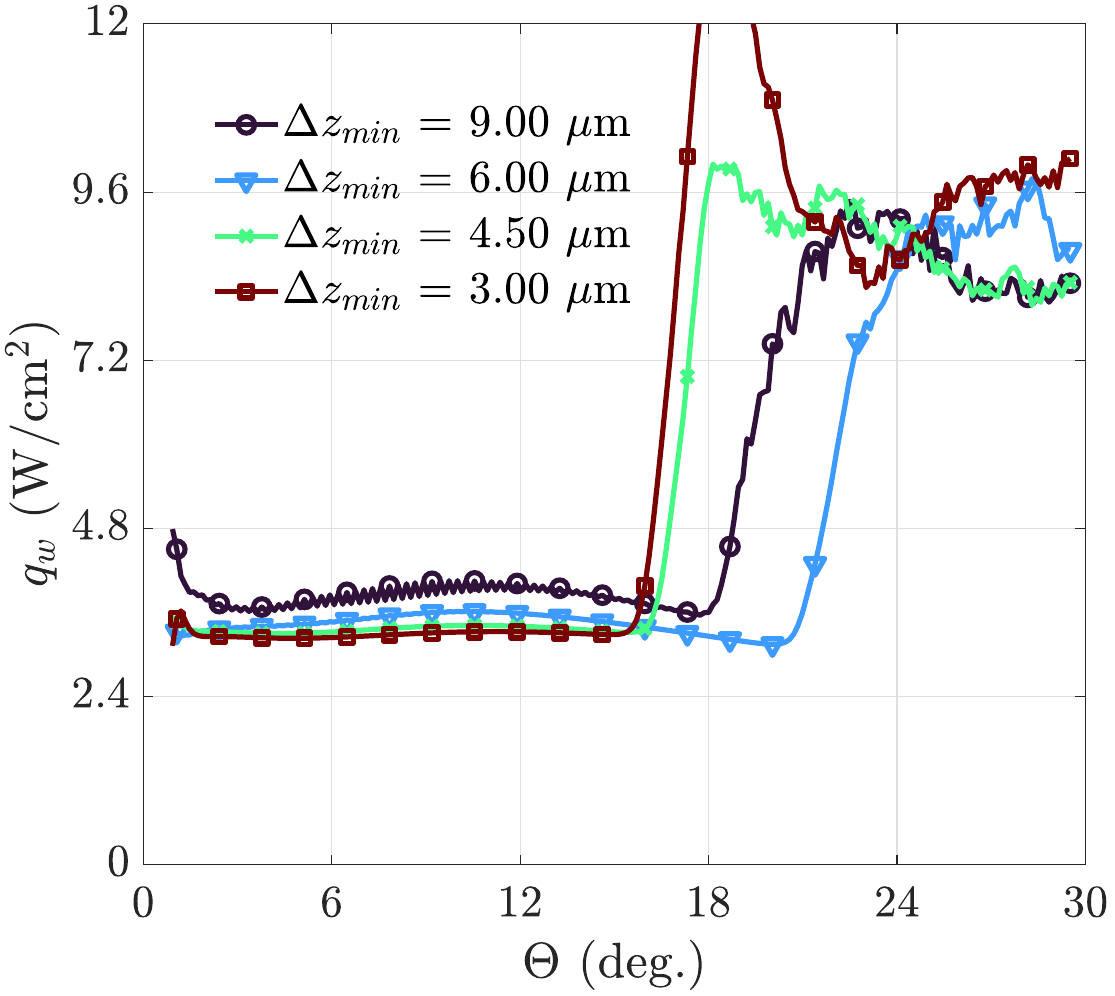}}
  \hfill
  \subfloat[]{%
    \includegraphics[width=0.32\textwidth]{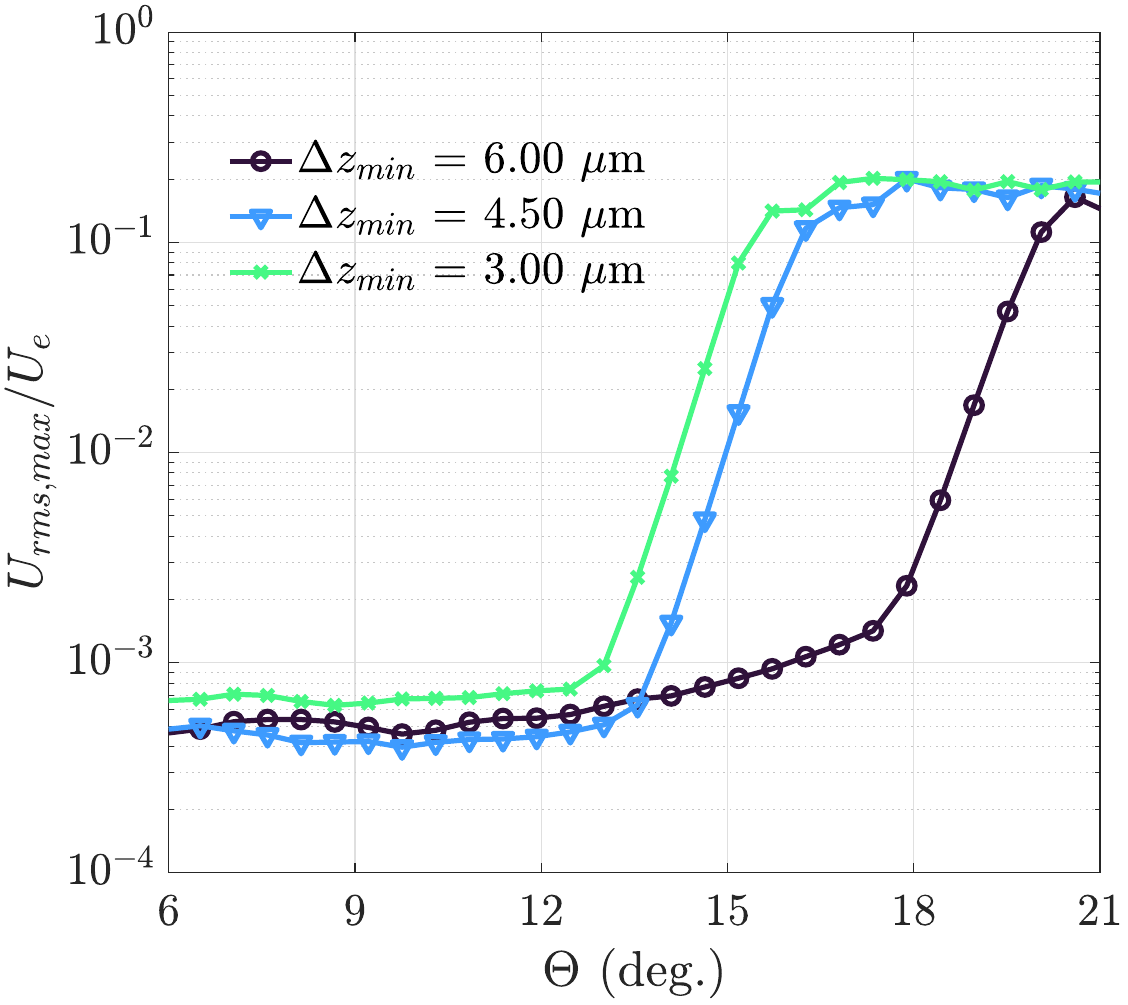}}
  \caption{Grid resolution study of surface heat flux for (a) variable wall-normal grid count and initial wall-normal grid spacing in 2D and (b) spanwise minimum grid spacing in 3-D. Sensitivity of streamwise velocity perturbation RMS to spanwise minimum grid spacing in 3-D is provided in (c).}
  \label{fig:grid_convergence_plots}
\end{figure}

A grid resolution study was first performed in 2D to ensure adequate resolution in the wall-normal direction. The number of nodes in the streamwise direction was held fixed at $n_{\xi}=601$, while the number of points in wall-normal direction, $n_{\eta}$, and grid spacing at the wall, $\Delta \eta_{w}$ were varied. These were varied such that the resolution at the shock remained constant, while the number of points in the shock layer and boundary layer increased with $n_{\eta}$. As can be seen in figure \ref{fig:grid_convergence_plots}(a), the heat flux at the wall (a sensitive quantity to grid resolution) is independent of increases to $n_{\eta}$ while $\Delta \eta_{w}$ is held fixed. There is a slight increase in heat flux when $\Delta \eta_{w}$ decreases from 2 $\mu$m to 1 $\mu$m. The relative difference is at maximum ($\Theta=37.4\degree)$ only $3.0\%$. Therefore, the $n_{\eta}=301$ and $\Delta \eta_{w}$ wall-normal dimensions were utilised for the DNS as the larger minimum spacing at the wall allowed for twice the physical simulation time than if $\Delta \eta_{w}$=1 $\mu$m were used.

Since the staggered roughness elements arrangement resulted in the sharpest spanwise gradients (recall vorticity fields in figure \ref{fig:mean_vorticity}), this configuration was subjected to a grid resolution study in that direction. The grid cell count in the streamwise and wall-normal directions were held fixed at $(n_{\xi},n_{\eta})=(\num{10400},\num{300})$, while the spanwise extent and number of cells were varied to provide increasing spanwise resolution through decreasing minimum spacings. All grids simulated 3 roughness wavelengths in the spanwise direction except the finest case which resulted in $\Delta z_{min}=3.0$ $\mu$m. This grid simulated half a roughness element with symmetry conditions imposed at the spanwise boundaries. This allowed for not only increased resolution but also to see if the imposed symmetry prevented breakdown to turbulence, which it did not. Nonetheless, the relative error in the steady, laminar heat flux obtained from using $\Delta z_{min}=4.5$ $\mu$m (corresponding to $n_z=120$) rather than $3.0$ $\mu$m resulted in, at most ($\Theta=8.9\degree$) a $2.9\%$ relative error (see figure \ref{fig:grid_convergence_plots}(b)). This was deemed sufficient for the present study. In addition, the resolution effect on the dominant instabilities present for each grid is shown in figure \ref{fig:grid_convergence_plots}(c) through the maximum (in the $\eta-z$ plane) in streamwise velocity RMS. Besides a constant offset due to slightly different levels of fluctuations upstream of peak instability growth, the $\Delta z_{min}=4.5$ $\mu$m and 3.0 $\mu$m grids exhibit virtually identical growth rates, thereby confirming the present results are essentially grid independent. For this reason, the grid dimensions listed in section \ref{subsec:num_approach} were utilised for all simulations presented, namely $(n_{\xi},n_{\eta},n_z) = (\num{10400},\num{300},\num{120})$ cells in the streamwise, wall-normal, and spanwise directions, respectively.

\section{Stagnation Flow Similarity Transformation} \label{appC}

For comparison, the velocity and temperature profiles are transformed making use of the Illingworth transformation~\citep{Illingworth50}. The streamwise and wall-normal similarity coordinates $\xi_{\mathrm{sim}}$ and $\eta_{\mathrm{sim}}$ are as follows:
\begin{equation}
    \xi_{\mathrm{sim}}(x) = \int_0^x\rho_e(x)U_e(x)\mu_e(x)dx \;\;\;\;\;\;\;\;\;\; \text{and} \;\;\;\;\;\;\;\;\;\; \eta_{\mathrm{sim}}(x,y) = \frac{U_e}{\sqrt{2\xi_{\mathrm{sim}}}}\int_0^y\rho dy
    \label{eq:illingworth-transformation}
\end{equation}
The similarity variables $f$ and $g$ are defined as follows:
\begin{equation}
    f'=\frac{u}{U_e} \;\;\;\;\;\;\;\;\;\; \text{and} \;\;\;\;\;\;\;\;\;\; g=\frac{T}{T_e}
\end{equation}

After transformation, the velocity and temperature profiles corresponding to streamwise locations upstream of the transition point all collapse to the similarity profile for plane stagnation flow. The plane stagnation flow is found by solving the compressible Falkner-Skan-type similarity solution, again making use of the Illingworth transformation.
\begin{align}
    \left(Cf''\right)'+ff''&=\beta\left(f'^2-g\right)\label{eq:momt-fs-ss}\\
    \left(Cg'\right)'+Prfg'&=-PrC(\gamma-1)M_e^2f''^2\label{eq:nrg-fs-ss}
\end{align}

Differentiation with respect to $\eta_{\mathrm{sim}}$ is denoted by $\cdot'$. The parameters $\beta$ and $C$ are the pressure gradient parameter and the Chapman-Rubesin parameter respectively defined as follows:
\begin{equation}
\beta = \frac{2\xi_{\mathrm{sim}}}{M_e}\frac{dM_e}{d\xi_{\mathrm{sim}}}
\label{eq:beta-pressure-gradient}
\end{equation}
\begin{equation}
    C = \frac{\rho\mu}{\rho_e\mu_e}
    \label{eq:chapman-rubesin}
\end{equation}
Thermodynamic properties are related using the ideal gas equation of state and $Pr$ denotes the Prandtl number which is set to $\Pran = 0.71$. For the cylinder, the pressure gradient parameter is set to $\beta=1$ to solve for plane stagnation flow. Edge conditions that vary along the streamwise direction are obtained from the following procedure:
\begin{itemize}
\item Along stagnation line:
    \begin{enumerate}
        \item Use the normal-shock relations to get the post-shock pressure, density, temperature, and Mach number.
        \begin{align}
            \frac{T_1}{T_\infty}&=\frac{[2\gamma M_\infty^2-(\gamma-1)][(\gamma-1)M_\infty^2+2]}{(\gamma+1)^2M_\infty^2}\\
            \frac{\rho_1}{\rho_\infty}&=\frac{(\gamma+1)M_\infty^2}{(\gamma-1)M_\infty^2+2}\\
            \frac{P_1}{P_\infty}&=\frac{2\gamma M_\infty^2-(\gamma-1)}{\gamma+1}\\
            M_1^2&=\frac{(\gamma-1)M_\infty^2+2}{2\gamma M_\infty^2-(\gamma-1)}
        \end{align}
        \item Use the conservation of total enthalpy to take the post-shock temperature (where the Mach number is non-zero, $U_1^2=M_1^2\gamma R T_1$) to the edge temperature (where the Mach number is zero). 
        \begin{equation}
            c_pT_1+\frac{U_1^2}{2} = c_pT_2+\cancelto{0}{\frac{U_2^2}{2}}
            \label{eq:conserve-tot-enthalpy}
        \end{equation}
        \item Once the post-shock and edge temperatures are known, use the isentropic relations to get the edge pressure and density based on the post-shock conditions.
        \begin{equation}
            \left(\frac{P_2}{P_1}\right)=\left(\frac{T_2}{T_1}\right)^{\frac{\gamma}{\gamma-1}};\;\;\;\frac{\rho_2}{\rho_1}=\left(\frac{T_2}{T_1}\right)^{\frac{1}{\gamma-1}}
            \label{eq:isentropic-relations}
        \end{equation}
    \end{enumerate}
\item Downstream, along boundary layer edge:
    \begin{enumerate}
        \item With the edge pressure known at the stagnation point, Modified Newtonian Theory is used to get the pressure variation downstream as a function of the local inclination of the surface to the flow. The equation is $P_e = P_{e,max}\sin^2\phi$, where $P_{e,max}$ is the static pressure at the boundary layer edge along the stagnation line. The local inclination, $\phi$, is given from the streamwise location, angle from the stagnation line $\Theta$, as $\phi = \frac{\pi}{2}-\Theta$.
        \item Again use the isentropic relations, this time along the boundary layer edge, to get the streamwise variation in density and temperature. The streamwise variation in viscosity can be found using any temperature dependent viscosity model such as Sutherland's Law or the Power Law. 
        \item Finally, conserving total enthalpy along the boundary layer edge, the edge velocity is given by $U_e=\sqrt{2c_p(T_{e,max}-T_e)}$, where $T_{e,max}$ is the static temperature at the boundary layer edge along the stagnation line. The edge Mach number is subsequently found from $U_e/\sqrt{\gamma RT_e}$.
    \end{enumerate}
\end{itemize}

\section{Local Linear Stability Analysis}\label{appLST}
\begin{figure}
  \centering
  \subfloat[Staggered]{%
    \includegraphics[width=0.32\textwidth]{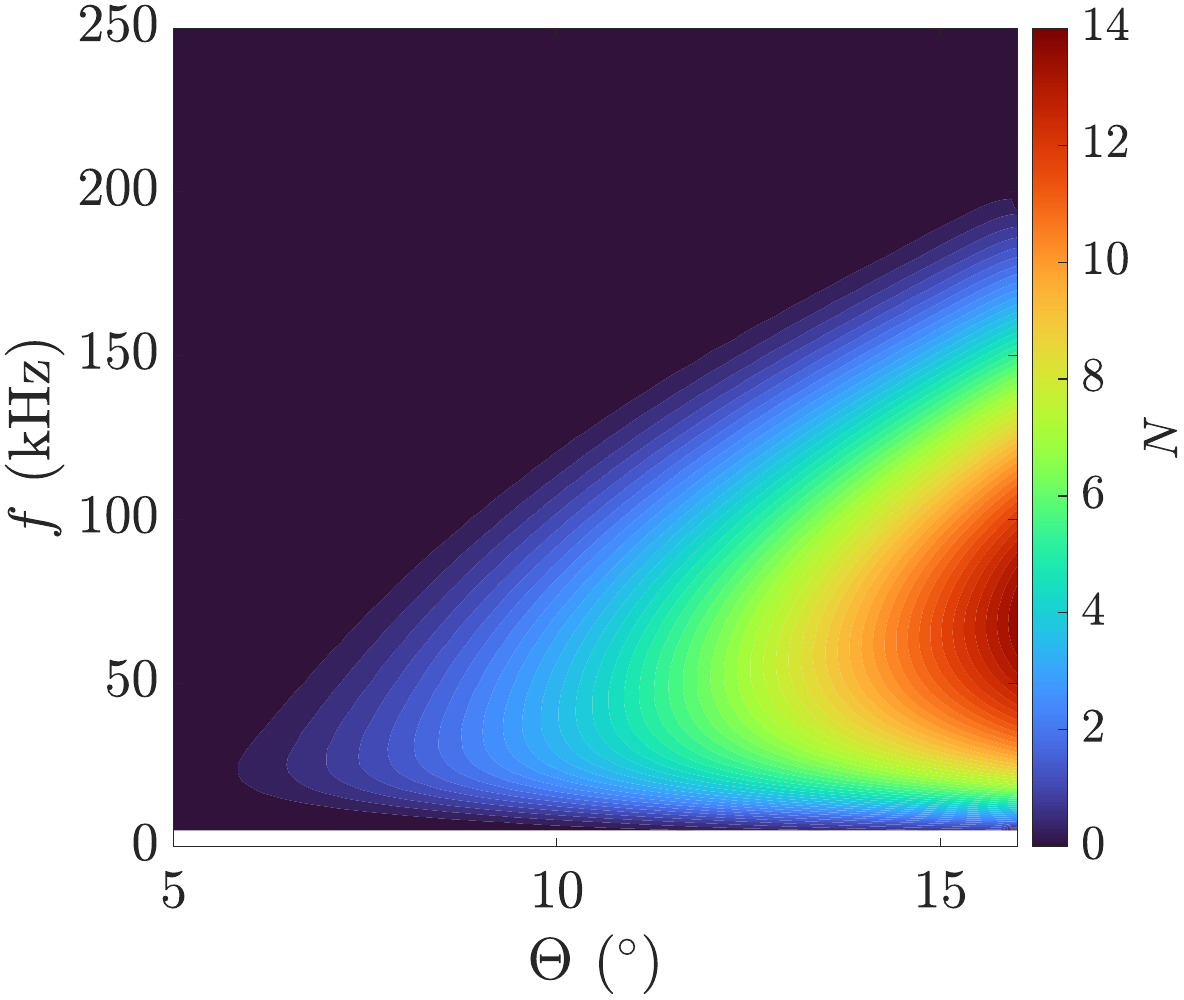}}
  \hfill
  \subfloat[Random]{%
    \includegraphics[width=0.32\textwidth]{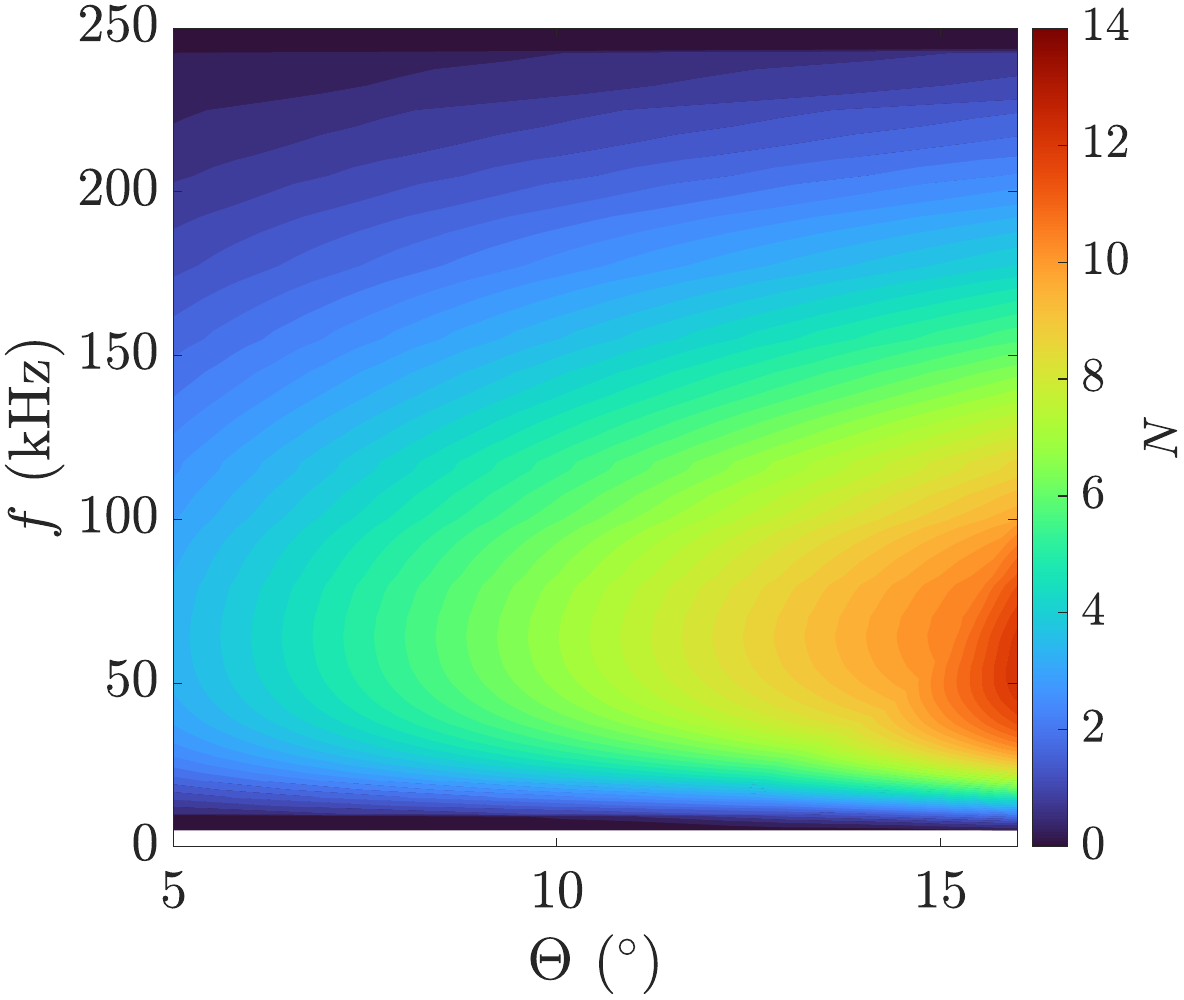}}
  \hfill
  \subfloat[Aligned]{%
    \includegraphics[width=0.32\textwidth]{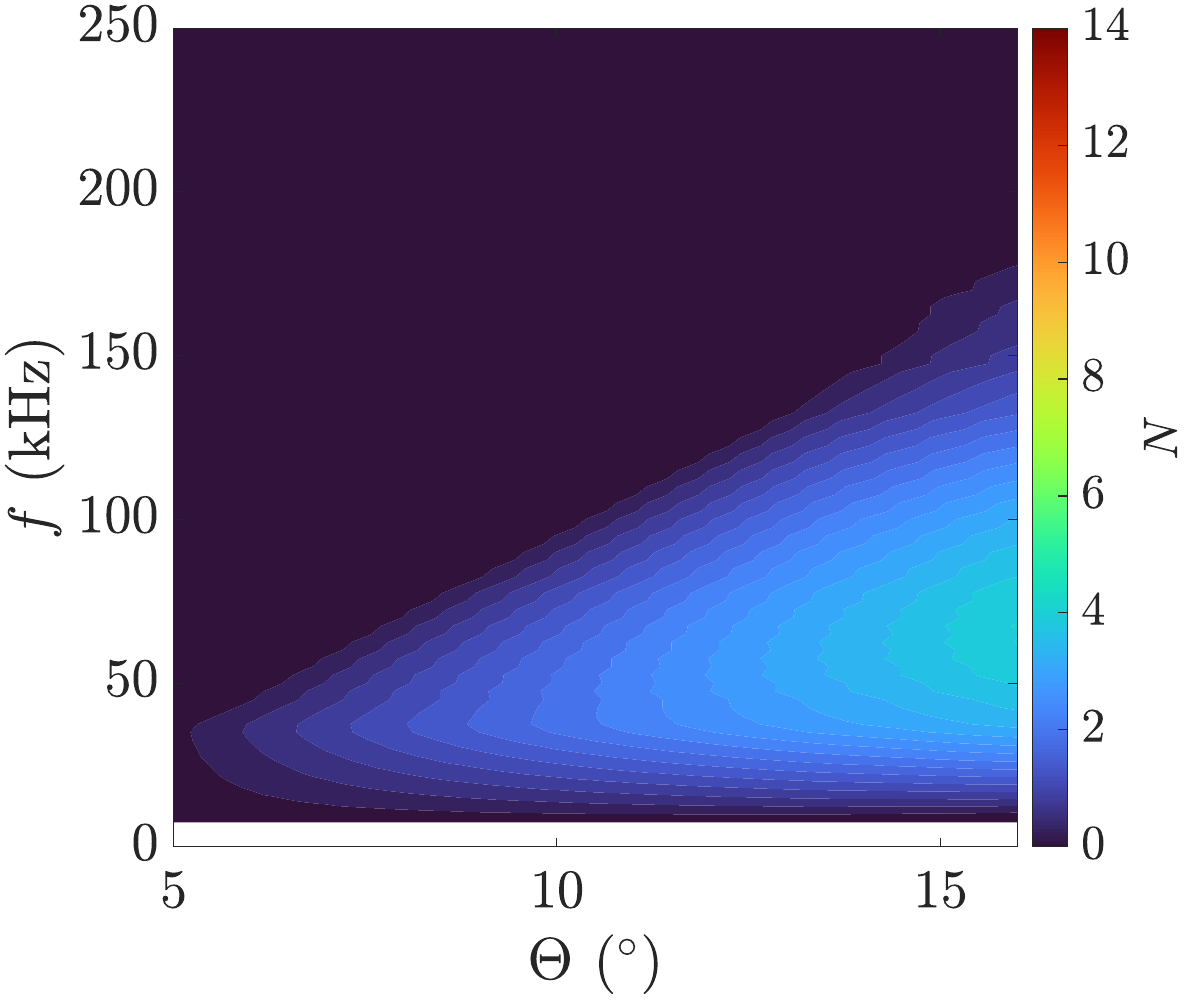}}
  \caption{$N$-factor contours obtained with local linear stability analysis on time- and roughness-averaged flow profiles using LASTRAC \citep{lastrac_manual}.}
  \label{fig:LST_N_comparison}
\end{figure}

Local linear stability theory was found to provide good agreement in terms of amplitude growth of T-S waves for the roughness cases when compared to the DNS and LDE simulation results. The local LST analysis was performed on the time- and wavelength-averaged flow profiles. Since the disturbance wavelengths are several times larger than the roughness wavelength (see figure \ref{fig:lde_wavelengths}), one can think about the LST analysis capturing the averaged effects of the roughness on the boundary layer. Of course, the local linear analysis with its null streamwise mean flow gradient assumption is strictly inappropriate for the case of finite height roughness. On the other hand, it is attractive to find ways to make use of the analysis given its inexpensive nature. We include N-factor contours for each roughness case in figure \ref{fig:LST_N_comparison} to show how the analysis is able to capture the trends of destabilisation of T-S waves being most pronounced for staggered, random, then aligned roughness elements.

\end{appendix}\clearpage

\bibliographystyle{jfm}
\bibliography{jfm}

\end{document}